\documentclass[plaindraft,letter]{jpp-AAS}

\usepackage{graphicx,amsmath,amssymb}
\usepackage{natbib}
\usepackage[pdftex,colorlinks,citecolor=blue]{hyperref}
\usepackage{cleveref}
\usepackage{float}
\usepackage{subcaption}
\usepackage{enumitem}
\usepackage[percent]{overpic}
\usepackage{xcolor}

\def\apjs{Astrophys.~J.~Supp.~Ser.}
\def\apj{Astrophys.~J.}
\def\apjl{Astrophys.~J.~Lett.}

\def\araa{Ann.~Rev.~Astron.~Astrophys.}
\def\mnras{Mon.~Not.~Roy.~Astron.~Soc.}

\def\pr{Phys.~Rev.}

\def\prx{Phys.~Rev.~X}
\def\prl{Phys.~Rev.~Lett.}
\def\jgr{J.~Geophys.~Res.}

\def\prx{Phys.~Rev.~X}

\def\physrep{Phys.~Rep.}
\def\pop{Phys.~Plasmas}
\def\pof{Phys.~Fluids}

\def\pnas{Proc.~Nat.~Acad.~Sci.}
\def\jcp{J.~Comput.~Phys.}
\def\jpp{J.~Plasma Phys.}

\def\ropp{Rev.~Plasma Phys.}

\def\njp{New J.~Phys.}
\ifCUPmtlplainloaded \else
  \checkfont{eurm10}
  \iffontfound
    \IfFileExists{upmath.sty}
      {\typeout{^^JFound AMS Euler Roman fonts on the system,
                   using the 'upmath' package.^^J}%
       \usepackage{upmath}}
      {\typeout{^^JFound AMS Euler Roman fonts on the system, but you
                   dont seem to have the}%
       \typeout{'upmath' package installed. JPP.cls can take advantage
                 of these fonts, if you use 'upmath' package.^^J}%
       \providecommand\upi{\pi}%
      }
  \else
    \providecommand\upi{\pi}%
  \fi
\fi

\ifCUPmtlplainloaded \else
  \checkfont{msam10}
  \iffontfound
    \IfFileExists{amssymb.sty}
      {\typeout{^^JFound AMS Symbol fonts on the system, using the
                'amssymb' package.^^J}%
       \usepackage{amssymb}%
         \let\leq=\leqslant
         
      }{}
  \fi
\fi

\ifCUPmtlplainloaded \else
  \IfFileExists{amsbsy.sty}
    {\typeout{^^JFound the 'amsbsy' package on the system, using it.^^J}%
     \usepackage{amsbsy}}
    {\providecommand\boldsymbol[1]{\mbox{\boldmath $##1$}}}
\fi

\defcitealias{kswhite19}{Kunz, Squire {\it et al.} 2019}

\newcommand{\pD}[2]{\frac{\partial #2}{\partial #1}}
\newcommand{\D}[2]{\frac{{\rm d} #2}{{\rm d} #1}}
\newcommand\bb[1]{\mbox{\boldmath{$#1$}}}
\newcommand\bs[1]{\boldsymbol{#1}}
\newcommand{\msb}[1]{\mathsfbi{#1}}

\newcommand{\rmd}{{\rm d}}

\renewcommand\bcdot{\,\bb{\cdot}\,}
\newcommand\btimes{\,\bb{\times}\,}
\newcommand\bdbldot{\,\bb{:}\,}
\newcommand\grad{\bb{\nabla}}

\newcommand{\ez}{\hat{\bb{z}}}
\newcommand{\eb}{\hat{\bb{b}}}

\newcommand{\ri}{r_{\rm L,i}}

\title[Collisionless, high-$\beta$ turbulence]{Self-organization in collisionless, \\ high-$\bb{\beta}$ turbulence}

\author[S.~Majeski, M.~W.~Kunz and J.~Squire]%
{S.~Majeski\ls$^{1,2}$%
  \thanks{Email address for correspondence: smajeski@princeton.edu}, 
M.~W.~Kunz$^{1,2}$, and J.~Squire$^{3}$}

\affiliation{$^1$Department of Astrophysical Sciences, Princeton University, Peyton Hall, Princeton, NJ 08544, USA\\[\affilskip]
$^2$Princeton Plasma Physics Laboratory, PO Box 451, Princeton, NJ 08543, USA\\[\affilskip]
$^3$Department of Physics, University of Otago, 730 Cumberland St, North Dunedin, Dunedin 9016, New Zealand\\[\affilskip]}

\pubyear{2024}
\volume{}
\pagerange{}
\date{\today}

\begin{document}

\maketitle

\begin{abstract}
The MHD equations, as a collisional fluid model that remains in local thermodynamic equilibrium (LTE), have long been used to describe turbulence in myriad space and astrophysical plasmas. Yet, the vast majority of these plasmas, from the solar wind to the intracluster medium (ICM) of galaxy clusters, are only weakly collisional at best, meaning that significant deviations from LTE are not only possible but common. Recent studies have demonstrated that the kinetic physics inherent to this weakly collisional regime can fundamentally transform the evolution of such plasmas across a wide range of scales. Here we explore the consequences of pressure anisotropy and Larmor-scale instabilities for collisionless, $\beta\gg 1$ turbulence, focusing on the role of a self-organizational effect known as `magneto-immutability'. We describe this self-organization analytically through a high-$\beta$, reduced ordering of the CGL-MHD equations, finding that it is a robust inertial-range effect that dynamically suppresses magnetic-field-strength fluctuations, anisotropic-pressure stresses, and dissipation due to heat fluxes. As a result, the turbulent cascade of Alfv\'enic fluctuations continues below the putative viscous scale to form a robust, nearly conservative, MHD-like inertial range. These findings are confirmed numerically via Landau-fluid CGL-MHD turbulence simulations that employ a collisional closure to mimic the effects of microinstabilities. We find that microinstabilities occupy a small (${\sim}5\%$) volume-filling fraction of the plasma, even when the pressure anisotropy is driven strongly towards its instability thresholds. We discuss these results in the context of recent predictions for ion-versus-electron heating in low-luminosity accretion flows and observations implying suppressed viscosity in ICM turbulence.
\end{abstract}


\section{Introduction}

\subsection{Motivation}

Rarely can a problem of astrophysical fluid dynamics be approached without any consideration for the effects of turbulence. In fact, many descriptions of fundamental astrophysical phenomena, such as transport in accretion flows and dynamo amplification of cosmic magnetic fields, are intrinsically reliant upon it. As a result, an abundance of literature exists analyzing the role of turbulence in environments from the solar wind to the intracluster medium (ICM) of galaxy clusters \citep[e.g.,][]{goldstein95,brandenburg05,schekochihin04}. Yet many of these studies employ theoretical or numerical methods founded upon the assumption that the plasma that pervades these systems is collisional. X-ray observations of the hot and dilute ICM in the Perseus and Coma clusters suggest otherwise, with implied Coulomb-collisional mean free paths typically only ${\sim}0.1$--$0.01$ times that of the large-scale gradients~\citep[e.g.,][]{kjz22}. Meanwhile, in-situ measurements of the plasma comprising the solar wind have long revealed that ion Coulomb mean free paths can reach nearly 1~au, allowing significant deviations from local thermodynamic equilibrium (LTE)~\citep{marsch06}. These plasmas are frequently modeled as collisional fluids because relaxation of the LTE assumption introduces myriad complications that make their theoretical description and simulation quite difficult. This is especially true in plasmas that possess significant scale separation between their dynamical gradient length scales and the kinetic length scales such as the ion Larmor radius ($\ri$). Even in high-$\beta$ plasmas where magnetic fields are energetically weak (with $\beta \doteq 8\upi p/B^2$ the ratio of the thermal and magnetic pressures), the vast length scales characteristic of astrophysical environments like the ICM mean that there can be as much as ten orders of magnitude separating the Coulomb mean free path from the Larmor radius. For fluctuation frequencies that are large compared to the collision frequencies, plasmas approximately conserve the double adiabatic invariants, $p_\perp/\rho B$ and $p_\|B^2/\rho^3$, where $p_\perp$ and $p_\|$ are the thermal pressures across and along the local magnetic-field direction. When the density ($\rho$) and magnetic-field strength ($B$) change, pressure anisotropy $\Delta \doteq \Delta p/p_\| \doteq p_\perp/p_\|-1$ results, in turn exciting a plethora of macrophysical and microphysical effects. From kinetic microinstabilities to plasma self-organization, these pressure anisotropy-mediated effects are crucial to our understanding of astrophysical turbulence. For that reason, they are the chief focus of this work.


\subsection{Consequences of pressure anisotropy}

Perhaps the simplest yet most consequential way in which pressure anisotropy modifies plasma dynamics is through its effect on Alfv\'en waves. The tension force responsible for these waves' propagation in a collisionless plasma is not only a function of $B$, but also of $\Delta$. As a result, the propagation speed of an Alfv\'enic disturbance is the effective Alfv\'en speed $v_{\rm A,eff} \doteq v_{\rm A}\sqrt{1+\beta\Delta/2}$, with the pressure anisotropy either enhancing or suppressing wave propagation depending on its sign. In high-$\beta$ plasmas, only a small amount of anisotropy is required to have a dramatic effect on Alfv\'enic motions. A notable example of large $\beta$ values enabling $\Delta$ to play this elevated role is in the `Alfv\'en wave interruption' process of \citet{squire16,squire17num,squire17}. Those authors found that, if the amplitude of a long-wavelength shear-Alfv\'en wave is sufficiently large, then the associated magnetic-field perturbation can adiabatically generate pressure anisotropy satisfying $\Delta \leq -2\beta^{-1}$, causing the Alfv\'en wave to self-interrupt and cease propagating. This pressure anisotropy also need not come from the Alfv\'en wave itself, but rather could be produced by other long-wavelength Alfv\'en waves and/or ion-acoustic waves that interact with the Alfv\'en wave \citep{mk23}. At Larmor scales, pressure anisotropy plays an additional role as a trigger of kinetic microinstabilities. At high $\beta$, these instabilities are typically the mirror and firehose, which can be excited when $\Delta \gtrsim \beta^{-1}$ and $\Delta \lesssim -2\beta^{-1}$, respectively~\citep{barnes66,hasegawa69,hm00}. Both instabilities grow rapidly on $\ri$-scales, causing sharp kinks in the magnetic field and thereby breaking conservation of the double adiabats. For this reason, their most notable effect on large-scale dynamics is their introduction of an effective collisionality, which isotropizes the thermal pressure back towards marginal instability. This effect has been shown to cause Braginskii-MHD-like behaviour in otherwise collisionless sound waves and suppress the nonlinear saturation of collisionless damping in compressive non-propagating modes, even when their wavelengths far exceed the Larmor scale \citep{kunz20,mks23}. It has also been suggested that, in the context of ICM turbulence, these microinstabilities might cause so much particle scattering that the effective Reynolds number could be increased to allow an extended turbulent cascade below the nominal (Coulomb-collisional) viscous scale \citep{St-Onge2020,zhuravleva19,kjz22}. 

On the other hand, a competing effect of pressure anisotropy has also been found to {\em suppress} viscous stress in ICM-relevant plasmas, while holding consequences for many other astrophysical environments. This effect, termed `magneto-immutability', involves the self-organization of Alfv\'enic turbulence in high-$\beta$  weakly collisional \citep{squire19} and collisionless \citep{squire23} plasmas to avoid changes in magnetic-field strength and thus the production of pressure anisotropy. That being said, exactly how this self-organization occurs and how robust it is remains somewhat mysterious.


\subsection{Magneto-immutability}\label{sec:introimm}

As its name suggests, magneto-immutability involves a tendency for a weakly collisional or collisionless turbulent plasma to organize in such a way that fluctuations in the magnetic-field strength become rare (e.g., as compared with fluctuations realized in collisional magnetohydrodynamic (MHD) turbulence). More specifically, it is defined as the suppression of field-aligned gradients in $u_\|$ and $\Delta p$ via self-organization (i.e., not through collisional or collisionless damping, or by choice of forcing). To understand how this definition  connects to the suppression of changes in $B$, we briefly review the work of \citet{squire19}, within which magneto-immutability was initially discovered. By studying weakly collisional, Braginskii-MHD turbulence, \citet{squire19} found that magneto-immutability occurs when the thermal pressure can affect the flow anisotropically. In Alfv\'enically driven turbulence where the flow is incompressibly forced perpendicular to the background magnetic field, the primary source of pressure anisotropy is through changes in $B$; thus, magneto-immutability inherently limits the production of pressure anisotropy by reorganizing turbulent motions to avoid those changes. By studying the incompressible induction equation, $\mathrm{d}_t \ln B = \eb\eb\bdbldot\grad \bb{u}$, \citet{squire19} related these effects to the suppression of field-aligned gradients of $u_\|$ and, by association, the suppression of field-aligned gradients of $\Delta p$. \citet{squire23} expanded the previous Braginskii-MHD investigation to the fully collisionless regime, finding that magneto-immutability is robust with respect to the closure employed for the pressure equations. This was accomplished by performing simulations using the Chew-Goldberger-Low (CGL; \citealt{cgl56}) MHD model of a collisionless fluid (referred to as `active-$\Delta$'), and comparing them with simulations using isothermal MHD that passively evolve $p_\perp$ and $p_\|$ in response to the density and magnetic-field fluctuations in the MHD turbulence \citep[referred to as `passive-$\Delta$' in][]{squire23}. Prior to these studies of magneto-immutability, it was thought that the only means for strong turbulence to extend to small scales in a weakly collisional, high-$\beta$ plasma was for the parallel viscous scale associated with the Coulomb collisionality to be significantly reduced via anomalous particle scattering (e.g., by plasma micro-instabilities). However, with magneto-immutability dynamically regulating the level of pressure anisotropy, and therefore the viscous stress, weakly collisional turbulence can form a surprisingly robust, MHD-like, approximately conservative inertial range.

\begin{figure}
    \centering
    \includegraphics[width=0.99\textwidth]{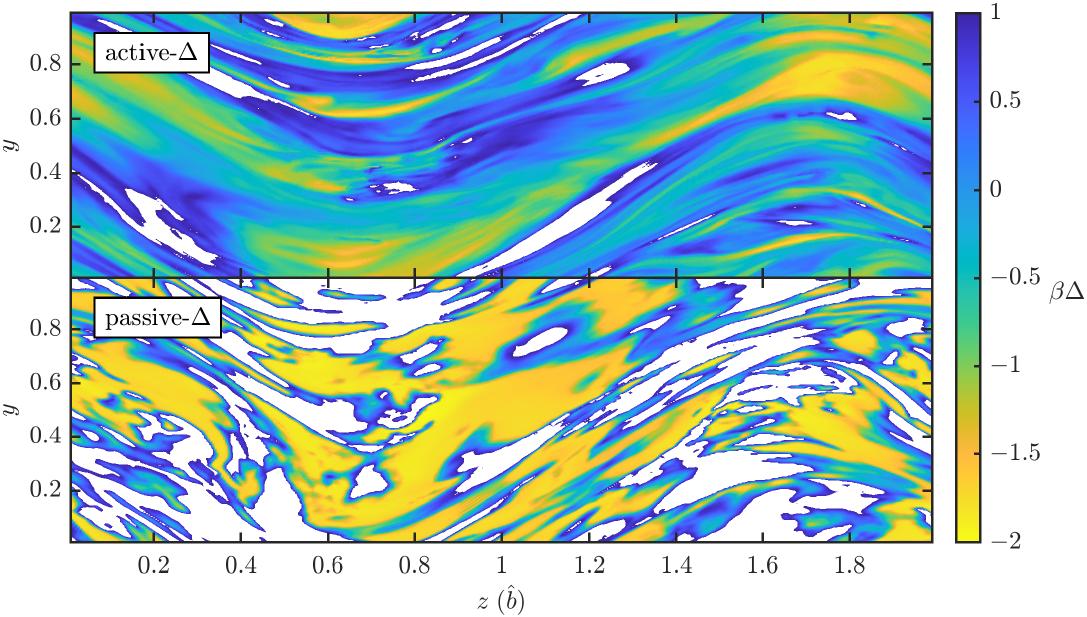}
    \caption{Dimensionless pressure anisotropy, multiplied by $\beta$, in an active-$\Delta$ and a passive-$\Delta$ simulation of driven turbulence. These simulations were performed with an initial background $\beta_0=10$, and forced Alfv\'enically such that the magnetic perturbation amplitude at the outer scale satisfies $\delta B_\perp \approx B_0/2$. White regions lie beyond the mirror and firehose thresholds, $\beta\Delta > 1$ and $\beta\Delta < -2$ respectively. Approximately half of the simulation domain is beyond these thresholds in the passive-$\Delta$ run, while only a small percentage is unstable in the active-$\Delta$ run, illustrating the effectiveness of magneto-immutability at reducing $\Delta$.}
    \label{fig:exaniso}
\end{figure}

To demonstrate what this looks like qualitatively, we present in figure~\ref{fig:exaniso} two snapshots taken from simulations of Alfv\'enically driven, high-$\beta$ turbulence: one using `active-$\Delta$' (top) and one using `passive-$\Delta$' (bottom), the numerical methodology for which is provided in \S\ref{sec:num}. These simulations are both performed at $\beta_0=10$, with the driven magnetic perturbation $\delta B_\perp \approx B_0/2$ at the outer scale. Given conservation of the double adiabats, this combination of forcing amplitude and high-$\beta$ is more than sufficient to generate values of pressure anisotropy that are well beyond the aforementioned mirror and firehose thresholds (the white regions). Yet, only in the `passive-$\Delta$' simulation do such unstable regions appear to be a regular occurrence. Discerning exactly why this occurs, and in what regimes we can expect this to hold, are the goals of this paper.


\subsection{Outline}

\citet{squire19} and \citet{squire23} demonstrated the profound effect that magneto-immutability can have on turbulence in high-$\beta$ plasmas. Consequently, it is important to understand just how robust is this self-organization before we can apply it more broadly to turbulent astrophysical plasmas. To that end, in this work we present a theory, based on a reduced model of the CGL-MHD equations, that improves our understanding of several key aspects of magneto-immutability and, in doing so, resolves some limits of their effects on high-$\beta$, collisionless turbulence. To prime the reader on how this is accomplished, the following outline summarizes the methodology and key topics of our investigation. In \S\ref{sec:dyneq} we introduce the Landau-fluid CGL-MHD equations, which are employed both analytically and numerically to describe collisionless plasma turbulence. The linear wave solutions of Landau-fluid CGL are discussed in \S\ref{sec:waves}, with particular attention paid to the difference in timescales between compressive modes in collisionless and collisional plasmas. In \S\ref{sec:ord} we introduce an ordering based on $\beta\gg 1$ and the principle of critical balance, which allows us to simplify the CGL-MHD equations and obtain the first analytical description of magneto-immutability in \S\ref{sec:redeq}. This `reduced CGL-MHD' model predicts, for example, that magneto-immutability does not depend on scale, that the associated self-organization process is unaffected to leading order in $\delta B_\perp/B_0$ by heat fluxes caused by field-aligned temperature gradients, and that the pressure anisotropy exhibits a passive-scalar-like $k$-space spectrum. In \S\ref{sec:num} we verify the assumptions used to develop our high-$\beta$ ordering and test several predictions made by the reduced CGL-MHD equations. This is accomplished using a numerical CGL-MHD solver built into the \textsc{Athena++} framework, described briefly within \S\ref{sec:code} and in more detail in Appendix~A of \citet{squire23}. We also explore additional aspects of magneto-immutable turbulence not predicted by the reduced equations, such as the volume-filling fraction of regions unstable to the firehose and mirror (\S\ref{sec:eos}), how magneto-immutability occurs through the misalignment of the velocity strain with the magnetic field (\S\ref{sec:org}, figure~\ref{fig:anglevk2d}), and the ability of micro-instability-induced scattering to interfere with self-organization (\S\ref{sec:nu}). In \S\ref{sec:conc}, we summarize the main findings of this work, and discuss the consequences they hold for turbulence in weakly collisional astrophysical environments such as low-luminosity black-hole accretion flows and the ICM.


\section{Theoretical description}\label{sec:theory}

\subsection{The dynamical equations}\label{sec:dyneq}

Because we wish to apply our theory to systems in which the ion Larmor radii are as much as ten orders of magnitude smaller than the collisional mean free paths, we choose to employ a collisionless model that neglects finite-Larmor-radius (FLR) effects, saving micro-scale considerations like the mirror and firehose instabilities for the numerical analysis of \S\ref{sec:num}. The two most suited models for describing a plasma under these assumptions are then the drift kinetic-MHD model of \citet{kulsrud83} and the CGL-MHD fluid approach of \citet{cgl56}. While the drift-kinetic approach is more accurate in that it self-consistently determines the heat fluxes via moments of the guiding-centre distribution function, we instead employ the simpler CGL-MHD model. This, of course, places some limits on which applications of magneto-immutability can be studied using our simulation technique. For example, if investigating the manner in which cosmic rays are accelerated or diffused by the turbulent field and flow at large scales, then the CGL-MHD model may suffice. On the other hand, if one wishes to study the shape of the plasma distribution function beyond the fluid moments, or probe the microphysics of dissipation at sub-ion-Larmor scales, then the approach we take here will be insufficient. Additionally, CGL-MHD model by itself provides no closure for the form of the heat fluxes, and so for lack of anything markedly better we adopt the `Landau-fluid' closure introduced by \citet{shd97}, which is designed to capture the effects of linear Landau damping on the fluid quantities. Fortunately, we find that the turbulence behaves in such a way that the exact form of the heat fluxes is relatively unimportant, suggesting that any errors introduced by applying these approximate heat fluxes are small (indeed, in \S\ref{sec:org} we demonstrate that magneto-immutability occurs in a hybrid-kinetic simulation of high-$\beta$ Alfv\'enic turbulence). With that in mind, the CGL-MHD equations are given by \citep{cgl56}:
\begin{subequations}\label{eq:fullcgl}
\begin{equation}\label{eq:cont}
\pD{t}{\rho}+\grad \bcdot(\rho \bb{u})=0 ,
\end{equation}
\begin{equation}\label{eq:moment}
\rho \D{t}{\bb{u}}=-\grad\left(p_{\perp}+\frac{B^2}{8\upi}\right)+\grad \bcdot\left[\eb\eb\biggl(p_\perp - p_\parallel + \frac{B^2}{4\upi}\biggr)\right] ,
\end{equation}
\begin{equation}\label{eq:induc}
\D{t}{\bb{B}}= \bb{B} \bcdot \grad \bb{u} - \bb{B} \grad \bcdot \bb{u}, 
\end{equation}
\begin{equation}\label{eq:cglpprp}
p_\perp \D{t}{} \ln \frac{p_\perp}{\rho B} = - \grad\bcdot(q_\perp\eb) - q_\perp\grad\bcdot\eb ,
\end{equation}
\begin{equation}\label{eq:cglpprl}
p_\parallel \D{t}{}\ln \frac{p_\parallel B^2}{\rho^3} = -\grad\bcdot(q_\parallel\eb) + 2q_\perp\grad\bcdot\eb ,
\end{equation}
\end{subequations}
where $\bb{B}$ is the magnetic field, $\bb{u}$ is the ion flow velocity, and $\rho$ is the mass density. Note that $p_\perp$ and $p_\parallel$ are defined with respect to the local magnetic-field direction $\eb=\bb{B}/B$. Here, $\rmd/\rmd t\doteq \partial/\partial t + \bb{u}\bcdot\grad$ is the convective time derivative. The equations \eqref{eq:cglpprp} and \eqref{eq:cglpprl} demonstrate how, in the absence of heat fluxes, the quantities $p_\perp/\rho B$ and $p_\| B^2/\rho^3$ are conserved in time along the flow of the plasma. Conservation of $p_\perp/\rho B$ and $p_\| B^2/\rho^3$ is the collective result of individual particles conserving both their magnetic moment $\mu \doteq mw_\perp^2/2B$ and their parallel action (i.e., bounce invariant) $\mathcal{J} = \oint \mathrm{d}w_\| \,mw_\|$, respectively (where $\bb{w}\doteq\bb{v}-\bb{u}$ is the particle velocity peculiar to the fluid frame). Because of the plasma's strong magnetization, the flows of perpendicular/parallel thermal energy, $q_{\perp/\|}$, occur exclusively along the local magnetic-field direction. For these quantities we adopt the `3+1 model' of \citet{shd97}:
\begin{subequations}\label{eq:fullhf}
\begin{equation}
q_{\perp}=-\frac{v_{\rm th \|}}{\sqrt{\upi} \left|k_{\|}\right|}\left[\rho \nabla_{\|}\left(\frac{p_{\perp}}{\rho}\right)-p_{\perp}\left(1-\frac{p_{\perp}}{p_{\|}}\right) \frac{\nabla_{\|} B}{B}\right],
\end{equation}
\begin{equation}
q_{\|}=-\frac{2v_{\rm th \|}}{\sqrt{\upi} \left|k_{\|}\right|} \rho \nabla_{\|}\left(\frac{p_{\|}}{\rho}\right),
\end{equation}
\end{subequations}
where $v_{\rm th,\|}=\sqrt{2 p_\|/\rho}$ is the parallel thermal speed, and $\nabla_\| = \eb \bcdot \grad$ is the field-aligned gradient. The field-aligned wavenumber $|k_\||$ in the denominators of \eqref{eq:fullhf} is meant to be representative of a characteristic parallel scale of the perturbations to $\rho$, $B$, and $p_{\perp/\|}$ \citep[e.g.,][]{sharma06}. This quantity is a stand-in for the magnitude of the magnetic-field-aligned gradient operator, which is generally difficult to calculate as it must be evaluated along the exact perturbed fields at each time step. For the majority of the numerical simulations presented in \S\ref{sec:num}, we take $|k_\parallel|=4\upi/L_\parallel$ where $L_\parallel$ is the field-parallel outer scale; for the remainder of this section, however, the exact value of $|k_\parallel|$ is unimportant. Because the heat fluxes \eqref{eq:fullhf} are designed to capture only linear collisionless damping, it is implicitly assumed that the perturbations being studied are small enough with respect to the background plasma that nonlinear damping effects can be ignored. We also assume that the background plasma pressure is isotropic, and that the electrons are cold. Effects of a background anisotropy on inertial-range kinetic  turbulence have been considered by \citet{kunz15} and could be straightforwardly implemented within our model, however they are not necessary for describing magneto-immutability and assessing its robustness. Similarly, electron pressure is not essential to the conclusions we will draw; as long as the electrons are sufficiently collisional and isothermal, a finite electron pressure has no qualitative effect on our results.\footnote{Evidence of this can be found already in the numerical simulations of \citet{squire23}, which showed that ion-to-electron temperature ratios of order unity had little effect on signatures of magneto-immutability. If the electrons were to be weakly collisional, however, certain effects caused by electron pressure anisotropy may need to be taken into account, a point we discuss in \S\ref{sec:conc}.} Finally, the CGL-MHD equations' inability to resolve $\ri$-scales inevitably implies that the growth of the mirror and firehose instabilities and their interaction with the plasma particles are not properly captured. In our numerical simulations (\S\ref{sec:num}), we model their effect on the plasma through an anomalous collisionality that isotropizes the pressure to marginally unstable values, but no effort is made to incorporate the effects of this micro-instability scattering into our analytic model for magneto-immutability.


\subsection{Relevant properties of collisionless hydromagnetic waves}\label{sec:waves}

Developing an intuition of the linear modes and their interactions is particularly beneficial towards understanding the scale-by-scale transfer of energy in a turbulent MHD cascade. It is therefore important to review the linear modes of the Landau-fluid CGL system, so that we can make assumptions and formulate our ordering in light of these fundamental behaviours. Many of these waves are discussed in greater detail within Appendix B of \citet{mks23}; here we restrict ourselves to aspects of the linear modes that are relevant to collisionless high-$\beta$ turbulence. 

Our discussion begins with those modes that are least affected by the transition from collisional to collisionless regimes. In particular, linear collisionless Alfv\'en waves are identical to their collisional counterparts when their wavelengths far exceed $\ri$, because neither density nor pressure perturbations are generated as they propagate. Meanwhile, the mode that experiences the most minor nonzero change is the fast mode, which has a nearly unchanged phase speed, although it is susceptible to collisionless damping depending on the propagation angle with respect to the background field, ($k_\|/k_\perp$). Consequently, most of the differences between high-$\beta$ CGL and collisional MHD turbulence do not originate from the behaviour of Alfv\'en or fast waves. The remaining collisionless hydromagnetic modes are non-propagating modes, ion-acoustic waves, and kinetic entropy modes, which are most easily compared to the slow-magnetosonic and pressure-balanced entropy modes of collisional MHD.\footnote{The non-propagating mode and kinetic entropy mode are equivalent to the `$+$' and `$-$' branches of the high-$\beta$ gyrokinetic dispersion relation from \citet[][\S 6.2.2]{schekochihin09}. In that case, those authors simply refer to non-propagating modes as `magnetic-field-strength fluctuations'.} Because there is one more dynamical equation in CGL-MHD than in collisional MHD, there is one additional linear mode solution. However, the kinetic entropy mode is both static and heavily damped at high $\beta$, with no clear collisional counterpart \citep[see, e.g.,][]{mks23}, so it can be excluded from this comparison.

We are then left with the one-to-one comparisons of non-propagating modes with pressure-balanced entropy modes, and ion-acoustic waves with slow magnetosonic waves. Both the non-propagating and pressure-balanced entropy modes have zero phase speed, with the entropy mode being fully static, and the non-propagating mode decaying due to transit-time damping at a rate $\gamma \sim |k_\||v_{\rm A}/\sqrt{\beta}$ that is small at high $\beta$. And while entropy modes satisfy perfect isotropic pressure balance (allowing them to remain static), non-propagating modes exist in a state of approximate pressure balance between $p_\perp$ and $B^2$ when $k_\| \ll k_\perp$ \citep{mks23}. For these reasons, at high $\beta$ these modes are rather similar, with the important feature that they are both approximately static. The comparison of ion-acoustic and slow waves, on the other hand, yields far fewer similarities, especially at high $\beta$. The most essential difference lies in their dispersion relations, with the ion-acoustic wave having a frequency $\omega \sim k_\|v_{\rm th}$, while the slow-wave frequency scales as $k_\| v_{\rm A}$ when $\beta \gg 1$ and $\omega \sim k_\|v_{\rm th}$ when $\beta \ll 1$. The ion-acoustic wave is also Landau damped at a rate $\gamma \sim |k_\||v_{\rm th}$, meaning that at high $\beta$ all characteristic timescales of ion-acoustic waves are much shorter than those of slow (and Alfv\'en) waves. These disparate timescales are fundamental to how collisionless waves interact with an Alfv\'enic cascade. 
For example, in collisional MHD turbulence, a timescale separation exists between fast and Alfv\'en waves because fast waves have a phase speed proportional to $k$ rather than just $k_\|$ (with $k_\| \ll k$ in the inertial range of Alfv\'enic turbulence, see \citet{gs95}). As a result, it is often argued that the rapid propagation of fast waves decouples them from any Alfv\'enic dynamics. In the next subsection (\S\ref{sec:ord}), we will argue that at $\beta \gg 1$, the ion-acoustic wave -- the only other compressive ($\grad \bcdot \bb{u} \neq 0$) wave -- decouples from the Alfv\'enic dynamics as well because $v_{\rm th} \gg v_{\rm A}$.


\subsection{An asymptotic ordering for high-$\beta$ collisionless turbulence}\label{sec:ord}

The equations \eqref{eq:fullcgl} are too complicated with which to work  analytically, therefore we seek an asymptotic ordering that distills the physics responsible for magneto-immutability. Before doing so, however, it is instructive first to consider the asymptotic ordering used in the collisional reduced MHD (RMHD) employed by \citet{zm92} and \citet[][\S 2]{schekochihin09}, from which our ordering borrows heavily:
\begin{equation}\label{eq:ordmhd}
    \frac{u_\perp}{v_{\rm A}} \sim \frac{u_\|}{v_{\rm A}} \sim \frac{\delta B_\perp}{B_0} \sim \frac{\delta B_\|}{B_0} \sim \frac{\delta \rho}{\rho_0} \sim \frac{\delta p}{p_0} \sim \frac{k_\|}{k_\perp}\doteq \epsilon \ll 1 \sim \beta, \quad \partial_t \sim k_\| v_{\rm A}.
\end{equation}
Here, all of the perturbations belong to a given $k_{\perp/\|}$ `cone' and interactions are taken to be local in $k$-space, meaning that as perturbations become smaller through the cascade so does the ratio $k_\|/k_\perp$. Arguably the most essential aspect of this ordering is that it enforces `critical balance' (CB) of the Alfv\'enic fluctuations, viz.~$k_\perp u_\perp \sim k_\| v_{\rm A}$. This is a statement that the nonlinear timescale associated with eddy deformation by the Reynolds stress $(k_\perp u_\perp)^{-1}$ is comparable to the linear propagation timescale of Alfv\'en waves $(k_\| v_{\rm A})^{-1}$, as is expected for strong turbulence \citep{gs95}. As a result, the evolutionary timescale at each wavenumber is expected to be roughly $\partial_t \sim k_\| v_{\rm A}$, since linear and nonlinear timescales are equivalent. This is also true of MHD slow modes, given that they propagate at $v_{\rm A}$ or slower, allowing for the formation of a similarly strong compressive cascade. As mentioned in \S\ref{sec:waves} however, the enhanced propagation speed of fast modes prevents them from being effectively coupled to the Alfv\'enic motions, and as such they are ordered out of RMHD. Our ordering shares most of these assumptions, yet differs in a few key ways.

First, we incorporate the largeness of the background $\beta$ directly into the ordering, so that $\beta \sim \epsilon^{-1}$. This may seem odd because $\beta$, being a background quantity, is constant across all scales even though $\epsilon$ becomes smaller as the fluctuations cascade anisotropically to larger $k$. However, we are concerned with plasmas in which perturbations near the outer scale often exceed $\beta^{-1}$ in relative amplitude. As a result, it would be incorrect to eliminate terms of order $(u_\perp/v_{\rm A})^2$ but keep those of order $u_\perp/\beta v_{\rm A}$. Ordering $\beta\sim\epsilon^{-1}$ allows us to retain both terms. We also assume that the Alfv\'enic component of our turbulence is critically balanced, as in collisional RMHD. This is not necessarily a given, however there is evidence from the work of \citet{squire23} that immutability does not interfere with critical balance, a feature that we reproduce through simulations of our own in \S\ref{sec:num}. Next, we assume that, as a result of the plasma being collisionless, $\delta p$ is replaced in the ordering by $\delta p_\perp$ and $\delta p_\|$, with similar amplitudes to \eqref{eq:ordmhd}. This is a straightforward assumption, yet its implications are more nuanced. Doing so replaces slow magnetosonic waves with ion-acoustic waves, and because we include $\beta$ in the overall ordering, this implies that nonlinear mixing of ion-acoustic waves by Alfv\'en waves is weak when local in $k$-space, since $k_\| v_{\rm th} \gg k_\perp u_\perp$. We will therefore assume that ion-acoustic waves, like fast waves, decouple from the Alfv\'enic turbulence. This assumption is supported by analytical calculations and simulations of the interaction of Alfv\'en and ion-acoustic waves in high $\beta$ plasmas, using the same CGL-MHD code employed in this work \citep{mk23}. Ordering out ion-acoustic waves therefore enables us to continue with the assertion that, for all variables, time derivatives are of order $\partial_t \sim k_\| v_{\rm A}$.

Our final major assumption, which follows from those before, is that density fluctuations can be neglected in determining the leading-order dynamics (i.e.~$\delta \rho/\rho_0 \sim \epsilon^2$ at most). In Alfv\'enic turbulence, high values of $\beta$ naturally inhibit the magnitude of density fluctuations by reducing the sonic Mach number ($u/v_{\rm th}$) of the forcing. Furthermore, any forcing must be associated with timescales that are sonic or faster if it is to produce substantial compressive fluctuations, a rather rare occurrence in the systems with which we are concerned. As a result, density fluctuations are expected to be small in amplitude at the outer scale. We are able to apply this assumption throughout the inertial range because the only linearly compressive waves -- ion-acoustic and fast modes -- have been ordered out of the dynamics. This assumption of small $\delta \rho$ allows us to consider only perturbations to the temperatures $\delta T_{\perp/\||}$ rather than the pressures $\delta p_{\perp/\|}$.

The collisionless, high-$\beta$ ordering that results from the above considerations and assumptions is
\begin{equation}\label{eq:ord}
    \frac{u_\perp}{v_{\rm A}} \sim \frac{u_\|}{v_{\rm A}} \sim \frac{\delta B_\perp}{B_0} \sim \frac{\delta B_\|}{B_0} \sim \frac{\delta T_\perp}{T_0} \sim \frac{\delta T_\|}{T_0} \sim \frac{1}{\beta} \sim \frac{k_\|}{k_\perp} \doteq \epsilon , \quad \partial_t \sim k_\| v_{\rm A}.
\end{equation}
Note that the choice $\delta T_{\perp/\|}/T_0 \sim \beta^{-1}$ means that $\beta \Delta \sim 1$, or in other words, the anisotropic pressure-stress is present at the same order as the flow inertia.\footnote{A separate, strictly Alfv\'enic ordering could be considered in which $(u_\perp/v_{\rm A})^2 \sim \Delta$, as one might expect from a pure shear-Alfv\'en eigenmode. With $\beta^{-1} \sim u_\perp/v_{\rm A}$, this would be relatively uninteresting as, with or without immutability, the $\Delta p$-stress would be too weak to affect the Alfv\'enic turbulence to leading order. Instead, it would be necessary to choose $\beta^{-1} \sim (u_\perp/v_{\rm A})^2$. However, this would introduce ambiguity into ordering $u_\|$ and $\delta B_\|$, which in our turbulence and that of \citet{squire23} appear to be comparable to their perpendicular counterparts.} This is important, because in order to demonstrate that magneto-immutability avoids significant $\Delta p$ stress, $\Delta p$ must first be made large enough to disrupt the turbulence in the absence of immutability.\footnote{This is akin to assuming that the Reynolds number $\mathrm{Re} \equiv \rho u L/\mu_{\rm visc}$ in hydrodynamic turbulence (with $\mu_{\rm visc}$ the fluid viscosity) satisfies $\mathrm{Re} \sim 1$, which prohibits the continuation of a turbulent cascade.} Some of the assumptions that yield \eqref{eq:ord}, such as $\delta \rho/\rho_0$ being negligible, are only justified qualitatively, therefore we test them against numerical simulations in \S\ref{sec:num} to establish confidence in the relevance of the ordering. We are now prepared to apply the ordering to \eqref{eq:fullcgl} and obtain the high-$\beta$ reduced CGL-MHD equations.


\subsection{The high-$\beta$ reduced CGL-MHD equations}\label{sec:redeq}

In this section we formulate equations that describe the leading-order evolution for each of the fields present in the ordering \eqref{eq:ord}. Doing so involves expanding these quantities in (fractional) orders of $\epsilon$, for example with $T_\|$:
\begin{equation}
    T_\| = T_0 + \delta T_\|^{(1)} + \delta T_\|^{(3/2)} + \delta T_\|^{(2)} + \dots,
\end{equation}
where the parenthetical superscript represents the order of each term in powers of $\epsilon$. Note that beyond the leading-order perturbation, which is of order $\epsilon$, half-integer orders must be used for all higher-order perturbations. This is required by the heat fluxes, which are proportional to $v_{\rm th} \sim v_{\rm A}/\sqrt{\epsilon}$. 

Keeping in mind the expectation that density fluctuations are $\mathcal{O}(\epsilon^2)$ at most and thus do not affect the leading order dynamics, the continuity equation \eqref{eq:cont} simplifies to $\grad \bcdot \bb{u} = 0$. If ordered according to \eqref{eq:ord}, this becomes
\begin{equation}\label{eq:ordcont}
    \grad \bcdot \bb{u} = \grad_\perp \bcdot \bb{u}_\perp + \nabla_\|u_\| \approx \grad_{\perp} \bcdot \bb{u}_\perp^{(1)} = 0.
\end{equation}
The divergence-free condition for $\bb{B}$ similarly yields $\grad_\perp \bcdot \delta \bb{B}_\perp^{(1)} = 0$ to leading order. These conditions on $\bb{u}_\perp^{(1)}$ and $\delta \bb{B}_\perp^{(1)}$ assert that the perpendicular flow and magnetic field perturbations are dominated by Alfv\'en waves, which are naturally incompressible. Given that large-scale collisionless Alfv\'en waves are linearly identical to those of collisional plasmas, it is no surprise that this aspect of the reduced equations is unchanged from standard RMHD. Continuing with the induction equation \eqref{eq:induc}, the leading-order parallel and perpendicular components become
\begin{subequations}
\begin{equation}\label{eq:prlind}
    \D{t^{(0)}}{\delta B_\parallel^{(1)}} = B_0\eb^{(0)} \bcdot \grad u_\|^{(1)} ,
\end{equation}
\begin{equation}
    \D{t^{(0)}}{\delta \bb{B}_\perp^{(1)}} = B_0\eb^{(0)} \bcdot \grad \bb{u}_\perp^{(1)} ,
\end{equation}
\end{subequations}
where
\begin{equation}
    \D{t^{(0)}}{} \doteq \pD{t}{} + \bb{u}_\perp^{(1)} \bcdot \grad_\perp \qquad \mathrm{and} \qquad \eb^{(0)} \doteq \ez + \frac{\delta \bb{B}_\perp^{(1)}}{B_0} ,
\end{equation}
with $\ez$ being the direction of the background magnetic field $\bb{B}_0$. Once again, these equations are equivalent to those obtained for the evolution of the magnetic field in standard RMHD. Equation~\eqref{eq:prlind} states that leading-order changes to the magnetic-field strength are generated through field-aligned shear in $u_\|$, which we will see is reduced in magneto-immutable turbulence. That this has not been ordered out directly by \eqref{eq:ord} hints that magneto-immutability must be achieved through self-organization.

Next we address the momentum equation. Beginning with the perpendicular direction, the first two orders are trivially
\begin{equation}\label{eq:notprp}
    \grad_\perp \delta T_\perp^{(1)} = 0 \qquad \mathrm{and} \qquad \grad_\perp \delta T_\perp^{(3/2)} = 0
\end{equation}
as a result of $\beta$ being large. Given that $k_\| \ll k_\perp$ at all scales, these perturbations can have no parallel gradients either, thus $\delta T_\perp ^{(1)} = \delta T_\perp^{(3/2)} = 0$. The next two orders give perpendicular pressure balance, 
\begin{equation}\label{eq:pbal}
    \frac{\delta T_\perp^{(2)}}{T_0} + \frac{\delta \rho^{(2)}}{\rho_0} = -\frac{2}{\beta} \frac{\delta B_\|^{(1)}}{B_0} \qquad \mathrm{and} \qquad \frac{\delta T_\perp^{(5/2)}}{T_0} +  \frac{\delta \rho^{(5/2)}}{\rho_0}= -\frac{2}{\beta}\frac{\delta B_\|^{(3/2)}}{B_0},
\end{equation}
analogous to that of RMHD and observed within the simulations of \citet{squire23}. Note that, although these equations involve the higher-order density perturbations, $\delta \rho^{(2,5/2)}$ need not be determined independently as we will show that the perpendicular temperature perturbations $\delta T_\perp^{(2,5/2)}$ do not enter into the leading-order equations. The subsequent order of the momentum equation dictates the evolution of $\bb{u}_\perp$:
\begin{align}\label{eq:prpmom}
    \rho_0 \D{t^{(0)}}{\bb{u}_\perp^{(1)}} &= -\grad_\perp P_{\rm total}^{(3)} + \frac{B_0}{4\upi}\frac{\partial \delta \bb{B}_\perp^{(1)}}{\partial z} + \frac{1}{4\upi}\delta \bb{B}_\perp^{(1)} \bcdot \grad_\perp \delta \bb{B}_\perp^{(1)} \nonumber\\* 
    \mbox{} &\quad -\frac{\beta}{8\upi} B_0\frac{\partial}{\partial z} \biggl(\delta \bb{B}_\perp^{(1)} \frac{\delta T_\|^{(1)}}{T_0} \biggr) - \frac{\beta}{8\upi}\delta \bb{B}_\perp^{(1)} \bcdot \grad_\perp \biggl(\delta \bb{B}_\perp^{(1)} \frac{\delta T_\|^{(1)}}{T_0} \biggr).
\end{align}
Here, $P_{\rm total}^{(3)}$ represents the combined thermal and magnetic pressures evaluated at third order, which can be determined as a whole by enforcing $\grad_\perp \bcdot \bb{u}_\perp = 0$. 
Because of perpendicular pressure balance \eqref{eq:notprp}, the leading-order pressure anisotropy is dominated by the contribution from $\delta T_\|^{(1)}/T_0$. The top line of \eqref{eq:prpmom} contains those terms that are already present in RMHD, while the bottom line incorporates feedback from the pressure anisotropy onto the perpendicular momentum. The contribution from pressure anisotropy remains because our ordering assumes ${\beta\Delta \sim 1}$, although this holds different meanings at different scales. Near the outer scale, the anisotropy is dominated by fluctuations of amplitude $\Delta \sim \beta^{-1}$. As the cascade continues to smaller scales however, fluctuations in $\Delta$ become smaller in amplitude and no longer compete with the magnetic tension, thus dropping out of the leading order of \eqref{eq:prpmom}. As a result, the pressure anisotropy that shows up in \eqref{eq:prpmom} is really only that of the largest scales, which the high-$k$ Alfv\'en modes feel as a nearly constant modified background $v_{\rm A,eff}$.

It is tempting to identify the pressure-anisotropy-related terms in \eqref{eq:prpmom} with magneto-immutability, however, equation~\eqref{eq:prlind} suggests that magneto-immutability will primarily be mediated by $u_\|$. Therefore the parallel component of \eqref{eq:moment} must be examined, for which the leading two orders are simply
\begin{equation}\label{eq:gradtprl}
    \eb^{(0)} \bcdot \grad \delta T_\|^{(1)} = (\eb \bcdot \grad \delta T_\|)^{(1)} = 0
\end{equation}
and
\begin{equation}\label{eq:gradtprl2}
    \eb^{(0)} \bcdot \grad \delta T_\|^{(3/2)} + \frac{\delta \bb{B}_\perp^{(3/2)}}{B_0} \bcdot \grad_\perp \delta T_\|^{(1)} = (\eb \bcdot \grad \delta T_\|)^{(3/2)} = 0.
\end{equation}
These equations result directly from ordering out ion-acoustic waves. If the ordering were such that $\partial_t u_\| \sim \epsilon k_\| v_{\rm th}^2$, then the inertial term would be of the same order as the field-aligned temperature gradient \eqref{eq:gradtprl}, which would then be non-zero. Linearization of the resultant reduced equations would yield an ion-acoustic-like eigenfrequency proportional to $k_\|v_{\rm th}$, however in our reduced equations no such eigenfrequency can be obtained. This is analogous to pressure balance in the perpendicular equations, which orders out fast modes by virtue of the assumption that the time derivative of $u_\perp$ is not proportional to $kv_{\rm th}$. Unlike \eqref{eq:notprp}, equations \eqref{eq:gradtprl} and \eqref{eq:gradtprl2} involve both parallel and perpendicular gradients and alignment with $\delta \bb{B}_\perp$. Hence, it is not necessarily true that $\delta T_\|^{(1)} = \delta T_\|^{(3/2)} = 0$ as is the case for $\delta T_\perp^{(1)}$ and $\delta T_\perp^{(3/2)}$, but only that the field-aligned gradients of $\delta T_\|$ are negligible. Also of note is that we have yet to make any mention of the heat fluxes, implying that this minimal variation of $\delta T_\|$ along field lines is dynamic, rather than diffusive.\footnote{Being dynamic, the suppression of $\eb\bcdot\grad u_\|$ is independent of the manner in which $\Delta p$ is generated. It is possible then that this approach might also apply to the weakly collisional model of Braginskii-MHD, in which magneto-immutability was initially discovered by \citet{squire19}. We investigate this possibility and some consequences of it within Appendix \ref{app:brag}.}

For our purposes, there is no need to obtain the higher-order contributions to the parallel momentum equation, and so we continue applying our ordering by examining the evolution equations for the double adiabats. The first step in doing so is to consider the heat fluxes \eqref{eq:fullhf} in light of \eqref{eq:notprp}, \eqref{eq:gradtprl}, and \eqref{eq:gradtprl2}, which can be rewritten in the following more transparent manner:
\begin{subequations}\label{eq:redhf}
\begin{equation}\label{eq:redhfprp}
q_{\perp}=-\frac{v_{\rm th \|}\rho }{\sqrt{\upi} \left|k_{\|}\right|}\biggl( \eb\bcdot\grad T_\perp + \frac{\beta_\perp\Delta}{2} \frac{\bb{B}\bcdot\grad B}{4\upi\rho}\biggr),
\end{equation}
\begin{equation}
q_{\|}=-\frac{2v_{\rm th \|}\rho}{\sqrt{\upi} \left|k_{\|}\right|} \Bigl(\eb\bcdot\grad T_\|\Bigr).
\end{equation}
\end{subequations}
Here $\beta_\perp \doteq 8\upi p_\perp/B^2$ and $\beta_\perp \Delta \sim 1$, so the right-most term in \eqref{eq:redhfprp} is smaller than $p_\perp \mathrm{d}_t \ln (T_\perp/B)$ by roughly $\epsilon^{3/2}$, and can therefore be neglected. What remains are heat fluxes proportional to the field-aligned temperature gradients $\eb \bcdot \grad T_{\perp/\|}$. However, these gradients are zero not just to leading order, but also to next order thanks to \eqref{eq:notprp}, \eqref{eq:gradtprl}, and \eqref{eq:gradtprl2}. As a result, the $q_{\perp/\|}$ can be neglected entirely to the leading two orders at which they would contribute to \eqref{eq:cglpprp} and \eqref{eq:cglpprl}. Thus, the surprisingly simple outcome is that
\begin{equation}\label{eq:imm}
    \D{t^{(0)}}{\delta B_\|^{(1)}} = \D{t^{(0)}}{\delta T_{\|}^{(1)}} = 0.
\end{equation}
This result is essentially the namesake of magneto-immutability. In order to satisfy both $\mu$ conservation and perpendicular pressure balance in a high-$\beta$ plasma, the flow {\em must} self-organize to enforce $(\eb\eb\bdbldot\grad \bb{u})^{(1)}=0$, so that \eqref{eq:prlind} and \eqref{eq:imm} are in agreement. As a result, the magnetic field only experiences convective changes in its strength to leading order.\footnote{Although the adoption of the ordering \eqref{eq:ord} may be viewed as restrictive, the combination of perpendicular pressure balance and $\mu$ conservation at high $\beta$ alone is sufficient to produce a magneto-immutable state. Such circumstances may come about naturally through other means, such as in the excitation of non-propagating modes. In such cases, the other components of \eqref{eq:ord} need not be satisfied to obtain ${\rm d}_t \delta B_\| \approx 0$.} In describing magneto-immutability qualitatively, \citet{squire23} also drew several comparisons with incompressibility in collisional MHD turbulence, and here we may draw another more direct comparison. In a collisional high-$\beta$ plasma, pressure balance requires the \textit{isotropic} pressure perturbation $\delta p$ be reduced in order to match the pressure of the magnetic perturbation $\delta B_\|$. With conservation of the single adiabat ${\rm d}_t(p/\rho^\gamma)=0$, the smallness of $\delta p$ then requires that ${\rm d}_t \rho^{-\gamma}={\rm d}_t \delta \rho=0$. Thus both magneto-immutability and incompressibility in high-$\beta$ magnetized turbulence result from a combination of pressure balance and adiabatic invariance, only here the involvement of $B$ in the collisionless invariants necessitates magneto-immutability. 

Provided proper initial and boundary conditions, we have successfully closed the system of equations. Collecting the full high-$\beta$, reduced CGL-MHD equations (and dropping the orders), we have
\begin{subequations}\label{eq:redeq}
\begin{equation}\label{eq:reddiv}
    \grad_\perp \bcdot \bb{u}_\perp = \grad_\perp \bcdot \delta \bb{B}_\perp = 0,
\end{equation}
\begin{equation}\label{eq:redind}
    \D{t}{\delta \bb{B}_\perp} = B_0\eb \bcdot \grad \bb{u}_\perp,
\end{equation}
\begin{equation}\label{eq:redmom}
    \rho_0 \D{t}{\bb{u}_\perp} = -\grad_\perp P_{\rm total} + \biggl( 1+\frac{\beta}{2}\frac{\Delta p}{p_0} \biggr)\biggl(\frac{B_0 \ez + \delta \bb{B}_\perp }{4\upi}\biggr) \bcdot \grad \delta \bb{B}_\perp,
\end{equation}
\begin{equation}\label{eq:dpevo}
    \D{t}{\Delta p} = \D{t}{\delta B_\|} = \eb \bcdot \grad u_\| = \eb \bcdot \grad \Delta p = 0 .
\end{equation}
\end{subequations}
In \eqref{eq:dpevo} and \eqref{eq:redmom}, we have replaced $\delta T_\|/T_0$ with $-\Delta p/p_0$ without consequence, because $\Delta p/p_0 = \delta p_\|/p_0$ up to order $\epsilon^{3/2}$. A homogeneous background pressure anisotropy $\Delta p_0$ can be straightforwardly included in these equations by appending it to the fluctuating $\Delta p$ in \eqref{eq:redmom}. As in collisional RMHD, the divergence of $\bb{u}_\perp$ can be employed to determine $P_{\rm total}$, and the convective derivative and field-parallel gradient become
\begin{equation}\label{eq:derivs}
    \D{t}{} = \pD{t}{} + \bb{u}_\perp \bcdot \grad_\perp \qquad \mathrm{and} \qquad \eb = \ez + \frac{\delta \bb{B}_\perp}{B_0},
\end{equation}
respectively. Note that the reduced order of $\nabla_\| \delta T_\|$ compared to $\delta T_\|$ itself allowed the anisotropic pressure to be pulled outside of the gradient operator in \eqref{eq:redmom} (since $\nabla_\| \Delta p$ becomes next order as well). In this sense, the pressure anisotropy is felt only as a modification to the effective Alfv\'en speed, thereby hindering its ability to interfere with the Alfv\'enic cascade as an anisotropic pressure-stress that could cause turbulent motions to damp into thermal energy. This is a central characteristic of magneto-immutability, and one of significant consequence. Without it, the turbulent cascade would cease long before reaching kinetic scales, dramatically modifying the transport properties of the plasma.


\subsection{Features of the reduced system}\label{sec:feats}

We now discuss some properties of the reduced system \eqref{eq:redeq}, beginning with the quantities that its turbulent cascade conserves. In the same manner as \citet{kunz15}, this analysis is made easier by defining Els\"asser variables that incorporate the modification to the Alfv\'en speed by the pressure anisotropy \citep{elsasser50}:
\begin{equation}
    \bb{z}^{\pm} = \bb{u}_\perp \pm v_{\rm A,eff} \frac{\delta \bb{B}_\perp}{B_0}.
\end{equation}
However, unlike in \citet{kunz15}, here $v_{\rm A,eff}$ is  understood to be a function of both position and time. It is a straightforward process to show that the reduced system~\eqref{eq:redeq} requires the cascades of Els\"asser energies to obey
\begin{equation}\label{eq:consaw}
    \pD{t}{W_{\rm AW}^{\pm}} \doteq \pD{t}{} \int \mathrm{d}^3 x \, \frac{1}{2} |\bb{z}^\pm|^2 = \mp \int \mathrm{d}^3 x \, \frac{P_{\rm total}}{\rho_0} \pD{z}{v_{\rm A,eff}}.
\end{equation}
This is notably distinct from the Els\"asser cascades of Alfv\'enic energy in collisional MHD and reduced kinetic-MHD (RKMHD), which satisfy $\partial_t W_{\rm AW}^\pm = 0$ individually \citep{schekochihin09,kunz15}. Here, only the total Alfv\'enic fluctuation energy $W_{\rm AW}^++W_{\rm AW}^-$ is conserved. Meanwhile, the cross helicity $W_{\rm AW}^+-W_{\rm AW}^-$, which measures the imbalance in forward-propagating and backward-propagating Alfv\'en waves, is not conserved. The fact that $W_{\rm AW}^++W_{\rm AW}^-$ is conserved proves that the pressure anisotropy, having reduced-order parallel gradients, cannot effectively thermalize the energy contained within Alfv\'enic fluctuations \citep[][also see figure \ref{fig:pstress} here]{squire23}. That being said, the non-conservation of cross-helicity means that $\Delta p$ is still able to redistribute energy between forward- and backward-propagating waves \citep{mk23}. The other conserved quantities of the magneto-immutable cascade are rather trivial to detect from \eqref{eq:redeq}:
\begin{equation}\label{eq:consnp}
    \pD{t}{} \int \mathrm{d}^3x \,\frac{B_0 \delta B_\|}{4\upi}  = 0 \qquad \mathrm{and} \qquad \pD{t}{} \int \mathrm{d}^3x\, \delta p_\|  = 0,
\end{equation}
where we have chosen to configure the constancy of $\delta B_\|$ in units of energy for consistency with the other conserved quantities. Equations~\eqref{eq:consnp} represent, respectively, cascades of non-propagating modes and kinetic entropy modes. The reason for the respective associations of each conserved quantity comes down to the eigenvectors of each mode. 
Non-propagating modes, being in a state of near-perpendicular-pressure balance, have $\delta B_\|/B_0 \gg \delta p_\perp/p_0$ at high $\beta$. However, they also exhibit $\delta p_\perp < \delta p_\|$, which is inconsistent with the fact that $\delta p_\| \gg \delta p_\perp$ in magneto-immutable turbulence. Thus, the passive cascade of $\delta p_\|$ can most easily be attributed to kinetic entropy modes, which vanish due to collisional damping in the MHD limit \citep{mks23}. To leading order, there is no conserved quantity for compressive fluctuations (namely those associated with $u_\|$), because ion-acoustic and fast waves have been ordered out of the dynamics.

To further highlight the features that make magneto-immutable turbulence unique, we can compare the reduced equations \eqref{eq:redeq} to the standard reduced MHD system, which is obtained by applying \eqref{eq:ordmhd} to the collisional MHD equations. The dynamics of the Alfv\'enic fluctuations $\delta \bb{B}_\perp$ and $\bb{u}_\perp$ are almost entirely unchanged from equations \eqref{eq:reddiv}--\eqref{eq:redmom}, with the only difference being that $\Delta p=0$ in collisional MHD \citep{zm92,schekochihin09}. This suggests that the differences between the two models originate in the compressive cascade, a component of the turbulence that plays a passive role in collisional RMHD. The equations of $\delta B_\|$ and $u_\|$ in RMHD turbulence are \citep{sc07}:
\begin{subequations}\label{eq:redmhd}
\begin{equation}\label{eq:mhdind}
    \D{t}{}\frac{\delta B_\|}{B_0} = \frac{1}{1+2/(\gamma\beta)} \,\eb\bcdot\grad u_\| \approx \eb\bcdot\grad u_\| \quad (\beta \gg 1),
\end{equation}
\begin{equation}\label{eq:mhdmom}
    \D{t}{}\frac{u_\|}{v_{\rm A}} = v_{\rm A} \eb\bcdot\grad \frac{\delta B_\|}{B_0},
\end{equation}
\end{subequations}
where $\gamma=5/3$ is the single adiabatic index for a monoatomic gas. These equations describe slow-magnetosonic waves, which are well known to comprise the passively advected compressive cascade in MHD turbulence, something that is clearly not present in \eqref{eq:redeq}. As discussed in the context of conserved quantities, the non-Alfv\'enic cascades can be attributed to non-propagating and kinetic entropy modes, with no truly compressive ($\grad \bcdot \bb{u} \neq 0$) cascade of any kind present to leading order.

When constructing the ordering \eqref{eq:ord} under the assumption $\delta B_\|/B_0 \sim \beta^{-1}$, we essentially guaranteed that realistic turbulent cascades eventually pass out of the parameter space within which the equations \eqref{eq:redeq} are strictly valid. Yet, these equations are nonetheless accurate even when $\delta B_{\|}/B_0 \ll \beta^{-1}$ (so long as $\beta \gg 1$). To prove this, in Appendix \ref{app:rkmhd} we show that the signatures of magneto-immutability, {\em viz.}~$\nabla_\|\Delta p$ and $\nabla_\| u_\|$ suppression, can be recovered via a subsidiary high-$\beta$ ordering of the RKMHD equations, which assume that $\delta B_\|/B_0 \ll \beta^{-1}$. Indeed, with the assumption of Landau-fluid heat fluxes we once again obtain the reduced system \eqref{eq:redeq}.\footnote{In this case, note that $\beta\Delta$ in the $u_\perp$ equation is not the local pressure anisotropy, but a background anisotropy set by large-scale $\beta\Delta \sim 1$ fluctuations, analogous to the anisotropy considered within \citet{kunz15}.} Note, however, that at these small scales where $\delta B_{\|}/B_0 \ll \beta^{-1}$, such lack of parallel gradients in $u_\|$ and $\Delta p$ need only be enforced passively, as they have already been suppressed by larger scales obeying the ordering \eqref{eq:ord}. In this sense, magneto-immutability exists in a somewhat weaker, watchdog-like state. If $k_\|$ were somehow introduced to $u_\|$ or $\Delta p$ at small scales -- for instance, by field-line wandering or magnetic reconnection \citep{meyrand19} -- it would be suppressed by magneto-immutable self-organization. However, if these parallel gradients were not created at small scales, they simply would remain negligible with $\Delta p$ cascading passively. To contrast this with collisional RMHD or $\beta \sim 1$ RKMHD, any re-introduced parallel gradients in $u_\|$ and $\Delta p$ would be allowed to persist in the absence of dissipative effects. Unlike $u_\|$, the evolution of small-scale $u_\perp$ in high-$\beta$ collisionless turbulence becomes quite similar to that of collisional RMHD. The $u_\perp$ fluctuations only feel a non-local modification to $v_{\rm A,eff}$ produced by large scale $\Delta p$ satisfying $\beta\Delta \sim 1$, thus $u_\perp$ and $\delta B_\perp$ effectively decouple from the $u_\|$ and $\delta B_\|$ cascades. Furthermore, the gradients in $v_{\rm A,eff}$ are negligible at these small scales, allowing the energetic coupling between $|\bb{z}^+|^2$ and $|\bb{z}^-|^2$ that appears in \eqref{eq:consaw} to weaken. Thus, the Alfv\'enic cascade approaches that of RKMHD with a constant background anisotropy, as described by \citet{kunz15}.

Equations~\eqref{eq:redeq} leave us with several new testable predictions. First and foremost is that magneto-immutability, as a reduction of $\eb\eb\bdbldot\grad \bb{u}$ and $\eb \bcdot\grad \Delta p$, is independent of scale. In the initial studies of \citet{squire19} and \citet{squire23}, the authors concluded that the pressure anisotropy being driven by turbulent fluctuations must be competitive with the background magnetic tension (i.e.~$\beta \Delta \sim 1$) in order for the suppression of $\eb\eb\bdbldot\grad \bb{u}$ and $\eb \bcdot\grad \Delta p$ to occur. However, both of these signatures of magneto-immutability persist throughout the cascade simply as a result of $\beta \gg 1$, even when $\beta\Delta_k \ll 1$ at some large wavenumber $k$. Note, however, that because the $\delta p_{\perp/\|}$ and $\delta B_\|$ are passively advected, the outer scale is important in that the compressive perturbations must be seeded with sufficiently large amplitude there.\footnote{For random forcing at large scales, this is a reasonable assumption. However, if one wished to simulate turbulence using \eqref{eq:redeq} and begin forcing somewhere in the inertial range, they would need to take care to properly initialize the amplitude of compressive fluctuations, or else the ordering may not be satisfied.} The second testable prediction is that magneto-immutability is, to leading order, independent of the heat fluxes. This is owed to the dynamical, rather than diffusive, manner in which minimal field-parallel variation of $\delta T_\|$ is achieved, forcing the heat fluxes to contribute only to the next-order pressure anisotropy. Both magneto-immutability's independence of the heat fluxes, and the field-parallel spreading of $\delta T_\|$ in the absence of heat fluxes can therefore be validated directly by well-constructed CGL-MHD simulations. Equation~\eqref{eq:imm} also predicts that $\delta T_\|$ and $\Delta p$ behave as passive scalars, thereby adopting the statistics of the flow that nonlinearly mixes them \citep{biskamp03}. Although not addressed analytically, it is also implied that a sufficiently large scattering frequency in the pressure equations could disable magneto-immutability. It is essential that the $\delta p_\perp$ and $\delta p_\|$ evolve independently enough that $\delta p_\perp$ can be suppressed by pressure balance to give $\eb\eb \bdbldot \grad \bb{u} \approx 0$, while $\eb\bcdot\grad \delta p_\|$ is suppressed by the $u_\|$ momentum equation. This arrangement can still be preserved in the presence of a small amount of scattering ($\nu \ll k_\|v_{\rm A}$), as the scattering term would remain next order in the $p_{\perp/\|}$ equations and the resultant reduced system would be unchanged. However, a scattering rate on the order of or larger than $k_\|v_{\rm A}$ would force the pressures to evolve together, and the two magneto-immutability criteria would not be able to be met independently (see Appendix~\ref{app:brag} for more). This point must be borne in mind when considering the effects of microinstabilities, because if their volume-filling fraction is large enough, then the consequent scattering may cause the plasma to become MHD-like.


\section{Numerical simulations}\label{sec:num}

In this section we describe simulations performed for the purpose of verifying the predictions of the reduced equations \eqref{eq:redeq}, organized in a manner that follows the layout of \S\ref{sec:theory}. As an overview, we begin by describing the numerical methods and the simulation setup in \S\ref{sec:code}, followed in \S\ref{sec:diag} by a summary of the key numerical diagnostics employed. Next, \S\ref{sec:eos} and \S\ref{sec:drive} provide evidence from the simulations that verify the assumptions that led to the ordering \eqref{eq:ord}, namely, that (i) the vast majority of the plasma possesses too little pressure anisotropy to trigger the mirror and firehose instabilities, and therefore should not be subject to their anomalous scattering; (ii) that ion-acoustic waves are weakly mixed by the Alfv\'enic turbulence; (iii) and that density fluctuations are sufficiently small that they may be neglected. In \S\ref{sec:comp}, we re-confirm other aspects of the turbulence that were seen previously by \citet{squire23} and which influenced our ordering, such as the perpendicular balance of the thermal and magnetic pressures, and the critically balanced scaling of the Alfv\'enic fluctuations. Once the assumptions that led to \eqref{eq:ord} are confirmed, we test the equations \eqref{eq:redeq} and their consequences in \S\ref{sec:org} and \S\ref{sec:hf}. This includes the misalignment of $\eb$ and the flow rate-of-strain (figure~\ref{fig:anglevk2d}), the predicted scale independence of magneto-immutability, the reduction of $\grad_\| \Delta p$, and the relative insensitivity of magneto-immutability to the magnitude of the heat flux. Lastly, in \S\ref{sec:nu} we discuss the ability of microinstability-induced scattering ($\nu_{\rm lim}$) to interfere in the self-organization process.


\subsection{Problem setup and method of solution}\label{sec:code}

To assess the claims set forth in \S\ref{sec:theory}, we perform a suite of driven CGL-MHD turbulence simulations with a variety of parameters tuned to address each individual prediction or assumption. We employ a modified version of the \textsc{Athena}++ MHD code \citep{stone20} that solves the CGL-MHD system \eqref{eq:fullcgl} closed with the Landau-fluid heat fluxes \eqref{eq:fullhf}. This code allows the parallel wavenumber $|k_\||$ in the Landau-fluid heat fluxes \eqref{eq:fullhf} to be specified freely, and incorporates the effects of mirror and firehose instabilities (when excited) through a collisional closure. Unless capped by some maximal value, the Landau-fluid $q_{\perp/\|}$ could grow very large at small scales because of the numerical simplification that $|k_\||$ is chosen to be constant. To prevent this from occurring, the heat fluxes are not allowed to surpass a maximal `free-streaming' value of ${\approx}v_{\rm th}p_{\perp/\|}$ \citep{hollweg74,cowie77}. Exact details of how this limitation is implemented in the code can be found in \S3.1.1 of \citet{squire23}. The collisional microinstability closure uses a limiting scattering frequency $\nu_{\rm lim}$ (also specified by the user), which activates only in regions of the domain that exceed the mirror ($\beta\Delta > 1$) or firehose ($\beta\Delta < -2$) thresholds \citep[e.g.,][]{sharma06}. Once activated, the pressures $p_{\perp/\|}$ are driven back towards the instability thresholds, rather than to complete pressure isotropy, at a rate set by $\nu_{\rm lim}$. The ability to choose $|k_\||$ and $\nu_{\rm lim}$ provides considerable freedom to explore how various collisionless effects change the behaviour of turbulence in this system. Further details about the CGL-MHD solver and microinstability closure can be found in Appendix~A of \citet{squire23}. As first described in \S\ref{sec:introimm}, we supplement these `active-$\Delta$' CGL-MHD simulations with a set of `passive-$\Delta$' simulations, which are performed in isothermal MHD but evolve the pressure anisotropy passively using equations \eqref{eq:cglpprp} and \eqref{eq:cglpprl} given the simulated MHD fields. Such passive-$\Delta$ simulations are useful for comparing MHD-like turbulence with CGL-MHD turbulence.

Given that the particles' Larmor scales are infinitesimally small in our model equations, the physical dimensions of our simulations are arbitrary. All simulations are performed in a fully periodic domain with dimensions $[L_x,L_y,L_z] = [1,1,2]$, where $L_z = L_\|$ is aligned with the background magnetic field $\bb{B}_0 = \ez$ and $L_x = L_y = L_\perp$ are the perpendicular dimensions. The `standard' resolutions employed for these simulations are $n_\perp = 192$ and $n_\| = 2n_\perp$, however higher- and lower-resolution simulations are performed (and explicitly referred to) for the sake of convergence and scale-dependence tests (see figure~\ref{fig:scalesep}). Although the initial magnetic field $\bb{B}_0$ remains the same in all simulations, the initially isotropic thermal pressure $p_0 = \beta_0 B_0^2/8\upi$ is varied by choosing $\beta_0$ to be either 1, 10, or 100. For $|k_\||$ and $\nu_{\rm lim}$, we make use of a set of `standard' values in which $|k_\|| = 4\upi/L_\|$ represents the wavenumber of compressive fluctuations near the outer scale and $\nu_{\rm lim} = 10^{10} v_{\rm A}/L_\perp$ yields a hard-wall limiter that prevents the pressure anisotropy from straying far beyond its microinstability thresholds. Specific instances in which $|k_\||$ and $\nu_{\rm lim}$ are modified from their standard values are noted on a case-by-case basis. All simulations are run until a final time of at least $t_{\rm f} = 10 L_\perp/v_{\rm A}$ to ensure that a steady-state fluctuation level is achieved within the turbulence.

The turbulence is forced exclusively through sinusoidal perturbations to the flow velocity $\bb{u}$ using an Ornstein--Uhlenbeck correlated process \citep{uhlenbeck30}, the strength of which is input numerically as the rate of change of the total kinetic energy in the domain $\mathrm{d}_t \mathcal{E}_{\rm K}$. Our fiducial runs employ $\mathrm{d}_t \mathcal{E}_{\rm K} = 0.32\rho_0 v_{\rm A}^2L_\perp^3$ and a correlation time of $t_{\rm corr} = L_\|/v_{\rm A}$. This choice of $t_{\rm corr}$ assumes Alfv\'enically correlated forcing, and the energy injection rate corresponds to an outer-scale magnetic perturbation amplitude of $\delta B_{\perp/\|} \approx B_0/2$ in steady state. For all simulations, the sinusoidal wavenumbers at which we force are limited to $k \in 2\upi/L_\| \times [1,3]$, over which the power distribution scales as $k^{-2}$. In this study, we frequently vary the mode of $\bb{u}$ forcing between Alfv\'enic and Random forcing. Random forcing is as it sounds: $\bb{u}$ is perturbed randomly in all directions without any special conditions relating to the directions of $\bb{B}$ and $\bb{k}$ (although it is still time correlated). In Alfv\'enic forcing however, we  perturb only $\bb{u} \perp \ez$, and do so in a manner that enforces incompressibility, $\grad_\perp \bcdot \bb{u}_\perp = 0$, at the outer scale (below the outer scale, however, $\bb{u}$ does become slightly compressible as a result of the nonlinear amplitudes).


\subsection{Numerical diagnostics}\label{sec:diag}

To analyze the simulations introduced in \S\ref{sec:code}, we make use of several numerical diagnostics, three of which we describe here because they are either used very frequently in our analysis or are particularly tailored to the subject of this work. These three are the Fourier spectra, the $\Delta p$ energy transfer function of \citet{squire23}, and a novel scale-by-scale $\eb\eb\bdbldot\grad\bb{u}$ alignment diagnostic based on the work of \citet{St-Onge2020}. 

The most frequently used diagnostic is a bin-averaged Fourier spectrum, defined for a field $\chi$ as
\begin{equation}
    E_\chi(k) = \frac{1}{\delta k} \sum_{|\boldsymbol{k}|=k} |\mathcal{F} [\chi](\bb{k}) |^2  ,
\end{equation}
where $\mathcal{F} [\chi](\bb{k})$ is the three-dimensional Fourier transform of $\chi$, and $k=\sqrt{k_x^2+k_y^2+k_z^2}$ is the magnitude of $\bb{k}$ (or $k_\perp = \sqrt{k_x^2+k_y^2}$ for $\bb{k}_\perp$). The spectra are calculated for $k$ bins, thus each bin contains contributions from several wavenumbers in the Fourier spectrum, and $\delta k$ represents the bin width in $k$ or $k_\perp$. Note that in figure~\ref{fig:rospres}($b$), when calculating the spectra of $\nabla_{\perp/\|}u_{\perp/\|}$, the gradients of $\bb{u}$ are calculated with respect to the \textit{local} (in configuration space) field, rather than using the same efficiency-motivated $\bb{k}_\perp = k_x\hat{\bb{x}} + k_y\hat{\bb{y}}$ assumption that the spectra employ.

Energy transfer functions illustrate the amount of turbulent kinetic energy transferred into or out of a given $k$-shell as a result of specific interaction terms in the model equations \citep[e.g.,][]{grete17}. In this work, we are chiefly interested in the effects of pressure anisotropy on the turbulent cascade, thus the transfer function we make use of is designed to capture the energy removed from the flow $\bb{u}$ at a given $k_\perp$ due to the anisotropic pressure-stress $\grad \bcdot (\eb\eb\Delta p)$. We use the same definition given in \citet{squire23}, where
\begin{equation}\label{eq:trans}
    \mathcal{T}_{\Delta p}(k_\perp) = \sum_{q_\perp} \int \mathrm{d}^3 \bb{x} \langle \sqrt{\rho}\bb{u} \rangle_{k_\perp} \bcdot \frac{\bb{B}}{\sqrt{4\upi\rho}} \bcdot \grad \left\langle \frac{\Delta p}{B^2}\bb{B} \right\rangle_{q_\perp},
\end{equation}
and $\langle\,\cdot\, \rangle_k$ indicates that the quantity has been Fourier transformed, filtered by wavenumber, and returned to real space. There are other definitions for this transfer function \citep[see][]{arzamasskiy23}, however, as remarked by \citet{squire23}, this particular formulation represents the $\Delta p$-stress as a damping of kinetic energy. This is well suited to the current study as we wish to understand how magneto-immutability permits the continuation of a turbulent cascade that would have otherwise been damped away near the viscous scale.

Finally, the $\eb\eb\bdbldot\grad\bb{u}$ alignment diagnostic is a method for visualizing the effects of magneto-immutability's organization on the turbulent flow. It designed to calculate the cosine of an angle $\theta$ that is representative of the alignment between the rate-of-strain tensor $\grad\bb{u}$ and the (spatially) local magnetic field direction $\eb$, on a scale-by-scale basis in $k_\perp$. Initially used by \citet{St-Onge2020} without separation by scale for a study of the incompressible fluctuation dynamo under the action of Braginskii viscous stresses, it is based off of analytical theory from \citet{kazantsev68}. In the fluctuation dynamo, changes to the magnetic-field strength are mediated by $\grad\bb{u}$ through its three eigenvalues, which are associated with field-line compressing motions, field-line stretching motions, and the incompressibility constraint (ensuring that the overall flow stretches as much as it compresses). The compressing and stretching motions result in changes to $|B|$, thus the angles between the compressing and stretching eigenvectors and the local magnetic-field direction dictate how efficient the flow is at changing the magnetic field strength.\footnote{The incompressibility eigenvalue, sometimes called the `null' eigenvalue, is distinguishable from the other two by having the smallest absolute magnitude.} In practice, we must ensure that the eigenvectors and eigenvalues are real before dotting into $\eb$, therefore we diagonalize $(\grad\bb{u} + \grad\bb{u}^\mathsf{T})/2$ with $\mathsf{T}$ denoting the transpose. Once the eigenvectors are obtained, the cosine of their angle with $\eb$ is calculated, yielding a compressing $\cos \theta$ and a stretching $\cos \theta$ at each grid point within the domain. These cosines are then compiled into a probability distribution $\mathcal{P}(|\cos\theta|)$, representing the likelihood that the alignment angle cosine takes on a specific value at a given point in time. We have found that the compressing and stretching distributions are qualitatively identical for all simulations performed in this work, which is expected for Alfv\'enic turbulence \citep[as opposed to the fluctuation dynamo studied in][]{St-Onge2020}. For that reason, it suffices to only show one of the two, which we choose to be the stretching angle. To exhibit the $\mathcal{P}(|\cos \theta|)$ as a function of scale, we first Fourier transform $\bb{u}$, filter it using masks in $k_\perp = \sqrt{k_x^2+k_y^2}$ bins, and then transform it back to real space before performing all of the aforementioned operations to obtain $\mathcal{P}(|\cos\theta|)$. The $\mathcal{P}(|\cos\theta|)$ are normalized to unity at each individual $k_\perp$, rather than $\mathcal{P}$ being a distribution in both $|\cos\theta|$ \textit{and} $k$, so as to avoid weighting larger wavenumbers less than smaller ones. When magneto-immutability is active, our expectation that the flow organizes to avoid changes in $|B|$ implies that we should see a small $|\cos\theta|$ between the eigenvectors of $\grad\bb{u}$ and $\eb$, whereas a non-immutable cascade should have $|\cos \theta| \sim 1$. Therefore, we can compare this cosine between passive- and active-$\Delta$ simulations to detect whether the flow behaves in a different fashion so as to produce a cascade with minimal variation of $|B|$.


\subsection{The effective equation of state}\label{sec:eos}

\begin{figure}
    \centering
    \mbox{\hspace{1em}${\color{black}(a)}$\hspace{0.48\textwidth}${\color{black}(b)}$\hspace{0.4\textwidth}}\\  \includegraphics[width=0.5\textwidth]{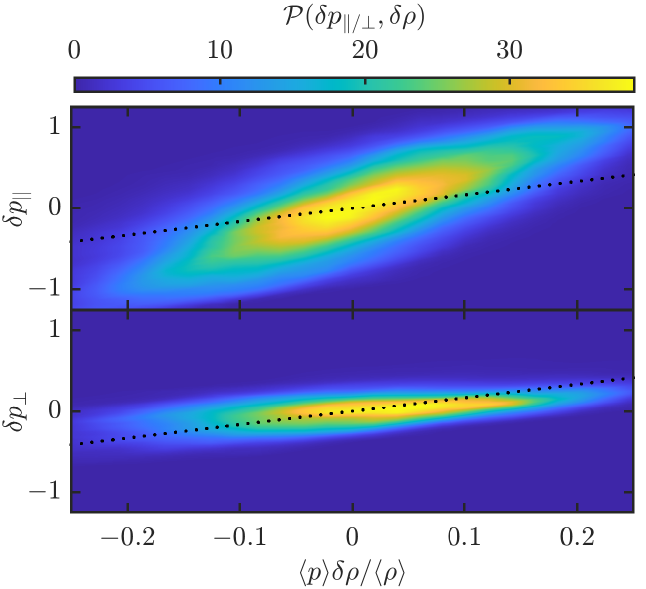}
    \includegraphics[width=0.48\textwidth]{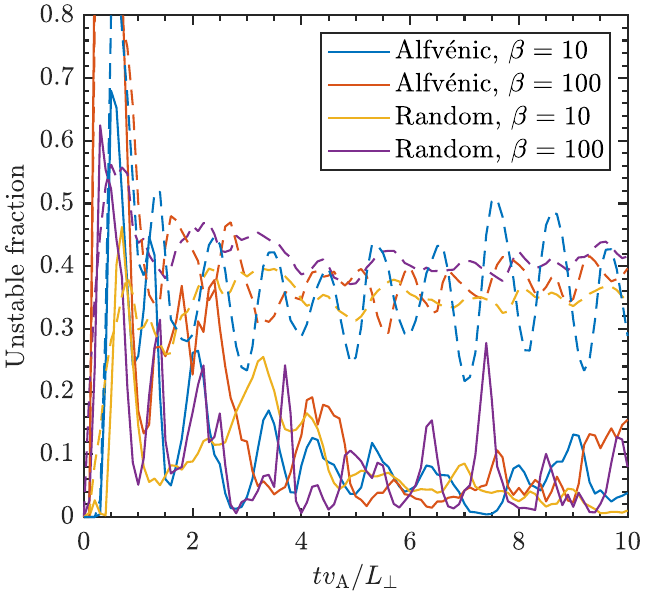}    
    \caption{($a$) Probability distributions of fluctuations in $p_\|$, $p_\perp$, and $\rho$, for $\beta=10$ Alfv\'enically driven turbulence. The black dotted lines represent a slope of $5/3$, the expectation for single-adiabatic, collisional MHD. Neither $\delta p_\|$ nor $\delta p_\perp$ appear to align with the MHD prediction or with each other. ($b$) The fraction of the domain whose pressure anisotropy lies beyond the microinstability thresholds as a function of time for all active-$\Delta$ (solid) and passive-$\Delta$ (dashed) simulations. The passive-$\Delta$ simulations typically have ${\sim}50\%$ of their volume unstable, whereas the unstable fraction of the active-$\Delta$ simulations typically remains below $10\%$.}
    \label{fig:eos}
\end{figure}

The most fundamental assumption underlying all of the conclusions of \S\ref{sec:theory} is that the plasma behaves in a sufficiently collisionless manner to be described by the CGL-MHD equations. For this to be true, the portion of the plasma having pressure anisotropy that is microphysically unstable, and therefore subject to an effective collisionality associated with particle scattering off Larmor-scale fluctuations (i.e. that with $|\Delta| \gtrsim \beta^{-1}$), must constitute a small fraction of the total volume. In their initial study of magneto-immutability in the CGL-MHD system, \citet{squire23} found that magneto-immutability suppresses the overall level of fluctuations in the pressure anisotropy, thereby reducing the fraction of the plasma that is unstable. To establish the impact that this suppression has on the plasma's effective collisionality, we plot in figure~\ref{fig:eos}($a$) the relationship between the fluctuations in the parallel and perpendicular pressures and the (relatively small) fluctuations in the density, averaged in time over the interval $tv_{\rm A}/L_\perp = [8,10]$. In an MHD-like plasma, both $\delta p_\perp$ and $\delta p_\|$ would scale with $\delta\rho$ with a slope of $5/3$, the single adiabatic index characteristic of a collisional plasma (black dotted lines). Neither is particularly well aligned with the single adiabatic index slope, with $\delta p_\|$ having a steeper slope, and $\delta p_\perp$ a shallower one. The significant differences between the pressure slopes suggest that the $\delta p_{\perp/\|}$ evolve in an independent manner as is expected for a predominantly collisionless plasma.

This non-MHD equation of state is explained by figure~\ref{fig:eos}($b$), which compares the fraction of the domain that is unstable to micro-instabilities between active- (solid curves) and passive-$\Delta$ (dashed curves) simulations as a function of time. In the steady state of the passive-$\Delta$ simulations, nearly half of the domain lies beyond the instability thresholds, meaning that much of the plasma volume is experiencing the large scattering rate $\nu_{\rm lim}$, and a collisionless model would not be a good description for a significant portion of the turbulent dynamics. Conversely, the active-$\Delta$ simulations are rarely beyond 10\% unstable, with an average falling closer to just 5\% in steady state. Importantly, this reduction in the unstable fraction appears to be independent of $\beta$, suggesting that higher values of $\beta$ are not likely to result in a significantly more collisional plasma. Infrequent excursions of the unstable fraction beyond 10\% do occur in all simulations, which could either be the result of intermittency or the randomness of the forcing.\footnote{The only occasion in which the volume-filling fraction exceeds $50\%$ in any simulation is during the first Alfv\'en crossing time of the domain. Although we do find enhanced heating during this initial period, the turbulence is still undeveloped, and there appears to be no trend relating the volume-filling fraction during this period to the steady state behaviour.} Though generally short lived, these brief events do have the ability to change the statistics of the turbulence. However, their cause is unlikely to originate in the inertial range, because the fluctuation amplitudes there are small and satisfy the ordering \eqref{eq:ord}. As we demonstrate in \S\ref{sec:org}, magneto-immutability is least active at the outer scales where the turbulence is being forced, and so these intermittent bursts are likely driven by large-scale motions. That being said, the anisotropic pressure stress is inherently non-local in $k$-space, so these bursts' effects on the turbulence are not limited to the largest scales. Therefore, all statistical measurements we report are obtained by averaging over a time interval of no less than $2L_\perp/v_{\rm A}$, taken beyond $tv_{\rm A}/L_\perp = 6$ to average over these bursts and ensure that the turbulence statistics have reached an approximate steady state.


\subsection{Compressive forcing and ion-acoustic fluctuations}\label{sec:drive}

\begin{figure}
    \centering
    \mbox{\hspace{1em}${\color{black}(a)}$\hspace{0.48\textwidth}${\color{black}(b)}$\hspace{0.4\textwidth}}\\  \includegraphics[width=0.49\textwidth]{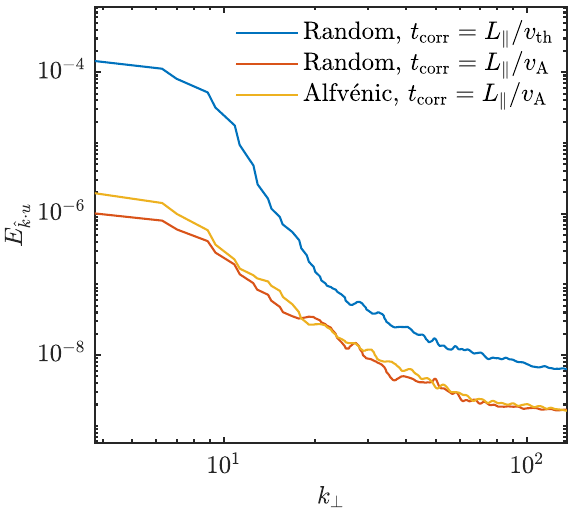}
    \includegraphics[width=0.465\textwidth]{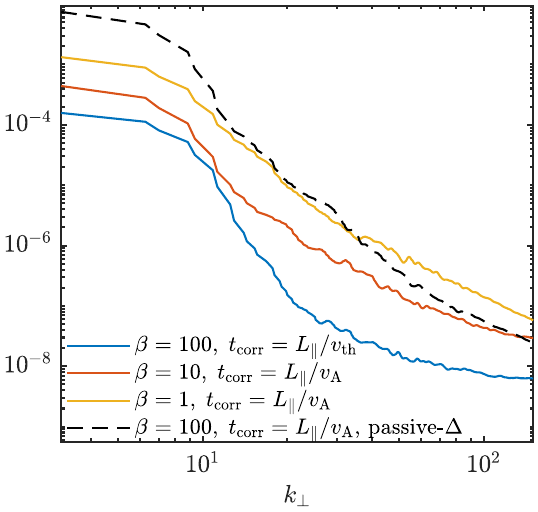}    
    \caption{Spectra of compressive velocity fluctuations versus perpendicular wavenumber. ($a$) The compressive spectra at $\beta=100$ are studied as a function of the type of forcing. When the correlation time is Alfv\'enic, random and Alfv\'enic forcing both produce similar results, with little energy in the compressive flow. If the correlation time is instead sonic, then an increase of two orders of magnitude is seen in the compressive flow energy. ($b$) Compressive spectra of randomly driven turbulence at different values of the plasma $\beta$. At $\beta=100$ the forcing is sonic, leading to significant compressive fluctuations at the outer scale, but the spectrum is extremely steep. As $\beta$ is reduced, the spectra become decreasingly steep, qualitatively approaching the passive-$\Delta$ result (black dashed).}
    \label{fig:compspec}
\end{figure}

With the knowledge that the turbulence is taking place within a predominantly collisionless plasma, we surmise that the compressive wave fluctuations will behave not as collisional fast and slow magnetosonic modes, but rather as collisionless fast and ion-acoustic waves. As such, we evaluate the feasibility of an ion-acoustic wave cascade at high $\beta$, given our expectation that ion-acoustic waves are not effectively mixed and cascaded by the Alfv\'enic fluctuations in such plasmas. Conveniently, the only waves that actively make $\grad \bcdot \bb{u}\ne 0$ in a collisionless plasma are the fast wave and the ion-acoustic wave (non-propagating modes are pressure balanced and approximately incompressible). Thus, the presence of ion-acoustic waves in the inertial range of these turbulence simulations can be diagnosed by the compressive flow spectrum of $\hat{\bb{k}} \bcdot \bb{u}_k$, which is provided in figures~\ref{fig:compspec}($a$) and ($b$). 

In figure~\ref{fig:compspec}($a$), we explore the dependence of the compressive flow spectrum on the correlation time of the forcing. When the correlation time of the forcing is Alfv\'enic (as is expected for the astrophysical turbulence we are concerned with here), very little energy is present in compressive fluctuations, and the power-law index of the $\grad \bcdot \bb{u}$ spectrum is rather steep. This is true in both the randomly driven and Alfv\'enically driven setups, with very little difference between the two spectra. Only when the correlation time of the forcing is decreased to be sonic ($t_{\rm corr} \sim L_\|/v_{\rm th}$), can a substantial amount of energy enter into compressive fluctuations. This is likely because, in Alfv\'enically correlated forcing, the timescale associated with the randomization of the forcing is too slow for the forced wavenumbers to significantly drive ion-acoustic waves. 
Indeed, the $\beta$ dependence of this conclusion is captured by figure~\ref{fig:compspec}($b$), where we plot these compressive spectra for only randomly driven simulations at $\beta=1,\;10$ and 100, with the last being sonically correlated. In effect, the $\beta=1$ simulation is also sonically correlated because $v_{\rm th} = v_{\rm A}$, but because a large difference between the Alfv\'en and ion-acoustic wave speeds is not present, the mixing of ion-acoustic waves by Alfv\'en waves is stronger, bringing the spectrum much closer to that of the dashed passive-$\Delta$ simulation, performed at $\beta=100$. The spectrum of the $\beta=100$ CGL-MHD simulation, although sonically correlated to generate substantial outer-scale compressive fluctuations, has an extremely steep spectrum. This steep spectrum implies that very little of the compressive mode energy driven at the outer scale penetrates into the inertial range, supporting our claim that ion-acoustic waves may be ordered out of the dynamics. At large wavenumbers the spectrum does become less steep. However, given the very small amount of energy contained in $E_{\bs{k\cdot u}}$ at such high $k$, this decrease in steepness is likely due to nonlinearities from Alfv\'enic fluctuations. The weaker cascade of compressive modes at high $\beta$ is also evident from the percentage of flow energy contained in compressive fluctuations. This percentage in the randomly forced $\beta=100$ simulations is $1\%$ for the Alfv\'enically correlated passive run, $0.3\%$ for the sonically correlated active run, and $0.007\%$ for the Alfv\'enically correlated active run. Clearly, sonic correlation is a requirement for any substantial amount of compressive flow to be generated, but the spectra in figure \ref{fig:compspec}(b) show this does not guarantee a strong turbulent cascade of that flow.

\begin{figure}
    \centering
    \mbox{\hspace{1em}${\color{black}(a)}$\hspace{0.48\textwidth}${\color{black}(b)}$\hspace{0.4\textwidth}}\\ 
    \includegraphics[width=0.485\textwidth]{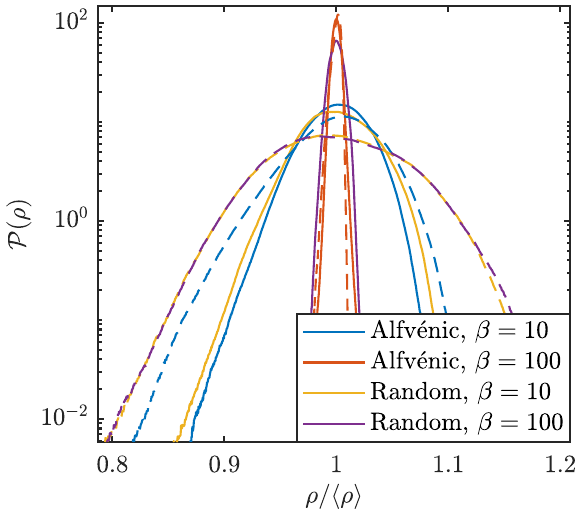}
    \includegraphics[width=0.49\textwidth]{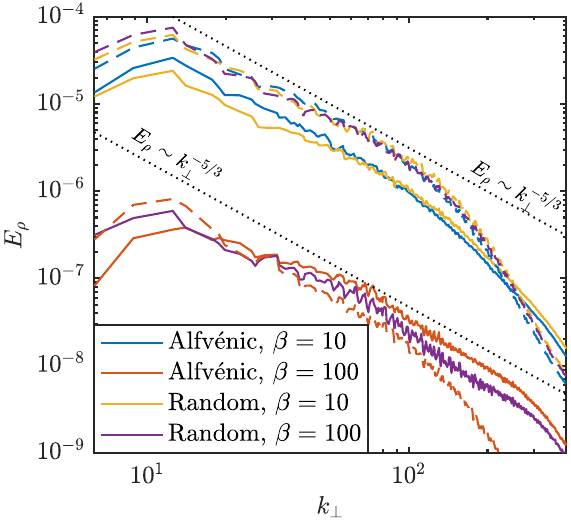}    
    \caption{($a$) Probability distribution of $\rho$ within the domain for all Alfv\'enically correlated simulations, active (solid) and passive (dashed). At higher $\beta$, the decrease in Mach number leads to weaker density fluctuations being driven at the outer scale, and less overall variation in $\rho$. Randomly driven simulations exhibit somewhat increased variation in $\rho$, but the dominant parameter is $\beta$. ($b$) Density fluctuation spectra for all Alfv\'enically correlated simulations. While the overall amplitudes are decreased with increasing $\beta$, the spectra remain strong with spectral indices near $-5/3$. This may indicate that the density is passively advected in the absence of fast and ion-acoustic wave cascades. The apparent break in power law behavior of the $\beta=100$ spectrum is a result of our particular choice for $\nu_{\rm lim}$, which is discussed in more detail within \S\ref{sec:nu}. In both panels ($a$) and ($b$), most all of the passive simulations exhibit larger density fluctuations, with only the $\beta=100$ Alfv\'enically driven passive run having $\delta \rho$ as small as its active counterpart.}
    \label{fig:density}
\end{figure}
The consequences of the lack of compressive modes in the turbulent inertial range can be seen in the statistics of the density fluctuations. Figure~\ref{fig:density}($a$) shows the probability distribution of various values of the density within the simulation domain for each of the Alfv\'enically correlated simulations, both active (solid) and passive (dashed). As $\beta$ is increased, the density takes on fewer values that deviate significantly from the background, with the Alfv\'enically driven simulations being only slightly narrower than the randomly driven runs. These narrow distributions are likely a result of the Mach number becoming smaller, given that our simulations drive an approximately fixed amplitude of $u/v_{\rm A}$ at the outer scale. How the level of these fluctuations depends on scale is more clear in the spectrum of density fluctuations, figure~\ref{fig:density}($b$). Interestingly, even though these fluctuations are clearly ${\lesssim} \epsilon^2 \rho_0$, making them too small to affect the dynamics to leading order, and ion-acoustic wave mixing by the Alfv\'enic cascade is weak, their spectra appear to follow a near $-5/3$ power law, as given by the dotted line. We expect that this is accounted for by the presence of kinetic entropy and non-propagating modes, which, given that they have no real frequency, should be well mixed by Alfv\'enic fluctuations (see \S\ref{sec:waves} and \S\ref{sec:redeq}). As they do not feed back on the overall dynamics due to their small amplitudes, their spectrum is probably captured by passive advection ($\mathrm{d}_t\delta \rho =0$), hence the $-5/3$ index.\footnote{Although none of the simulations we performed produced density fluctuations exceeding $\delta \rho \sim \epsilon^2\rho_0$, it would certainly be possible to achieve $\delta \rho \sim \epsilon^{3/2}\rho_0$ or $\epsilon \rho_0$ by explicitly seeding larger amplitude kinetic entropy or non-propagating modes at the outer scale. To explore this possibility, the consequences of such significant density fluctuations for immutability are discussed in depth within Appendix \ref{app:dens}.} The $\beta=100$ simulations do appear to show a break from the single power law behavior of the other simulations, although this is likely an effect of our choice of $\nu_{\rm lim}$, which is discussed in more detail within \S\ref{sec:nu}. Note that in figures~\ref{fig:density}($a$) and ($b$), all but one of the passive simulations possess density fluctuations larger than their active counterparts; only the $\beta=100$ Alfv\'enically driven passive simulation possesses $\delta \rho$ as small as its active counterpart. The fact that the choice between random and Alfv\'enic forcing makes a difference in the passive simulations at $\beta=100$, but not the active simulations, highlights the separation of compressive timescales between MHD and CGL. In CGL, the type of forcing makes little difference unless it is sonically correlated so that it may excite ion-acoustic modes. On the other hand, the correlation time need not be sonic to excite MHD slow modes, so random forcing can in fact drive larger density fluctuations in the passive simulation.


\subsection{Comparison with previous work}\label{sec:comp}

Certain aspects of our ordering \eqref{eq:ord} and reduced equations \eqref{eq:redeq} explain features of high-$\beta$ CGL turbulence that were already observed in the Alfv\'enically driven simulations of \citet{squire23}, but not yet fully understood. Here we reproduce some of the key results of \citet{squire23}, confirming that our simulations explore the same effects and qualifying the extent to which the compressibility of forcing matters in a system where ion-acoustic and fast modes are not effectively cascaded. All simulations shown within this section are performed at $\beta = 10$ with Alfv\'enically correlated forcing and standard resolution (as defined in \S\ref{sec:code}).

\begin{figure}
    \centering
    \mbox{\hspace{1em}${\color{black}(a)}$\hspace{0.48\textwidth}${\color{black}(b)}$\hspace{0.4\textwidth}}\\ 
    \includegraphics[width=0.475\textwidth]{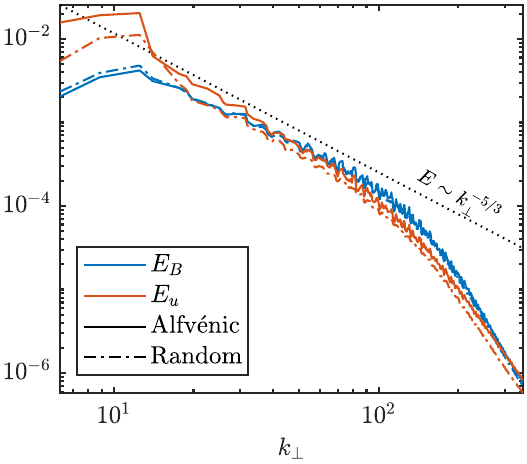}\;
    \raisebox{0.01\height}{\includegraphics[width=0.485\textwidth]{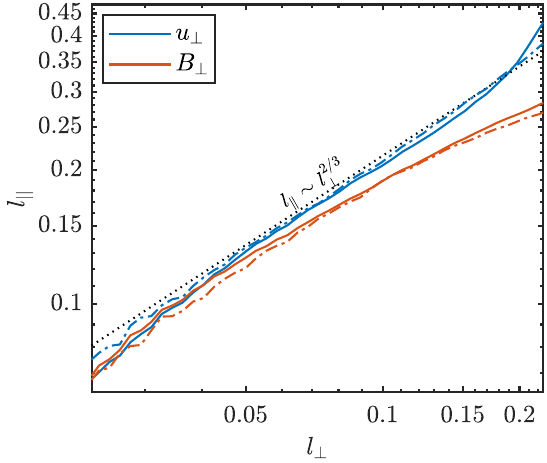}}   
    \caption{($a$) Kinetic energy spectra of $\beta=10$, Alfv\'enically correlated simulations with random (solid) and Alfv\'enic (dash-dotted) forcing. Only slight qualitative differences are visible between each type of forcing, reflecting the lack of a strong ion-acoustic cascade regardless of forcing. ($b$) Characteristic turbulent eddy sizes along and across the local magnetic field for Alfv\'enic variables. Both the randomly and Alfv\'enically driven simulations closely follow the scaling relationship predicted by critical balance in standard MHD, $(l_\|/l_0) \sim (l_\perp/l_0)^{2/3}$, represented by the black dotted line.}
    \label{fig:simspec}
    \centering
    \vspace{1em}
    \mbox{\hspace{3em}${\color{black}(a)}$\hspace{0.45\textwidth}${\color{black}(b)}$\hspace{0.4\textwidth}}\\ 
    \includegraphics[width=0.5\textwidth]{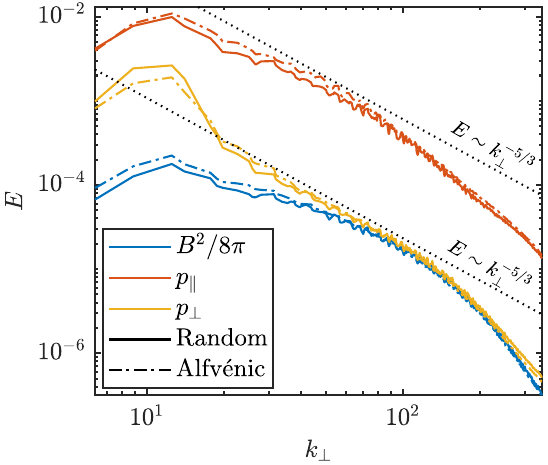}\;
    \includegraphics[width=0.45\textwidth]{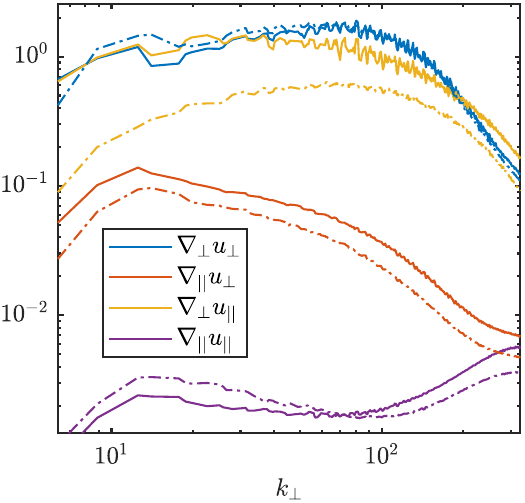}    
    \caption{($a$) Spectra of the perpendicular-thermal, parallel-thermal, and magnetic pressures for randomly (solid) and Alfv\'enically (dash-dotted) driven $\beta=10$ simulations. The difference between $p_\|$ and $p_\perp$, combined with the rough equivalence of $B^2$ and $p_\perp$, reflects perpendicular pressure balance \citep{squire23}. ($b$) The rate-of-strain spectra, showing suppression of $\nabla_\| u_\|$ as predicted by \eqref{eq:redeq}. The most noticeable difference between forcing modes occurs here in the spectra of $\nabla_\perp u_\|$.}
    \label{fig:rospres}
\end{figure}

In figure~\ref{fig:simspec}($a$), the kinetic and magnetic energy spectra from the randomly (solid) and Alfv\'enically (dash-dotted) forced simulations reveal inertial ranges that are close to the $-5/3$ power law expected in MHD turbulence. Although the individual spectral slopes deviate very slightly above or below this exact value, overall there is little qualitative difference between the turbulence resulting from the two modes of forcing. This indicates that the pressure-anisotropy stress does not effectively remove energy from the cascade, at least not to the extent that would na\"{i}vely be expected when $\beta \Delta \sim 1$ \citep{squire23}. Upon careful inspection, the power-law index of the kinetic energy spectrum in the randomly driven simulation is steeper than that of the Alfv\'enically driven simulation, although the difference is small, likely originating from next-order effects not captured by \eqref{eq:redeq}. In figure~\ref{fig:simspec}($b$), the field-perpendicular and parallel scales of the Alfv\'enic flow and magnetic-field perturbations are given. For both $u_\perp$ and $B_\perp$, the perpendicular direction is defined with respect to the average magnetic field of all points being used in the difference equation of the related structure functions \citep[e.g.,][]{chen2011}. The dash-dotted and solid coloured lines once again represent the Alfv\'enic and randomly driven simulations, while the black dotted line represents the $(l_\|/l_0) \sim (l_\perp/l_0)^{2/3}$ relationship predicted for critically balanced MHD turbulence. This appears to indicate that cascades of $\delta B_\perp$ and $u_\perp$ are critical balanced, another foundational assumption of our ordering \eqref{eq:ord}. At the largest scales, some disagreement occurs, but this is to be expected. The definition of $B_\perp$ becomes vague in the presence of larger-amplitude fluctuations, and $u_\perp$ defined with respect to the averaged field will differ substantially from $u_\perp$ defined with respect to the local field at each point in the structure function. Additionally, larger scales may still be somewhat influenced by the forcing, which does not exclusively drive critically balanced fluctuations. Further details of how these characteristic eddy sizes are measured can be found within \S3.2.2 of \citet{squire23} or \S6.3 of \citet{cho09}.

The most easily verified predictions of \eqref{eq:redeq} are that the parallel pressure dominates the pressure anisotropy, and that this results from the perpendicular pressure being balanced by magnetic pressure. In figure~\ref{fig:rospres}($a$), the thermal and magnetic pressure spectra are shown, depicting both of these features.\footnote{Note that the magnetic pressure spectrum is the spectrum of $B^2$ having units of energy squared times length, not to be confused with the magnetic fluctuation spectrum, which is the spectrum of $B$, and has units of energy times length.} The perpendicular pressure, although larger at the outer scale, approaches and ultimately matches the magnetic pressure in the inertial range, as predicted by \eqref{eq:pbal}; as stipulated in the reduced model, this empirical result requires the leading-order $\delta p_\perp$ perturbation to be $\mathcal{O}(\epsilon^2)$. By contrast, fluctuations in $p_\|$ are much larger in amplitude, and all appear to follow the spectral index of $-5/3$ typical of passive advection, with some minor variation. The inertial ranges of these pressure spectra bear a striking resemblance to those of the $\beta=16$, hybrid-kinetic simulation of \citet[figure 8a]{arzamasskiy23}. This is particularly important given that the dynamics of $p_\perp$ and $p_\|$ are generally reliant on the heat fluxes, which the CGL-MHD and hybrid-kinetic approaches model in very different ways. For such agreement to exist between the two models, it is likely that the heat fluxes are reduced by the field lines becoming nearly isotherms. This supports a key prediction of our reduced system \eqref{eq:redeq} (see equations \eqref{eq:gradtprl} and \eqref{eq:gradtprl2}). This suppression of $q_{\perp/\|}$ is also verified specifically for our CGL-MHD simulations within \S\ref{sec:hf}. The $p_\|$ spectrum appears to have a slightly shorter inertial range than the $p_\perp$ and $B^2$ spectra, a feature which may be due to the fact that the heat fluxes are stronger for $p_\|$ than for $p_\perp$. Although the $q_{\perp/\|}$ are nominally ordered out by the reduction in $\nabla_\| \delta T_{\perp/\|}$, it is possible that as the turbulence approaches grid scales magneto-immutability weakens somewhat from the effects of finite resolution. We find that in higher resolution simulations this steepening in $E_{p_\|}$ trends with the grid scale, therefore it is likely a numerical artifact and has no impact on our physical understanding of the $p_\|$ cascade. 
Figure~\ref{fig:rospres}($b$) depicts the rate-of-strain spectra of the turbulent flow, broken up into field-perpendicular and parallel gradients, as well as perpendicular and parallel flows. As expected from \eqref{eq:redeq}, parallel gradients of $u_\|$ are dramatically suppressed with respect to all other elements of the rate-of-strain tensor \citep{squire23}. Importantly, this includes $\nabla_\perp u_\|$, which emphasizes that it is not simply that $u_\|$ itself is being reduced, only its gradients along the magnetic field. Between figures \ref{fig:simspec} and \ref{fig:rospres}, the most apparent difference between random and Alfv\'enic driving exists in the spectra of $\nabla_\perp u_\|$. All other gradients of the flow velocity are similar between each mode of forcing. However, the randomly driven simulation features a spectrum of $\nabla_\perp u_\|$ that is roughly constant with $k_\perp$, unlike in the Alfv\'enically driven run where it is increasing with $k_\perp$, in accordance with standard MHD scalings \citep{squire23}. This highlights the fact that, even though the forcing has a clear effect on $u_\|$, it has no apparent effect on the suppression of $\nabla_\|u_\|$, given that it is equally suppressed for both random and Alfv\'enic forcing.

\begin{figure}
    \centering
    \includegraphics[width=0.9\textwidth]{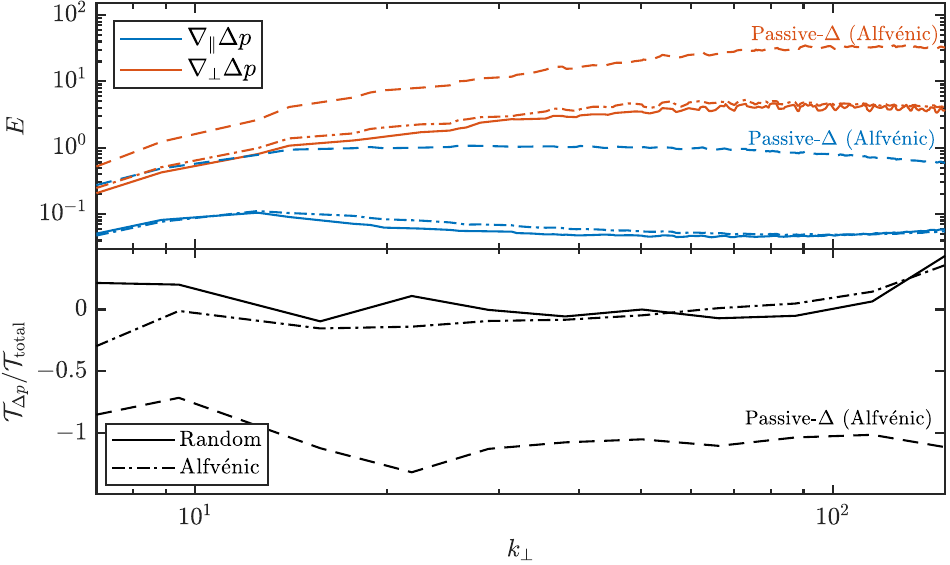}  
    \caption{Top panel: Spectra of the field-parallel and -perpendicular gradients of the pressure anisotropy for $\beta=10$ simulations that are either active and Alfv\'enically forced, active and Randomly forced, or passive and Alfv\'enically forced. Both of the active-$\Delta$ simulations show a significant decrease in $\nabla_\| \Delta p$ with respect to the passive simulation, as well as greater difference between $\nabla_\perp \Delta p$ and $\nabla_\| \Delta p$. Bottom panel: Transfer rate into and out of the turbulent flow due to the anisotropic pressure-stress, normalized to the total cascade rate, for the same simulations as the top panel. Both active simulations show significant suppression of the pressure stress resulting from the reduction of $\nabla_\|\Delta p$ seen in the top panel. The passive simulation predicts a stress that is capable of damping turbulent motions entirely, since $\mathcal{T}_{\Delta p} \sim \mathcal{T}_{\rm total}$ across the full inertial range.}\label{fig:pstress}
\end{figure}
One of the most significant consequences of magneto-immutability is the reduction of the anisotropic pressure stress via suppression of $\nabla_\| \Delta p$. To compare the degree to which this is achieved in random versus Alfv\'enically forced turbulence, we plot the $\nabla_{\perp/\|}\Delta p$ spectra and the $\Delta p$ transfer function $\mathcal{T}_{\Delta p}(k_\perp)$ of our $\beta=10$ simulations in figure~\ref{fig:pstress}. In the top panel, the active-$\Delta$ simulations exhibit clear suppression of $\nabla_\| \Delta p$, especially with respect to $\nabla_\| \Delta p$ from the Alfv\'enically driven, $\beta=10$ passive run. That being said, the difference between the ratio $\nabla_\parallel \Delta p/\nabla_\perp \Delta p$ in the passive and active runs, which is only a factor of ${\approx}2$, is more subtle. Although the reduced equations only explicitly constrain the parallel gradient of $\Delta p$ through \eqref{eq:gradtprl}, the overall production of pressure anisotropy is also reduced by suppressing changes in the magnetic-field strength. As a result, perpendicular gradients of $\Delta p$ are also smaller simply because of the smaller overall magnitude of $\Delta p$ fluctuations.

The results of $\nabla_\| \Delta p$ suppression are seen in the energy transfer due to the pressure anisotropy $\mathcal{T}_{\Delta p}(k_\perp)$ on the bottom panel of figure \ref{fig:pstress} (equation \eqref{eq:trans}). $\mathcal{T}_{\Delta p}(k_\perp)$ is normalized to the total energy transfer rate, which we approximate by the Kolmogorov cascade rate $\mathcal{E}_{\rm K}/\tau_{k_0} \sim \mathcal{E}_{\rm K} (2\upi u_{\rm rms}/L_\perp)$ where $\tau_{k_0}$ is the outer-scale turnover time. For the passive run, $\mathcal{T}_{\Delta p}/\mathcal{T}_{\rm total}$ being $\sim 1$ means that if the turbulence cascaded according to MHD, the anisotropic pressure-stress would remove all of the energy from the turbulent fluctuations, ending the inertial range as soon as it begins. However with immutability, this pressure stress is much smaller, permitting the inertial range to be relatively conservative. Overall, very little difference exists between the randomly and Alfv\'enically driven simulations.


\subsection{Organizing the turbulence}\label{sec:org}

A key aspect of why magneto-immutability constitutes self-organization is that \eqref{eq:ord} does not explicitly order $\nabla_\| u_\|$ as small; rather it becomes necessary for $\nabla_\| u_\|$ to be suppressed in order to satisfy both perpendicular pressure balance and $\mu$ conservation. Given that $\eb$ is always $\mathcal{O}(1)$, and figure~\ref{fig:rospres}($b$) demonstrates that $u_\|$ is significant, it can only be the alignment angle between the magnetic field and the flow rate-of-strain tensor that suppresses variations in the magnetic-field strength. To probe this organization, we analyze our simulations using the alignment angle diagnostic described in \S\ref{sec:diag}.
\begin{figure}
    \centering
    \includegraphics[width=0.95\textwidth]{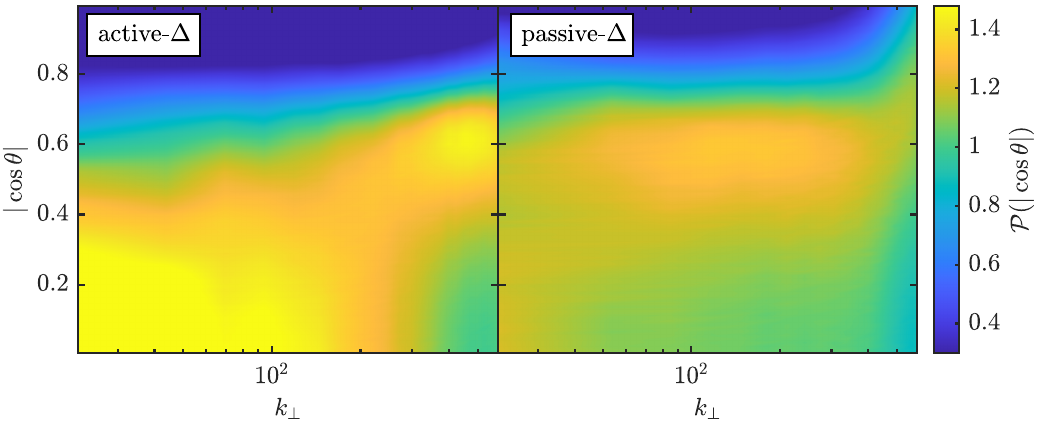}
    \caption{Probability distributions of the cosine of the angle between the rate-of-strain stretching eigenvector and the local magnetic-field direction, for $\beta=10$, Alfv\'enically driven turbulence. Distributions are calculated and normalized individually within each $k_\perp$ bin. The distributions of the compressing eigenvector cosines are qualitatively indistinguishable from their stretching counterparts for both the active and passive runs. The active-$\Delta$ simulation yields $\cos\theta \approx 0$ throughout the inertial range, indicating that motions in the flow that would normally lead to magnetic-field growth are misaligned with $\eb$, rendering them incapable of significantly perturbing $|B|$. By contrast, the passive-$\Delta$ simulation has its peak probability around $\cos\theta \approx 0.6$, which produces an $\mathcal{O}(1)$ dot product between the two vectors and allows significant changes in the magnetic-field strength.}\label{fig:anglevk2d}
\end{figure}
Examples of the scale-dependent $\mathcal{P}(|\cos\theta|)$ for both passive and active Alfv\'enically driven simulations at $\beta=10$ are shown in figure~\ref{fig:anglevk2d}. Recall that the dynamics in the passive simulations are simply isothermal MHD, and so the passively evolved pressure anisotropy does not affect $\mathcal{P}(|\cos\theta|)$. In figure~\ref{fig:anglevk2d}($a$), it is clear that the peak of the distribution from the active run closely follows $\cos\theta \approx 0$ until dissipation scales are reached at large $k_\perp$. This indicates a near-complete misalignment of the rate-of-strain eigenvectors and $\eb$. By contrast, the passive (MHD) run distribution in figure \ref{fig:anglevk2d}($b$) is peaked near $\cos\theta\approx 0.6$, which yields an $\mathcal{O}(1)$ dot product between $\eb$ and $\grad \bb{u}$, thereby permitting more change to $|B|$. In figure \ref{fig:anglevk}, we show the dominant alignment angle as a function of $k_\perp$ for all active (solid) and passive (dashed) simulations by tracking the peak of $\mathcal{P}(|\cos\theta|)$ across $k_\perp$. Without exception, all active simulations exhibit distributions of $\cos \theta$ that are concentrated between 0 and ${\approx}0.2$, while all passive simulations fall between $\cos \theta \approx$ 0.5 and 0.7. In some cases, this misalignment begins very close to the outer scale of the simulation where $u_\perp/v_{\rm A}\sim 1/2$. This might not be expected given that our ordering \eqref{eq:ord} assumes $u_\perp /v_{\rm A} \ll 1$, however, figures \ref{fig:rospres}(a) and \ref{fig:pstress} indicate that it is not required for the fluctuation amplitude to be very small (e.g. $\lesssim 10\%$) before qualitative features of the reduced (magneto-immutable) system are detectable.
\begin{figure}
    \centering
    \includegraphics[width=0.65\textwidth]{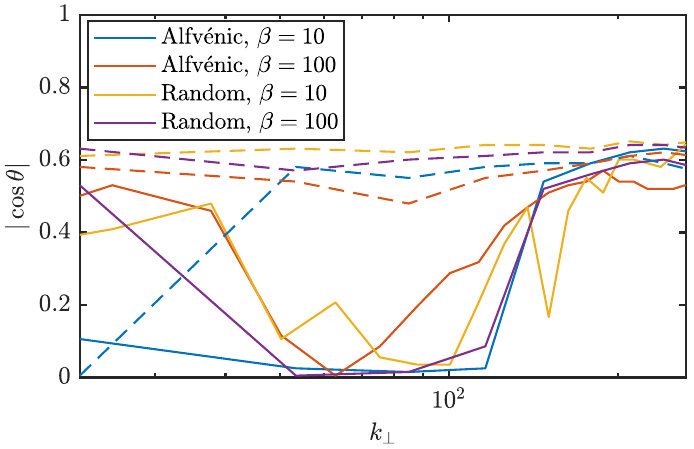}
    \caption{Peak cosines of the $|\cos\theta|$ probability distributions versus $k_\perp$ for various Alfv\'enically correlated simulations, both passive (dashed) and active-$\Delta$ (solid). All active simulations exhibit large misalignment in the inertial range, while all passive simulations show greater alignment, permitting larger changes to $|B|$. There is no clear trend with $\beta$ or forcing mode in the scale at which the active-simulation misalignment ends and begins to resemble the passive simulations. For most this occurs near numerical dissipation scales, which is equivalent for all simulations shown. The lack of misalignment at larger scales is likely due to the forcing not respecting magneto-immutability.}\label{fig:anglevk} 
\end{figure}

Many of the diagnostics shown elsewhere within this work and in \citet{squire23} demonstrate magneto-immutability by comparing with passive simulations that possess exactly identical parameters, such as forcing and resolution. Unfortunately, for particle-in-cell simulations (or reality), there exist no corresponding passive-$\Delta$ simulations that can so accurately represent the MHD equivalent turbulence. With the $\eb\eb\bdbldot\grad\bb{u}$ alignment diagnostic, however, there is no need to compare with an exactly analogous passive-$\Delta$ simulation to determine whether magneto-immutability is at work, as small values of $|\cos \theta |$ alone are sufficient indication. This allows us to search for magneto-immutability within a kinetic framework that self-consistently determines the heat fluxes, which the Landau-fluid heat fluxes can only approximate.
\begin{figure}
    \centering
    \includegraphics[width=0.85\textwidth]{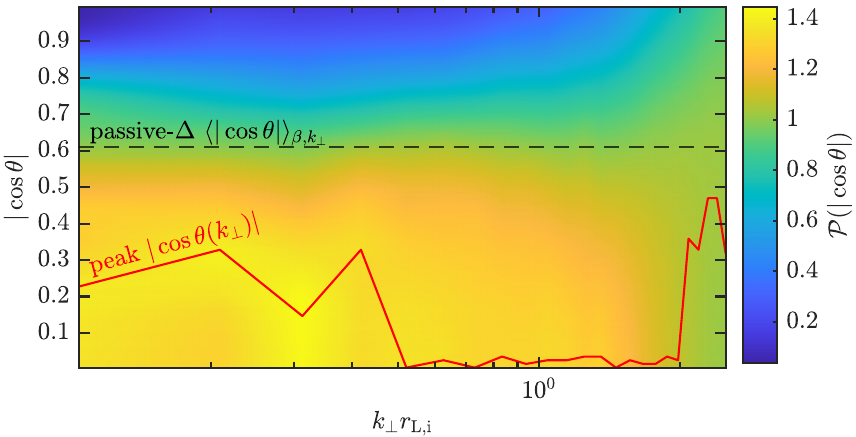}  
    \caption{Distribution of $\eb\eb\bdbldot\grad\bb{u}$ alignment angle $\theta$ as a function of $k_\perp$ for the $\beta=16$, Alfv\'enically driven and correlated simulation of \citet{arzamasskiy23}. This simulation was performed within a $(120.5 \ri)^2 \times 241 \ri$ box, at a resolution of $384^2 \times 768$. The peak alignment angle cosine (red line) is seen to be very close to 0 in the inertial range, in contrast to the average cosine measured from our passive-$\Delta$ (isothermal MHD) simulations (black dashed line). As with our CGL simulations, this misalignment is weaker near the outer scale, where immutability struggles to coexist with the forcing.} 
    \label{fig:picimm}
\end{figure}
In figure \ref{fig:picimm}, we show the calculation of $|\cos\theta|$ for a $\beta=16$, Alfv\'enically driven hybrid-kinetic turbulence simulation initially described within \citet{arzamasskiy23}. The code used to perform this simulation was the \texttt{Pegasus++} hybrid kinetic-ion fluid-electron particle-in-cell code \citep{kunz14}. The simulation shown resolves scales both above and below the ion Larmor scale, additionally incorporating a nonzero electron temperature $T_{\rm e} = T_{\rm i}$. The peak of the $|\cos\theta|$ distribution is traced by the red line, while a comparison with the inertial-range average of all of our passive simulations is provided by the dashed black line. The misalignment is distinctly stronger for the hybrid-kinetic simulation than for the passive-$\Delta$ simulations, lasting until just past the ion Larmor scale. While at scales satisfying $k_\perp r_{\rm L,i} \lesssim 0.4$ the misalignment is weaker, the peak value of the cosine is still significantly less than the MHD average, and the vast majority of the probability density falls well below $|\cos(\theta)|=0.62$. Such imperfect misalignment is also seen in our simulations in figure \ref{fig:anglevk}, and is perhaps not surprising given that these scales are closest to the forcing scales. The peak in alignment at $k_\perp r_{\rm L,i} \sim 0.4$ specifically is likely explained by the presence of oblique firehose modes, which \citet{arzamasskiy23} note grow most rapidly at $k_\perp r_{\rm L,i} \sim 0.4$ and are expected to enhance $\eb\eb\bdbldot\grad\bb{u}$. Those authors also reported significant viscous dissipation near the outer scale, perhaps further evidence that immutability is unable to coexist with the forcing, a fact reflected in the imperfect misalignment demonstrated at the largest scales within figure \ref{fig:picimm}. This outer-scale heating can also be seen to some degree in our simulations in figure~\ref{fig:pstress}, as well as in most of the simulations of \citet{squire23} (who used a different forcing scheme to that used here). Figure 18 of \citet{arzamasskiy23} demonstrates that energy transfer from the viscous stress appears to be suppressed in the inertial range relative to the outer scale, instead being dominated by the Maxwell stress. Additional evidence hinting at the presence of magneto-immutability in \citet{arzamasskiy23} includes the tendency of the turbulence to avoid the instability thresholds (their figure 5), and significant suppression of the spectrum of $p_\perp$ fluctuations in order to maintain pressure balance with the electron and magnetic-field contributions (their figure 8).

\begin{figure}
    \centering
    \includegraphics[width=0.85\textwidth]{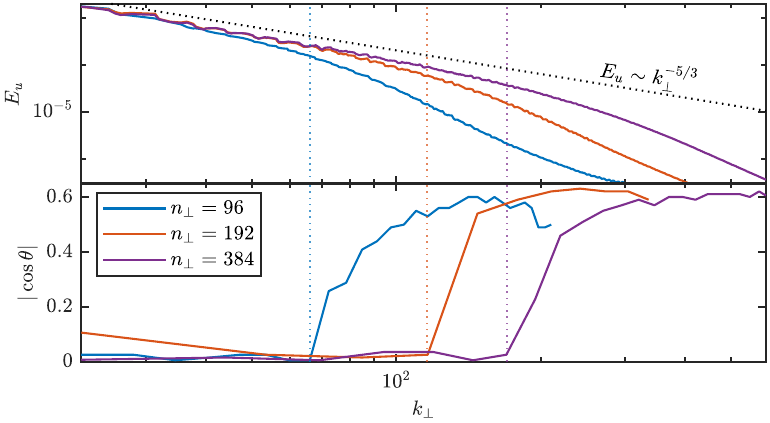}  
    \caption{Top panel: Kinetic energy spectra for Alfv\'enically driven $\beta=10$ simulations at 3 different resolutions across $\bb{B}_0$, all with $n_\|=2n_\perp$. The dotted black line represents a $k^{-5/3}$ power law. Bottom panel: Alignment angle cosines as a function of $k_\perp$ for each of the three resolutions, with vertical dotted lines marking the transition away from misalignment of $\grad \bb{u}$ with $\eb$. It appears that the transition trends with the dissipation scale of the turbulence, likely as a result of departure from the ordering \eqref{eq:ord}.}
    \label{fig:scalesep}
\end{figure}

In the CGL simulations depicted by figures \ref{fig:anglevk2d} and \ref{fig:anglevk}, there is no obvious trend of the scale (in $k_\perp$) at which immutability ceases to affect $\cos\theta$; most simulations trend toward MHD-like alignment at roughly the same scale. To understand the meaning of this scale, we compare the alignment angle cosines with the kinetic energy spectra of the $\beta=10$, Alfv\'enically driven simulation at several different resolutions in figure \ref{fig:scalesep}. The resolutions are given for the coordinates perpendicular to $\bb{B}_0$ of each run, with $n_\| = 2n_\perp$. A clear trend with resolution exists, meaning that the alignment angle cosine only changes when the flow reaches dissipation scales and the turbulence no longer follows the ordering \eqref{eq:ord}. Notably, this misalignment persists to smaller scales than those satisfying $\delta B_\|/B_0 \gtrsim \beta^{-1}$, where the change in the magnetic field strength due to individual turbulent fluctuations is large enough to generate $\beta\Delta \gtrsim 1$ (in the absence of immutability). This implies that, at least as far as the resolutions we can probe go, magneto-immutability is not an outer-scale effect but rather persists throughout the inertial range regardless of how small $\beta\Delta$ is at a given $k_\perp$, albeit in the somewhat modified sense discussed in \S\ref{sec:feats}.


\subsection{The role of heat fluxes}\label{sec:hf}

Not only is the $\Delta p$ stress suppressed by the reduction of $\nabla_\| \delta T_\|$ and $\nabla_\| \Delta p$, the heat fluxes are as well. Importantly, as predicted by \eqref{eq:redeq}, the $\nabla_\|\delta T_{\perp/\|}$ reductions are not diffusive (i.e., caused by the strong heat fluxes at high-$\beta$), but rather dynamical, originating from the momentum equation. For that reason, we should be able to artificially enhance or suppress the heat fluxes in our simulations -- for example, by adjusting the parameter $|k_\parallel|$ -- and observe little effect on the signatures of magneto-immutability. 
\begin{figure}
    \centering
    \includegraphics[width=0.87\textwidth]{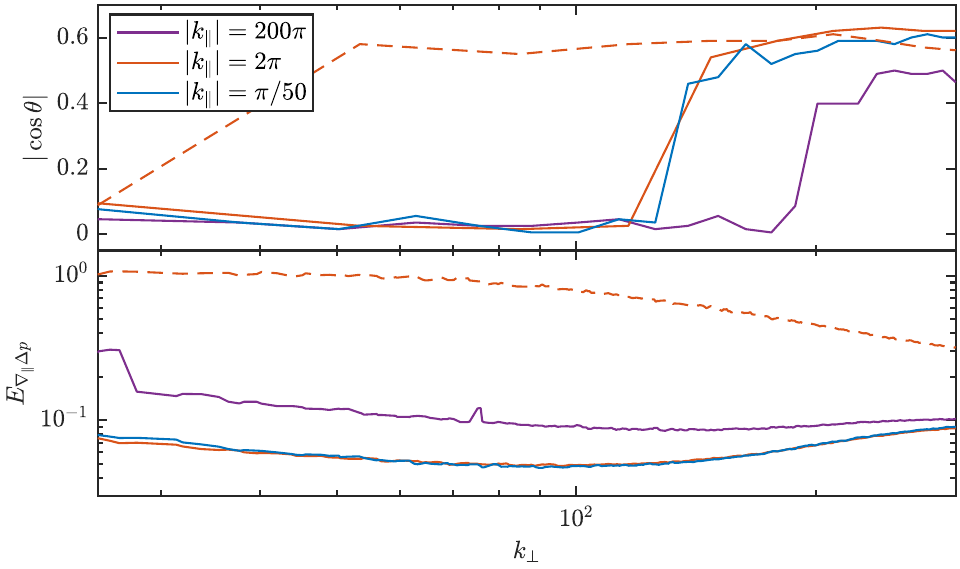}  
    \caption{Top panel: Alignment angle cosine versus $k_\perp$ for $\beta=10$, Alfv\'enically driven simulations with $|k_\||$ in \eqref{eq:fullhf} increased (purple), unmodified (orange), and decreased (blue) by a factor of 100. Given a change in heat flux strength of $10^4$, the difference between strong (blue) and weak (purple) heat fluxes appears to have little effect on $\theta$. An orange dashed line represents the passive-$\Delta$ equivalent of the unmodified heat flux run. Bottom panel: The spectra of $\nabla_\| \Delta p$ for each run. The large suppression of parallel gradients in the pressure anisotropy reflect the suppression of $\nabla_{\|} T_{\|/\perp}$, as predicted by \eqref{eq:redeq}. This limits the ability of heat fluxes to play a role in the turbulence, and importantly does not rely on their strength, with even the $k_\| = 200\pi$ simulation, which is effectively double-adiabatic (purple), showing a comparable reduction. The reduced $\nabla_\| \Delta p$ also implies that the heat fluxes do not interfere in the avoidance of significant $\Delta p$-stress.} 
    \label{fig:hf}
\end{figure}
In figure~\ref{fig:hf} we gather plots of $\cos\theta$ and the $\nabla_\| \Delta p$ spectra, expressed as a function of $k_\perp$, for simulations where $|k_\||$ is increased or decreased by a factor of 100. In all runs depicted, $\beta=10$ and the forcing is Alfv\'enic, with a corresponding passive simulation given as an orange dotted line, given that it was performed with the nominal Landau wavenumber of $|k_\|| = 2\upi$. The simulation with $|k_\||$ increased by 100 (purple curve in each plot) is effectively double-adiabatic MHD because of how weak the heat fluxes are. 

In the top panel of figure \ref{fig:hf}, little variation is seen in the alignment angle cosine as a function of $k_\perp$, with only the double-adiabatic run being misaligned slightly further in $k_\perp$ than the others. Considering that the heat fluxes are $10^4$ times stronger for the blue curve than the purple curve, the extension of misalignment by less than a factor of 1.5 in $k_\perp$ is an extremely small difference. In the bottom panel, the field-parallel gradients of the pressure anisotropy are shown, featuring strong suppression of $\nabla_\| \Delta p$ in the active-$\Delta$ simulations as compared to the passive run. Although the active simulations are all more similar to each other than the passive one, parallel gradients of the double adiabatic run are moderately larger, by a factor of $\sim 2$. The comparable reduction of $\nabla_\| \Delta p$ for all values of $|k_\||$ shown implies that the heat fluxes have little effect on whether or not immutability is able to effectively avoid $\Delta p$-stress. Note that the results of figure \ref{fig:picimm} suggest that that this heat flux suppression in figure \ref{fig:hf} extends to the kinetic simulations of \citet{arzamasskiy23}, where the $q_{\perp/\|}$ are not approximated via the Landau-fluid form. In order for $\eb \bcdot \grad u_\|$ to be relegated to next order in \eqref{eq:imm}, the heat fluxes must be negligible, a fact also made clear by \eqref{eq:rkmhdmu} of Appendix \ref{app:rkmhd}. Therefore, to observe the kind of misalignment seen in figure \ref{fig:picimm}, it must be that the fully kinetic heat fluxes, like the Landau-fluid approximations, have little effect on the leading-order dynamics. Unfortunately, unless a closure for $q_{\perp/\|}$ is assumed this cannot be proven in general. Nonetheless, these results still further the favourable comparison between Landau-fluid CGL and kinetic simulations of high-$\beta$ collisionless turbulence.


\subsection{The role of micro-instabilities}\label{sec:nu}

The only portions of this study that are not described by our ordered equations \ref{eq:redeq} are the role of microinstability limiters. Within the CGL model, dependence on this physics arises via our choice of $\nu_{\rm lim}$. As discussed in \S\ref{sec:eos}, the sizes and distributions of micro-unstable patches are highly intermittent and difficult to predict; however, by varying $\nu_{\rm lim}$ we can obtain useful information about how they interact with the turbulence. For all other simulations outside of this section, the instability scattering rate is fixed to $\nu_{\rm lim} = 10^{10} v_{\rm A}/L_\perp$, essentially providing a hard wall on the maximum possible pressure anisotropy. Given that the microinstabilities being accounted for grow at $\ri$ scales, which are formally zero in the CGL-MHD system, it seems reasonable that they might scatter particles at a rate much faster than any of the dynamics being studied. For this reason, and for simplicity, such `hard-wall' limiters have been employed frequently in past studies of pressure-anisotropic turbulence far above $\ri$ scales \citep[see \textit{e.g.},][]{sharma06,squire23,squire19,santoslima14}. However, studies focused on the mirror and firehose instabilities in the absence of background turbulence have found that the scattering rate induced typically follows the relationship $\nu \sim \beta \eb\eb \bdbldot\grad\bb{u}$ \citep{kunz14,riquelme18}, which in a turbulent environment takes on different values at different scales. Unfortunately, this is expensive to implement directly as it requires measuring the flow shear locally at each time step. To investigate the consequences of different choices for $\nu_{\rm lim}$, we perform Alfv\'enically driven simulations at $\beta=100$, where the relative forcing of $\Delta p$ is larger than at $\beta=10$, and vary $\nu_{\rm lim}$ between 20, 200, and $10^{10} v_{\rm A}/L_\perp$. 

\begin{figure}
    \centering
    \mbox{\hspace{1em}${\color{black}(a)}$\hspace{0.7\textwidth}}\\ 
    \includegraphics[width=0.78\textwidth]{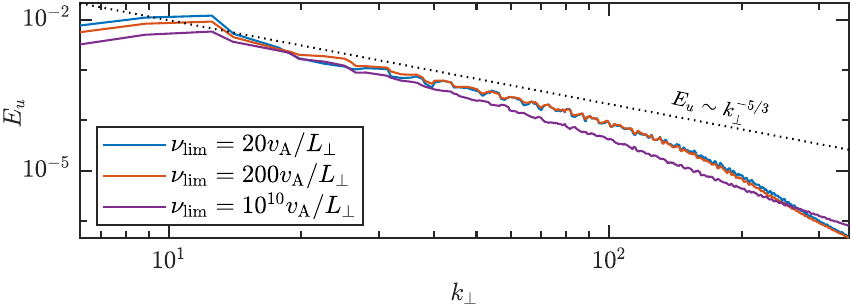}\\
    \mbox{\hspace{6em}${\color{black}(b)}$\hspace{0.35\textwidth}${\color{black}(c)}$\hspace{0.4\textwidth}}\\
    \includegraphics[width=0.405\textwidth]{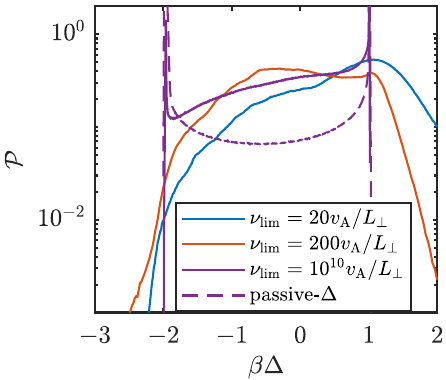}
    \raisebox{-0.015\height}{\includegraphics[width=0.375\textwidth]{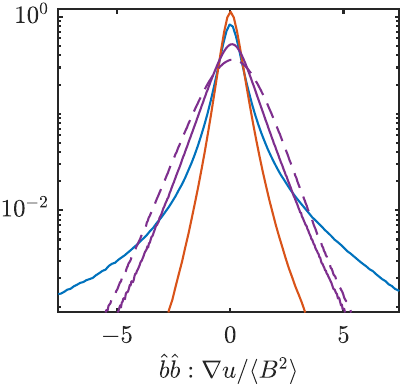}}\\    
    \mbox{\hspace{1em}${\color{black}(d)}$\hspace{0.7\textwidth}}\\ 
    \includegraphics[width=0.78\textwidth]{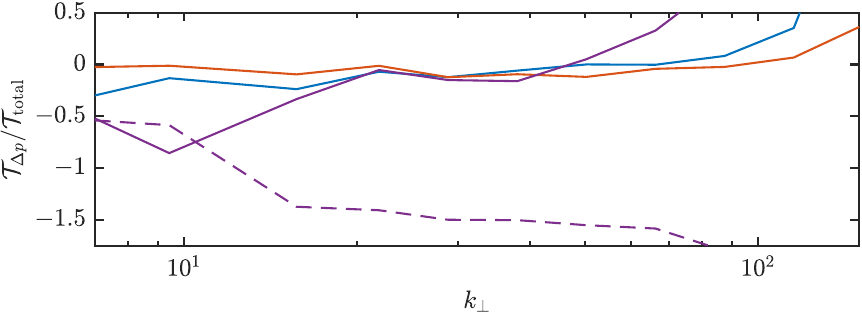}
    \caption{($a$) Kinetic energy spectra for $\beta=100$, Alfv\'enically driven simulations with $\nu_{\rm lim} \in [20, 200, 10^{10}]v_{\rm A}/L_\perp$. Significant spectral steepening is observed for the simulation with hard-wall limiters ($\nu_{\rm lim} = 10^{10}v_{\rm A}/L_\perp$, the default in all other simulations), as well as a slightly earlier apparent dissipation scale. ($b$) Probability distribution of the values of $\beta\Delta$ for each simulation. The hard-wall-limited simulation exhibits significant peaks with cutoffs near the mirror and firehose thresholds, while lower $\nu_{\rm lim}$ simulations extend beyond the cutoffs. The $\nu_{\rm lim} = 200 v_{\rm A}/L_\perp$ simulation is the only run having a distribution with a global maximum between the instability thresholds. ($c$) Probability distribution of the values of $\eb\eb \bdbldot \grad \bb{u}$ for each simulation, demonstrating that the width of the distribution reflects the strength of magneto-immutable organization. The distribution from the hard-wall-limited simulation is the closest of the active simulations to the passive, non-immutable simulation, with $\nu_{\rm lim} = 200v_{\rm A}/L_\perp$ being the narrowest, suggesting that there is an intermediate value of $\nu_{\rm lim}$ that allows magneto-immutability to act most effectively. ($d$) The $\Delta p$-stress transfer functions for each value of $\nu_{\rm lim}$. There is clearly less dissipation resulting from pressure anisotropy at $\nu_{\rm lim}=200v_{\rm A}/L_\perp$ than in any other run, with the hard-walled simulation by far experiencing the most viscous dissipation.
}
    \label{fig:nu}
\end{figure}
The effects of these variations in $\nu_{\rm lim}$ on the kinetic spectra are shown in figure~\ref{fig:nu}($a$). Interestingly simulations with lower scattering rates, where the pressure anisotropy is allowed to make larger excursions beyond the microinstability thresholds, lead to less steep spectral indices that are much more in line with the $-5/3$ expectation of our ordering and a conservative cascade. On the other hand, the simulation with a hard-wall limiter $\nu_{\rm lim}$, as used in all other simulations throughout this work, exhibits a spectrum that is slightly steeper and appears to be dissipated earlier in $k_\perp$. This steepening implies an increase in damping from the $\Delta p$ stress, as some of the energy in the turbulent motions is being removed from the cascade.\footnote{The steepening in the hard-wall-limited simulation flow spectrum is also the most probable cause of the features observed in the density spectra of figure~\ref{fig:density}($b$) at $\beta=100$, since the density fluctuations are likely passively cascaded.} The reason for this steepening and apparent dissipation can be seen in figures~\ref{fig:nu}($b$) and ($c$), which present PDFs of the measured values of $\beta\Delta$ and $\eb\eb \bdbldot \grad \bb{u}$ in each run. These diagnostics essentially represent how effective magneto-immutability is at regulating the overall magnitude and production rate of pressure anisotropy. In the hard-wall-limited passive simulation (purple dashed), the distribution of $\eb\eb \bdbldot \grad \bb{u}$ is less peaked in the stable regions ($-2<\beta\Delta<1$) than any of the active runs, with values of $\beta\Delta$ that are effectively pinned to the mirror and firehose limiters, and no local peak in $\beta\Delta$ between the limiters. The hard-wall-limited active simulation is the least peaked of the active simulations at $\eb\eb \bdbldot \grad \bb{u} = 0$, also having no local peak in $\beta \Delta$ between the instability limiters. By comparison, lower values of $\nu_{\rm lim}$ appear to better focus the distribution of $\eb\eb \bdbldot \grad \bb{u}$ around 0, and at least in the case of $\nu_{\rm lim} =200v_{\rm A}/L_\perp$, maximize the proportion of $\beta\Delta$ that lies between the instability limiters. Note that, in the process of pushing $\Delta p$ between the instability thresholds, a slight background ($k=0$) pressure anisotropy develops in the $\nu_{\rm lim} = 200v_{\rm A}/L_\perp$ run, with the most probable value of $\beta\Delta$ in figure \ref{fig:nu}(b) being $\sim-0.5$ rather than 0. This is simply an effect of the asymmetry of the instability thresholds, and is expected to occur whenever the outer-scale amplitude is sufficiently large that fluctuations can reach both microinstability thresholds. If the simulation were initialized with a homogeneous, background (undriven) pressure anisotropy, the eventual steady state would still resemble that shown here, as the microinstability limiters alone set the location of this peak in the PDF. The consequences of the varying strengths of magneto-immutable self-organization are shown in figure~\ref{fig:nu}($d$), via the $\Delta p$-stress transfer functions of each simulation. It is clear that the $\nu_{\rm lim}=200v_{\rm A}/L_\perp$ simulation, where magneto-immutability appears to enforce most strictly the suppression of $\eb\eb\bdbldot\grad\bb{u}$, experiences the least amount of viscous dissipation from the pressure anisotropy. The $\nu_{\rm lim}=20v_{\rm A}/L_\perp$ run dissipates slightly more, and the hard-walled simulation experiences significant stress far into the inertial range. This likely explains why the hard-walled simulation exhibits kinetic energy spectra that are both steeper and shorter than the other two runs (panel $a$).

We hypothesize that this weakening of magneto-immutability occurs because whenever a strong limiter scattering rate $\nu_{\rm lim}$ is activated, the magneto-immutable orderings of \eqref{eq:cglpprl} and \eqref{eq:cglpprp} are broken. In an MHD plasma with $\nu^{-1}$ being much less than any dynamical timescale, this does not lead to any dissipation because the scattering drives the anisotropy to zero, and hence there is no anisotropic pressure stress. However, when the scattering is induced by microinstabilities, it only drives the anisotropy to $\beta\Delta \sim 1$ levels. This is visible in the hard-wall-limited curve of figure~\ref{fig:nu}($b$), where the distribution of $\beta\Delta$ is sharply peaked at the instability thresholds (although the proportion of points $\mathcal{P}(\Delta)\rmd\Delta$ at the thresholds is still not large). As a result, $\Delta$ still remains dynamically important, but it is no longer capable of organizing to avoid $\Delta p$-stress, because the collisional term makes the dynamical equations different from those that support magneto-immutability. Therefore, the more severely the ordering is violated by the instability limiters, the more the anisotropic pressure-stress is allowed to affect the cascade of turbulent energy. Interestingly then, the results of figure \ref{fig:nu} imply that throughout this work, many of the signatures of magneto-immutability we detect would become even stronger if a lower (and more realistic) value of $\nu_{\rm lim}$ were used. That being said, it appears that if $\nu_{\rm lim}$ is too small, then the organization can again become somewhat less efficient. This drop in efficiency may be a result of the amplitude of $\beta\Delta$ exceeding the values expected by the ordering, possibly upsetting the $\beta\Delta \sim 1$ assumption.

While there is clearly a choice for $\nu_{\rm lim}$ that maximizes magneto-immutability, the question remains of whether such a choice is physically motivated or not. An instructive comparison to make in this pursuit is between our optimal $\nu_{\rm lim}$ and the effective scattering rate measured in the high-$\beta$, hybrid-kinetic simulations of \citet{arzamasskiy23}. Those authors found that microinstability scattering in their simulations appeared to follow the relationship $\nu \sim \beta (\eb\eb \bdbldot \grad \bb{u})_L$, where the $L$ subscript denotes that the quantity is estimated at the outer scale. We can estimate this rate for the simulations of figure~\ref{fig:nu} from the RMS velocity $\langle u\rangle_{\rm RMS} \approx 0.75 v_{\rm A}$ to find that  $\nu \sim 235v_{\rm A}/L_\perp$. Interestingly, this nearly coincides with the value of $\nu_{\rm lim}$ in our survey that is minimally disruptive to magneto-immutability.\footnote{This scattering rate would ideally depend on the local value of the parallel rate of strain, rather than an average estimate. To visualize the range of values that $\beta(\eb\eb\bdbldot\grad\bb{u})_L$ could take on then, the horizontal axis of figure \ref{fig:nu}(c) can be multiplied by $p_0$, which in our code units equals 100 for each of the simulations shown. The result is a PDF of the potential values of $\nu_{\rm lim}$ driven by each simulation. Note that the fluctuations that generally determine the scattering rate are not those at the core of the distribution, but those near the tails. This is because microinstabilities are driven in our simulations through intermittent, larger-amplitude fluctuations rather than in the bulk of the plasma, given our small volume-filling fractions. We propose that the source of this agreement comes from our understanding of how magneto-immutability behaves in weakly collisional, Braginskii-MHD plasmas: when a pitch-angle scattering rate is included in the equations governing the anisotropic pressures and the ordering \eqref{eq:ord} is satisfied, we find that magneto-immutable self-organization can occur at all scales satisfying (see Appendix \ref{app:brag} for derivation)}
\begin{equation}
    \beta \frac{k_\|v_{\rm A}}{\nu} \gg 1
\end{equation}
If we naively substitute the result of \citet{kunz14} that, for pressure anisotropy driven by a flow shear, $\nu \sim \beta \eb\eb \bdbldot \grad \bb{u} \sim \beta k_\|u_\|$, then the condition for magneto-immutability subject to microinstability scattering is just that $u_\| \ll v_{\rm A}$. This condition is automatically satisfied by virtue of our assumed ordering. The only effect that can then interrupt magneto-immutability in sub-Alfv\'enic turbulence is the forcing, hence why \citet{arzamasskiy23} obtained a scattering rate based on the parallel rate-of-strain at the outer scale. Indeed, our figure \ref{fig:anglevk} also demonstrates how magneto-immutable self-organization tends only to fail at the forcing scales for different values of  and types of forcing. Thus, the value of $\nu_{\rm lim}$ in our simulations should also be estimated via $\nu \sim \beta (\eb\eb \bdbldot \grad \bb{u})_L$. In short, it is therefore likely that the most physically motivated choice of $\nu_{\rm lim}$ is also approximately that which best supports a strong, magneto-immutable cascade.

In considering our choice of $\nu_{\rm lim}$, it is worthwhile to also discuss alternative sources of anomalous scattering, such as that resulting from the ion-cyclotron instability or distinct versions of the firehose.\footnote{Another class of kinetic microinstability altogether is associated with significant ion heat fluxes \citep[see, e.g.][]{schekochihin10,bott24}. Given the suppression of field-aligned ion temperature gradients, these instabilities are unlikely to be driven by inertial-range magneto-immutable fluctuations, although they could be driven near the forcing scales. While estimates of anomalous scattering have been verified for electron-scale heat-flux instabilities, no such analyses exist yet for their ion-scale counterparts. For this reason, we cannot comment with certainty on how they would affect the results presented here.} In \citet{bott21} and \citet{arzamasskiy23}, the limiting negative pressure anisotropy is $-1.4/\beta$ corresponding to the oblique firehose instability, rather than the $-2/\beta$ employed in our simulations which represents the parallel (or “fluid”) firehose. If the limiters in our simulations activated at the oblique threshold, some minor quantitative results of the turbulence would change, but we do not expect any significant qualitative changes. In particular, the peak value of the PDF of $\beta \Delta$, which takes on $\approx-0.5$ in figure \ref{fig:nu}(b), would likely shift to $\approx-0.2$ due to there being a new centre between the mirror/firehose thresholds. Additionally, we expect a slightly larger fraction of microinstabilities given the more stringent threshold. To estimate this fraction, we calculated the average microinstability filling fraction with the oblique firehose threshold (rather than the parallel) for the final $\delta t=2L_\perp/v_{\rm A}$ of the $\beta=100$, $\nu_{\rm lim}=200v_{\rm A}/L_\perp$ simulation featured in figure \ref{fig:nu}. We find that $18.5\%$ of the domain is unstable with the oblique threshold, compared to $10.4\%$ with the parallel threshold. While significant quantitatively, this increase is not nearly sufficient to isotropize the entire plasma. Moreover, this estimated increase is almost certainly an overestimate, given that in this simulation no anomalous scattering ($\nu_{\rm lim}$) was turned on until the anisotropy reached $-2/\beta$. By comparison, the volume-filling fraction in the beta=16 hybrid-kinetic simulation of \citet[figure 10]{arzamasskiy23}, which appears to obey the $-1.4/\beta$ threshold, was $17.9\%$. Finally, in their broad parameter survey, \citet{squire23} performed a simulation adopting the oblique firehose threshold, finding in their figure 4 that there was little qualitative difference from the simulations employing the parallel threshold.

With regards to the positive $\Delta p$ instabilities, \citet{ley24} found that a secondary ion-cyclotron instability can be triggered if the mirror threshold is exceeded substantially, causing the positive pressure anisotropy to be limited instead at ${\approx}0.5/\sqrt{\beta}$. However, that threshold is not likely to be relevant to our study for the following reasons. In order for this softer threshold to come into play, the pressure anisotropy must be able to overshoot the more restrictive $1/\beta$ mirror threshold by a substantial amount. Dedicated kinetic studies of the mirror instability in a driven system indicate that the overshoot is proportional to $(\eb\eb\bdbldot\grad\bb{u}/\Omega_{\rm i})^{1/2}$ \citep[where $\Omega_{\rm i} = eB/m_{\rm i}c$ is the ion-cyclotron frequency]{kunz14}. In well-magnetized plasmas like the ICM or black-hole accretion flows, where the separation between the scales driving the pressure anisotropy (which are comparable to the outer scale when magneto-immutability is active) and the ion-Larmor scale is extremely large, this overshoot is predicted to be extremely small. In this case, the mirror instability should reach its saturated state and scatter particles to keep the pressure anisotropy bound within ${\approx}1/\beta$ before the pressure anisotropy can reach the ion-cyclotron threshold at ${\approx}0.5/\sqrt{\beta}$.


\section{Summary and Discussion}\label{sec:conc}

This work has investigated high-$\beta$ collisionless turbulence through analytical and computational means, with special attention paid to the self-organization of the magnetic field and bulk flow via `magneto-immutability'. We introduced a new asymptotic ordering that explicitly makes use of $\beta \gg 1$, and yields a set of reduced CGL-MHD equations that not only reproduces previously found characteristics of magneto-immutability, but also makes several new predictions. Numerical simulations of the full CGL-MHD equations with Landau-fluid heat fluxes and microinstability limiters were then employed to verify the assumptions of our ordering and to test the new predictions. The most important conclusions drawn by this study are: 
\vspace{0.5em}
\begin{enumerate}[align=left,leftmargin=2em,itemindent=0em,labelsep=0.5em,labelwidth=1.5em]
  \setlength\itemsep{0.5em}
\item Magneto-immutability, as defined by the suppression of both $\eb\eb\bdbldot\grad\bb{u}$ and $\nabla_\|\Delta p$ through self-organization, is an inertial-range effect in high-$\beta$ turbulence that satisfies the ordering \eqref{eq:ord}.
\item By suppressing $\eb\eb\bdbldot\grad\bb{u}$, magneto-immutability reduces the fraction of the plasma that otherwise would have had its pressure anisotropy well beyond the mirror/firehose thresholds, in turn preventing the plasma from behaving in an entirely collisional manner through microinstability-induced scattering.
\item The suppression of heat fluxes, $\Delta p$-stresses, and micro-instabilities by magneto-immutable self-organization allows high-$\beta$ collisionless turbulence to evolve in a manner that is largely determined by fluid moments of the plasma particle distribution. This is remarkable given that this parameter regime of plasma physics is particularly susceptible to collisionless effects.
\item Magneto-immutability is relatively insensitive to the strength of heat fluxes, because of the dynamical reduction of the field-aligned temperature gradients $\nabla_\| \delta T_{\perp/\|}$. 
\item The suppression of $\eb\eb\bdbldot\grad\bb{u}$ is achieved through a local misalignment between $\eb$ and the eigenvectors of the $\grad\bb{u}$ tensor, rather than through a suppression of the overall rate-of-strain (e.g.,~figure~\ref{fig:anglevk2d}). Importantly, this misalignment is shown to extend beyond CGL-MHD to the hybrid-kinetic simulations of \citet{arzamasskiy23} (see figure \ref{fig:picimm}).
\item No strong cascade of ion-acoustic waves exists in high-$\beta$ collisionless turbulence; these modes can be driven effectively only at the outer scale by sonically correlated (or faster) forcing. This means that there is little dependence of the turbulence on details of the forcing in these plasmas, so long as its correlation time remains sufficiently slow. The spectrum of density fluctuations is determined by non-propagating modes, which adopt the statistics of the Alfv\'enically turbulent flow that mixes them.
\end{enumerate}
\vspace{0.5em}

In their study of magneto-immutability within the Landau-fluid CGL system, \citet{squire23} drew comparisons with the hybrid-kinetic simulations of \citet{arzamasskiy23}, finding that these comparisons reflected well on the ability of the fluid model to capture the essential dynamics of collisionless high-$\beta$ turbulence. The theory and simulations presented in this work further solidify those conclusions, especially with respect to the rate-of-strain alignment diagnostic, for which we have demonstrated qualitatively similar results between CGL-MHD and hybrid-kinetic simulations (\S\ref{sec:org}). That being said, this work has touched on new aspects of high-$\beta$ collisionless turbulence that could benefit from further comparison with well-tuned kinetic simulations. One such aspect is the microinstability scattering rate $\nu_{\rm lim}$, and how that compares to one that minimally disrupts magneto-immutability while still regulating the overall anisotropy (figure~\ref{fig:nu}). Calculations of $\nu_{\rm eff}$ from kinetic simulations of Alfv\'enic turbulence have been performed in \citet{arzamasskiy23}, finding good agreement with a Braginskii-based estimate that depends on the outer-scale rate of strain. Future investigations may then benefit from a comparison between CGL-MHD simulations with an evolving, rather than fixed, $\nu_{\rm lim}$, and kinetic simulations that capture the scale dependence of $\nu_{\rm eff}$. Not only could the effectiveness of the collisional microinstability closure be studied more directly, but it would also permit a comparison of the relationship between intermittency and the unstable fraction. Figure~\ref{fig:eos}($b$) demonstrates the highly intermittent nature of the instability volume-filling fractions in our CGL simulations, especially as compared to the smoothly varying, inferred volume-filling fractions in the passive runs. This intermittency is perhaps better quantified by the broad, non-Gaussian tails visible in the PDF of $\eb\eb \bdbldot \grad \bb{u}$ from the $\nu_{\rm lim}=20v_{\rm A}/L_\perp$ run in figure~\ref{fig:nu}($c$). In fact, such non-Gaussian tails are present in every CGL simulation, but not in the passive simulations. The most likely cause of this intermittency is then the feedback of pressure anisotropy on the flow. Understanding exactly how pressure anisotropy mediates intermittency in collisionless plasmas would be a particularly useful extension of this work, with broad implications for subjects like cosmic-ray propagation in high-$\beta$ turbulence \citep{reicherzer23}. 

Our results may serve to explain certain features of turbulence that are relevant to the understanding of plasma dynamics in the ICM of galaxy clusters. Of the 17 galaxy clusters for which \citet{zhuravleva19} and \citet{heinrich24} inferred the effective viscosity from the spectra of (de-projected) density fluctuations, all measurements pointed to a viscosity that is smaller than the Spitzer value \citep{spitzer62}. It has been suggested that this may be the result of the growth of microinstabilities, which, through their anomalous scattering of particles, effectively decrease the plasma viscosity below its Coulombic value. However, the same conditions that allow the weakly collisional, high-$\beta$ ICM to become microphysically unstable also render it susceptible to the effects of magneto-immutability. The most notable difference between turbulence in the ICM and that which is studied in this manuscript, is that ICM turbulence appears to feature outer-scale fluctuations satisfying $M_{\rm A}\doteq u/v_{\rm A}> 1$. These large-scale super-Alfv\'enic fluctuations are not guaranteed to be in critical balance, thus from our ordering they cannot be expected to self-organize \`a la magneto-immutability. In this case, microinstability scattering is expected to be driven much more strongly than in the sub-Alfv\'enic turbulence of our CGL simulations \citep[e.g.,][]{kjz22}. The results of \S\ref{sec:nu} and Appendix \ref{app:brag}, however, imply that realistic microinstability scattering does not prevent magneto-immutable self-organization from occurring when the ordering \eqref{eq:ord} is satisfied. We therefore hypothesize that viscous stresses in the ICM are suppressed through a combination of two effects. At large scales where $M_{\rm A}> 1$, microinstabilities regulate the pressure anisotropy via an enhanced anomalous scattering rate. Following the conclusions of \citet{kjz22}, this reduces the viscous scale to the Alfv\'en scale where $M_{\rm A}\sim 1$. Below this scale, the turbulence becomes critically balanced and efficient magneto-immutable self-organization can occur, resembling that seen in this manuscript, but with a larger volume-filling fraction of slowly scattering micro-instabilities. An important consequence of this would be the prediction that the cascade should continue down to kinetic scales -- a roughly ten-decade extension of the length of the inertial range of ICM turbulence from current predictions. From an observational standpoint, if turbulent fluctuations can be observed to cascade well below the Alfv\'en scale, then magneto-immutability must be active, given the limitations on micro-instability scattering rates \citep{kjz22}. Unfortunately, we expect that many of the other methods employed in this work for disentangling the two effects may be difficult to apply to observations of the ICM. For example, current observational capabilities are unable to calculate $\eb\eb\bdbldot\grad\bb{u}$ or the alignment angle as a function of scale below the Alfv\'en scale. On the other hand, large-amplitude turbulence could be studied at very high resolution within our numerical framework given a more ICM-like setup, so as to resolve the trans-Alfv\'enic transition and any other revealing features. We reserve such a study for a separate publication.

While we have made frequent reference to the importance of magneto-immutability in interpreting ICM observations, there are numerous yet-to-be-investigated ways in which magneto-immutability might affect turbulence in other high-$\beta$ plasmas. For example, \citet{kempski19} showed that incompressible turbulence driven by the magnetorotational instability (MRI; \citealt{balbus91,hawley91}) when subject to Braginskii viscosity \citep{balbus04} can self-organize so as to reduce the total (fluctuation- plus Keplerian-shear-produced) parallel rate of strain, thereby reducing the average pressure anisotropy in the plasma despite efficient angular-momentum transport by robust Reynolds and Maxwell stresses. That study could be extended using our Landau-fluid CGL-MHD model, exploring further the impact of magneto-immutability on the transport and turbulent cascade while making contact with previously published studies of collisionless MRI turbulence and transport that used Landau-fluid CGL-MHD \citep{sharma06}, hybrid-kinetic \citep{kunz16}, and pair-plasma kinetic \citep{bacchini22,sandoval24} simulations. A particularly timely extension would be to investigate the compressive part of this magnetorotationally driven cascade in its inertial range and its role in plasma heating and angular-momentum transport. For example, a recent study by \citet{kawazura22} used a set of reduced-MHD equations tailored for the `shearing sheet' to perform a local study of a magnetorotationally unstable accretion disc having a predominantly azimuthal mean magnetic field. Those authors found that compressive modes comprise a larger portion of the bulk kinetic energy than Alfv\'enic fluctuations, a result that they have recently confirmed via large-scale incompressible MHD simulations \citep{kawazura24}. They drew a link between this dominant compressive component and the heating of particles through the putative Landau damping of these fluctuations. However, their reduced model was based on the MHD equations, and it is known that the MRI in weakly collisional and collisionless plasmas (such as protogalaxies and low-luminosity accretion flows) is different than its MHD counterpart \citep[e.g.,][]{qdh02,balbus04,squire17_mri}. With the MRI driving turbulence on Alfv\'enic (rather than sonic) timescales, and given the results of figures~\ref{fig:compspec}($a$) and ($b$), it is not obvious that greater-than-unity ratios of compressive to Alfv\'enic energy could be achieved in high-$\beta$, collisionless accretion flows.

In constructing the reduced system \eqref{eq:redeq}, we made the simplifying assumption that the electrons are cold and isothermal. For the purpose of understanding how magneto-immutability behaves in various astrophysical environments where this is not necessarily the case, it is worth exploring the consequences of relaxing these assumptions. If the isothermal electron assumption held but we allowed them to be warm, say $T_{\rm e}\sim T_{\rm i}$, magneto-immutability would be unaffected so long as the density fluctuations remained smaller than $\mathcal{O}(\epsilon)$, which is the most likely case for high-$\beta$ turbulence \citep[this was tested using the same CGL-MHD code in][with the authors finding little effect]{squire23}. If however the density fluctuations were larger (approaching $\sim \epsilon\rho_0$), some modifications would be made to the signatures of immutability. First, the perpendicular pressure balance would not lead to the suppression of $\delta p_\perp$, but rather the sum of $\delta p_\perp + T_{\rm e} \delta \rho/m_{\rm i} \approx 0$. The fluctuation $\delta p_\perp/p_0$ could then remain $\mathcal{O}(\epsilon)$, and suppression of $\eb\bcdot\grad u_\|$ would not be guaranteed. On the other hand, the $u_\|$ momentum equation would still yield $\eb \bcdot \grad \Delta p \approx 0$ to leading order, thus immutability's tendency to reduce anisotropic pressure stress would be unaffected. The picture is much less straightforward if we also relax the assumption of isothermal electrons. Collisionless electrons are necessary for modelling environments such as high-$\beta$ radiatively inefficient accretion flows \citep{quataert03,sharma07}, but the effects of electron pressure anisotropy and microinstabilities on magneto-immutability are beyond the scope of this work. Future efforts on this topic would not only yield interesting results on high-$\beta$ turbulence, but also motivate cost-effective ways to model fully collisionless astrophysical plasmas, akin to what has been done between the Landau-fluid CGL approach and hybrid-kinetic particle-in-cell~\citep{squire23,arzamasskiy23}.

\vspace{2ex}
\noindent{\bf Acknowledgments}
\vspace{1ex}

The authors are grateful to A.~Bott, S.~Cowley, P.~Kempski, E.~Quataert, and A.~Schekochihin for valuable conversations, as well as the three anonymous referees for comments that helped sharpen the presentation.

\vspace{2ex}
\noindent{\bf Funding}
\vspace{1ex}

S.M. and M.W.K. were supported in part by NSF CAREER Award No.~1944972. J.S.~was supported by Rutherford Discovery Fellowship RDF-U001804 and Marsden Fund grant MFP-U002221, which are managed through the Royal Society Te Ap\={a}rangi. High-performance computing resources were provided by the PICSciE-OIT TIGRESS High Performance Computing Center and Visualization Laboratory at Princeton University.

\vspace{2ex}
\noindent{\bf Declaration of Interests}
\vspace{1ex}

The authors report no conflict of interest.

\appendix


\section{Comparison with reduced kinetic MHD}\label{app:rkmhd}

The ordering \eqref{eq:ord} that yields the reduced high-$\beta$ CGL-MHD equations is based heavily upon that employed by \citet{schekochihin09} and \citet{kunz15} to derive the reduced kinetic MHD (RKMHD) equations, often used to describe collisionless Alfv\'enic turbulence at long wavelengths. As a result, key signatures of immutability, like $\eb \bcdot \grad \Delta p$ and $\eb\eb \bdbldot \grad \bb{u}$ suppression, can in fact be obtained by applying a high-$\beta$ subsidiary ordering to RKMHD. However, other aspects of the reduced system \eqref{eq:redeq}, such as the non-local influence of $\Delta p$ and coupled Els\"asser energy cascades, cannot be recovered. To better understand the relationship between RKMHD and our reduced equations, then, in this appendix we explore just how well magneto-immutability can be recovered from RKMHD, discussing the key differences from \eqref{eq:redeq} as a result of primary versus subsidiary ordering of $\beta$.

The RKMHD equations, as given by equations (155)--(160) of \citet{schekochihin09}, are simplified by our assumptions of zero electron temperature and collision frequency to yield the following:
\begin{subequations}\label{eq:appind}
\begin{equation}
    \pD{t}{\Psi} = v_{\rm A} \eb \bcdot \grad \Phi,
\end{equation}
\begin{equation}\label{eq:appu}
    \D{t}{} \grad_\perp^2 \Phi  = v_{\rm A}\eb \bcdot \grad \grad_\perp^2 \Psi,
\end{equation}
\begin{equation}\label{eq:appkin}
    \D{t}{} \biggl( \delta f - \frac{v_\perp^2}{v_{\rm th}^2} \frac{\delta B_\|}{B_0} F_0 \biggr)  + v_\| \eb \bcdot \grad  \delta f = 0,
\end{equation}
\begin{equation}\label{eq:appn}
    \frac{\delta n}{n_0} = -\frac{1}{2n_0} \int \mathrm{d}^3\bb{v}\, \biggl( \frac{v_\perp^2}{v_{\rm th}^2}\frac{\beta}{\beta+1}  - 2 \biggr)\biggl(\delta f - \frac{v_\perp^2}{v_{\rm th}^2} \frac{\delta B_\|}{B_0} F_0 \biggr) ,
\end{equation}
\begin{equation}\label{eq:appb}
    \frac{\delta B_\|}{B_0} = -\frac{\beta}{\beta+1} \frac{1}{2n_0} \int \mathrm{d}^3\bb{v}\, \frac{v_\perp^2}{v_{\rm th}^2}\biggl(\delta f - \frac{v_\perp^2}{v_{\rm th}^2} \frac{\delta B_\|}{B_0} F_0 \biggr) ,
\end{equation}
\end{subequations}
where the full particle distribution function is $f(v_\perp,v_\|) = F_0(v_\perp,v_\|) + \delta f(v_\perp,v_\|)$ with $F_0$ a background Maxwellian distribution, $v_{\perp/\|}$ the particle velocities, and 
\begin{equation}
\bb{u}_\perp = \ez \btimes \grad_\perp \Phi \quad \mathrm{and}  \quad \frac{\delta \bb{B}_\perp}{\sqrt{4\upi\rho_0}} = \ez \btimes \grad_\perp \Psi.
\end{equation}
The convective derivative $\mathrm{d}/\mathrm{d}t$ and field-aligned gradient $\eb \bcdot \grad$ are the same as in \eqref{eq:derivs}. As a result of our assumptions, equations~\eqref{eq:appn} and \eqref{eq:appb} have both become equivalent to perpendicular pressure balance, which is more easily seen when the velocity integrals are performed:
\begin{equation}\label{eq:rkmhdpbal}
    \frac{\delta p_\perp}{p_0} = -\frac{2}{\beta} \frac{\delta B_\|}{B_0}.
\end{equation}
Just like in \eqref{eq:pbal} then, $\beta \gg 1$ dictates that $\delta p_\perp$ does not contribute to the leading-order pressure anisotropy $\Delta p$. Next we derive the parallel momentum equation, which in high-$\beta$ reduced CGL-MHD, leads to the suppression of viscosity. Taking $\int \mathrm{d}^3 \bb{v}\, m_{\rm i}v_\|$ of \eqref{eq:appkin} leads to
\begin{equation}
    \D{t}{}(\rho u_\|) = -\eb \bcdot \grad \delta p_\|.
\end{equation}
Given that RKMHD has time derivatives that scale as $\mathrm{d}_t \sim k_\| v_{\rm A}$, the leading order of this equation is $\eb \bcdot \grad \delta p_\| \approx 0$; as a result, the anisotropic pressure stress is zero to leading order once again. This should come as no surprise because the CGL-MHD parallel momentum equation is exactly the $v_\|$ moment of the full kinetic-MHD equation for $f$ (indeed, the only difference between the fluid CGL-MHD model and kinetic MHD is in the higher moments of $f$, such as $v_{\perp/\|}^2$). This is clear if we take $\int \mathrm{d}^3\bb{v}\, m_{\rm i}v_\|^2/2$ of \eqref{eq:appkin} to get the equation for the $\mu$ adiabat:
\begin{equation}\label{eq:rkmhdmu}
    \D{t}{\delta p_\perp} - 2 \frac{p_0}{B_0} \D{t}{\delta B_\|} = - \eb \bcdot \grad \biggl[ \frac{1}{2}\int \mathrm{d}^3\bb{v}\, m_{\rm i} v_\|v_\perp^2 \delta f \biggr].
\end{equation}
In our model, the right-hand-side of \eqref{eq:rkmhdmu} is approximated by the heat flux $q_\perp$ of \eqref{eq:fullhf}. If we were to ignore $q_\perp$, the reduction of $\delta p_\perp$ from \eqref{eq:rkmhdpbal} would yield $\mathrm{d}_t \delta B_\| \approx 0$, and thus $\eb \bcdot \grad u_\| \approx 0$, another signature of magneto-immutability. If the heat fluxes were nonzero, we could still achieve this reduction in $\eb \bcdot \grad u_\|$ if $q_\perp \propto \eb \bcdot \grad T_\perp$, or $q_\perp \propto \eb \bcdot \grad p_\perp$ given small density fluctuations from low Mach number forcing (as in our reduced CGL-MHD model).

However, this reduction of $\eb \bcdot \grad u_\|$ and $\eb \bcdot \grad \Delta p$ is of little consequence to the turbulent evolution, because in this model, the pressure anisotropy stress on the turbulent flow has already been ordered out. The momentum equation for $\bb{u}_\perp$ is written in terms of the potential $\Phi$ in \eqref{eq:appu}, where it is clear that only one characteristic velocity is present -- the background Alfv\'en speed. As RKMHD does not include $\beta^{-1} \sim \epsilon$ in the primary ordering, the effects of anisotropic pressure on the evolution of $\bb{u}_\perp$ are lost and cannot be recovered through a subsidiary ordering, and so there is no way to enforce $\beta \Delta \sim 1$. This subsidiary ordering would then certainly misrepresent the outer scale of our CGL-MHD turbulence simulations. However, would RKMHD suffice when the fluctuations in $\Delta p_k$ at some large wavenumber $k$ far from the outer scale become too small to satisfy $\beta \Delta_k \sim 1$? It may seem reasonable to apply the \citet{schekochihin09} RKMHD to the deep inertial range of such turbulence, however as discussed in \S\ref{sec:feats}, this would still miss a possibly important effect: $\Delta p$ can act very non-locally in $k$-space through its modification of $v_{\rm A,eff}$. Small-scale Alfv\'enic fluctuations in the high-$\beta$ reduced CGL-MHD model are subject to a background effective Alfv\'en speed set by the turbulence and consequent pressure anisotropies at the largest scales. Because of this, patches of the turbulence may evolve somewhat uniquely, or may vary in their ability to interact with cosmic rays, for example \citep{mmp21}. In this situation, the \citet{kunz15} model of RKMHD that includes pressure anisotropy in the background particle distribution could more accurately capture these effects. The slowly-evolving, large-scale motions would provide the background pressure anisotropy upon which the anisotropic RKMHD could be evolved, and as the model otherwise includes the same assumptions that lead to immutability signatures in isotropic RKMHD, it also captures the reduction of $\eb\eb\bdbldot\grad\bb{u}$ and $\eb\bcdot\grad\Delta p$ at high $\beta$.


\section{Magneto-immutability and the Braginskii viscous stress}\label{app:brag}

The initial discovery of magneto-immutability in \citet{squire19} came from an investigation of weakly collisional Braginskii-MHD turbulence, rather than turbulence with a collisionless model as is studied in this work. While the Braginskii closure for $\Delta p$ differs dramatically from that of our collisionless Landau-fluid CGL model,\footnote{Note that the Landau-fluid CGL-MHD equations, given a uniform scattering rate, do reproduce the Braginskii-MHD model in the collisional limit $\nu \gg k_\|v_{\rm th}$. This scattering rate must not only isotropize the pressures, but also suppress the heat fluxes, using the approach given in \citet{sharma03}.} the magneto-immutable suppression of the $\Delta p$-stress in our reduced CGL approach comes only from the momentum equation, which is shared by both models. It is therefore within reason to suspect that the mechanism for viscosity suppression also originates from the momentum equation in Braginskii-MHD. For that reason, in this appendix we derive the condition for viscous stress reduction in Braginskii-MHD by assuming that the cause is the same as that leading to \eqref{eq:gradtprl} (i.e., $\eb\bcdot\grad\delta T_\parallel = 0$ to leading order), essentially obtaining a threshold for realizing magneto-immutable behaviour in the weakly collisional limit. 

Unlike the CGL-MHD model, Braginskii-MHD does not evolve the pressure anisotropy directly from conservation of the double adiabatic invariants. Instead, it assumes that the rate of scattering is sufficiently rapid ($\nu \gg k_\|v_{\rm th}$) that a balance is struck between production of anisotropy via changes in $B$ and $\rho$ and its depletion through pitch-angle scattering. As a result, the leading-order perturbation to $p$ is isotropic, and $\Delta p$ only arises at next order in $k_\|v_{\rm th}/\nu$. This allows the double-adiabatic equations \eqref{eq:cglpprp} and \eqref{eq:cglpprl} to be replaced with \citep{braginskii65}:\footnote{Another regime of Braginskii-MHD can be obtained by instead assuming that $\nu \gg k_\|v_{\rm A}$, which is a considerably weaker criterion than $\nu \gg k_\| v_{\rm th}$ at high $\beta$. In this limit, however, the heat fluxes are in the collisionless regime and not ordered out of the system, and so they must be taken into account. It may be the case that their effects are suppressed by magneto-immutability regardless, although further investigation would be needed to confirm such a conclusion.}
\begin{equation}\label{eq:brag}
    \D{t}{} \ln \frac{p}{\rho^{5/3}} = \frac{2}{\nu}\biggl[\biggl(\eb\eb-\frac{\msb{I}}{3}\biggr)\bdbldot\grad\bb{u}\biggr]^2 \qquad \mathrm{and} \qquad \Delta p = \frac{3p}{\nu}\biggl(\eb\eb-\frac{\msb{I}}{3}\biggr)\bdbldot\grad\bb{u}.
\end{equation}

To achieve suppression of parallel viscous forces through magneto-immutability in the same manner as realized in our reduced CGL-MHD model, we seek to ensure that the leading order of the $u_\|$ momentum equation becomes $\eb\bcdot\grad\Delta p \approx 0$. As with with the reduced CGL-MHD model, we make the simplifying assumption that both density fluctuations and $\grad\bcdot\bb{u}$ are negligible, and apply the ordering \eqref{eq:ordmhd} to the Braginskii-MHD equations. Note, however, that we make no assumption regarding the size of $\beta$, as we will instead derive a $\beta$-dependent criterion for magneto-immutability to take effect in Braginskii-MHD. Therefore, we will simply have to assume that $\Delta p$ cannot be neglected in the momentum equation, so that it can later inform us of how large $\beta$ needs to be in order to suppress $\Delta p$'s parallel gradients (otherwise magneto-immutability would be impossible to recover). Reducing the momentum equation produces the following leading-order equation for $u_\perp$, which still describes pressure balance, but in this case the isotropic pressure dominates to leading order
\begin{equation}
    \frac{\delta p}{p_0} \approx -\frac{\beta}{2}\frac{\delta B_\|}{B_0},
\end{equation}
only in this case it is struck with the isotropic pressure perturbation. The leading order of the parallel momentum equation can then be written as
\begin{equation}\label{eq:appprlmom}
    \rho_0 \D{t}{u_\|} \approx -\eb\bcdot\grad\Delta p + B_0\eb\bcdot\grad\delta B_\|.
\end{equation}
To compare the sizes of each term, we now substitute for $\Delta p$ the weakly collisional closure \eqref{eq:brag}, which, if ordered according to \eqref{eq:ordmhd}, yields $\Delta p \approx (3p_0/\nu) \eb \bcdot \grad u_\|$. Substituting this into \eqref{eq:appprlmom} yields
\begin{equation}\label{eq:bragprlmom}
    \rho_0 \D{t}{u_\|} \approx -\frac{3\beta}{2\nu}B_0^2\eb\bcdot\grad(\eb\bcdot\grad u_\|) + B_0\eb\bcdot\grad\delta B_\|.
\end{equation}
In \eqref{eq:bragprlmom}, the $\Delta p$-stress takes on its familiar viscous form. The left-hand side and the final term on the right-hand-side are both of order ${\sim}\epsilon k_\| B_0^2$, while the viscous term is ${\sim}\epsilon k_\|B_0^2(\beta k_\|v_{\rm A}/\nu)$. Thus, if the viscosity is to dominate, forcing the plasma to self-organize in order to avoid it, we require
\begin{equation}\label{eq:bragcond}
    \beta \frac{k_\|v_{\rm A}}{\nu} \gg 1.
\end{equation}
If this criterion is met, then $\eb\bcdot\grad\Delta p$ will be suppressed and the Alfv\'enic cascade will not be strongly damped by viscous stress. By design, this is precisely the regime within which magneto-immutability was studied in \citet{squire19} and \citet{squire23}. Equation \eqref{eq:bragcond} also implies that magneto-immutability in Braginskii-MHD, unlike its CGL-MHD counterpart, \textit{is} scale dependent. Consider a scenario in which outer-scale motions are too slow to produce $\Delta p$ faster than it can be eroded by the scattering rate, and the criterion \eqref{eq:bragcond} is not satisfied. As the cascade progresses, the eddy turnover times get shorter at smaller scales, and the production of pressure anisotropy occurs at a faster rate. Eventually, when the generation of pressure anisotropy occurs on timescales small enough that it competes with the scattering, magneto-immutability can step in to regulate its magnitude. At some point, however, the scales will become sufficiently small that the collisional assumption of $\nu/k_\|v_{\rm th} \gg 1$ becomes inadequate and a fully collisionless model must be used.


\section{High-$\beta$ reduced CGL with large density fluctuations}\label{app:dens}

One of the fundamental assumptions we make within this work is that density fluctuations are $\mathcal{O}(\epsilon^2)$ or smaller. Physically, this is motivated by the difficulty of driving both high Mach number and sonically correlated turbulence in astrophysical high-$\beta$ plasmas. Indeed for all of the simulations performed within the scope of this work, in no circumstances did $\delta \rho$ exceed $\epsilon^2 \rho_0$, a necessary condition for obtaining the excellent agreement between our predictions and the simulation results. Nonetheless, it is worthwhile to at least consider the consequences of $\delta \rho \sim \epsilon^{3/2}\rho_0$ or $\epsilon\rho_0$, and how that would affect the conclusions we have reached so far.

We begin with $\delta \rho \sim \epsilon^{3/2}\rho_0$, after which $\delta \rho \sim \epsilon\rho_0$ is a relatively simple extension. All non-$\delta \rho$ oriented aspects of the ordering \eqref{eq:ord} can be used once again, although we will drop $\delta T_{\perp/\|}$ in favor of $\delta p_{\perp/\|}$ given the enhancement of $\delta \rho$. Among other things, this means that the pressure balance of equation \eqref{eq:pbal} becomes 
\begin{equation}\label{eq:pprppbal}
    \frac{\delta p_\perp^{(2)}}{p_0} = -\frac{2}{\beta}\frac{\delta B_\|^{(1)}}{B_0} \qquad \mathrm{and} \qquad
    \frac{\delta p_\perp^{(5/2)}}{p_0} = -\frac{2}{\beta}\frac{\delta B_\|^{(3/2)}}{B_0},
\end{equation}
and equations \eqref{eq:gradtprl} and \eqref{eq:gradtprl2} are the same but with $\delta p_\|$ swapped for $\delta T_\|$:
\begin{subequations}
\begin{equation}\label{eq:gradpprl}
    \eb^{(0)} \bcdot \grad \delta p_\|^{(1)} = (\eb \bcdot \grad \delta p_\|)^{(1)} = 0,
\end{equation}
\begin{equation}\label{eq:gradpprl2}
    \eb^{(0)} \bcdot \grad \delta p_\|^{(3/2)} + \frac{\delta \bb{B}_\perp^{(3/2)}}{B_0} \bcdot \grad_\perp \delta p_\|^{(1)} = (\eb \bcdot \grad \delta p_\|)^{(3/2)} = 0.
\end{equation}
\end{subequations}
This shows that the suppression of the anisotropic pressure-stress, being independent of the magnitude of density fluctuations, is a particularly robust aspect of immutability. Continuing, the equations that evolve $\delta \bb{B}_\perp$ and $\bb{u}_\perp$ are unaffected, as is the continuity equation. Note that although we are allowing the density fluctuations to be larger, we will still employ the assumption that $\mathrm{d}_t \delta \rho^{(3/2)}$ is negligible, owed to the fact that these density fluctuations are still likely the product of non-propagating modes. As such, in evaluating the $p_\perp$ and $p_\|$ equations, we can approximate
\begin{equation}
\D{t}{} \ln \frac{p_\perp}{\rho B} \approx -\D{t}{} \ln B \qquad \mathrm{and} \qquad
\D{t}{} \ln \frac{p_\|B^2}{\rho^3} \approx \D{t}{} \ln p_\|B^2,
\end{equation}
where we have again used the fact that $p_\perp$ has no $\mathcal{O}(\epsilon)$ perturbation. Starting with equation \eqref{eq:cglpprp}, the heat flux $q_\perp$ is still 0 at order $\epsilon^{1/2}$, since the density fluctuations are only $\mathcal{O}(\epsilon^{3/2})$ and $\eb \bcdot \grad \delta p_\|^{(1,3/2)} = 0$. However, at order $\epsilon$ where left-hand-side of \eqref{eq:cglpprp} first appears, we have
\begin{equation}\label{eq:bdpprp}
    \D{t}{} \frac{\delta B_\|^{(1)}}{B_0} = \frac{v_{\rm th}}{\sqrt{\upi}|k_\||} (\eb^{(0)}\bcdot \grad)^2 \frac{\delta \rho^{(3/2)}}{\rho_0}.
\end{equation}
As a result, instead of finding that $\hat{b}\bcdot \grad u_\| = 0$ by comparing this with parallel induction (as we found when $\delta \rho \lesssim \epsilon^2 \rho_0$), we find
\begin{equation}
     \eb^{(0)}\bcdot \grad \biggl(u_\|^{(1)} - \frac{v_{\rm th}}{\sqrt{\upi}|k_\||} \eb^{(0)}\bcdot \grad \frac{\delta \rho^{(3/2)}}{\rho_0} \biggr) = 0.
\end{equation}
Note that we cannot use this equation to fully determine $u_\|$, given that taking $\delta \rho$ to be smaller would imply $u_\|^{(1)} = 0$, which is not the same as the misalignment we predicted and measured in \S\ref{sec:org}. The equation for $\delta p_\|^{(1)}$ is obtained with ease given \eqref{eq:bdpprp}:
\begin{equation}\label{eq:bdpprl}
    \D{t}{} \frac{\delta p_\|^{(1)}}{p_0} = -\frac{4v_{\rm th}}{\sqrt{\upi}|k_\||} (\eb^{(0)}\bcdot \grad)^2 \frac{\delta \rho^{(3/2)}}{\rho_0}.
\end{equation}
Therefore, although the suppression of the $\Delta p$-stress is preserved by the increase in density fluctuation amplitude, the misalignment of $\eb\eb\bdbldot\grad\bb{u}$ and passive advection of $\Delta p$ are not necessarily. Instead, the anisotropic pressure fluctuations are expected to collisionlessly damp as they are mixed passively by the Alfv\'enic turbulence. The exact rate of damping would then be determined by the passively advected density fluctuations that are evolved according to $\mathrm{d}_t\delta \rho^{(3/2)} = 0$.

The extension of the above to $\delta \rho \sim \epsilon\rho_0$ is rather simple, with only the following modifications: Instead of the $\mathcal{O}(\epsilon^{1/2})$ contribution to the heat fluxes $q_{\perp/\|}$ being 0, they will remain nonzero due to the density perturbation, with \eqref{eq:cglpprl} and \eqref{eq:cglpprp} becoming $\eb^{(0)}\bcdot \grad \delta \rho^{(1)}$ to leading order. Following this, \eqref{eq:bdpprl} and \eqref{eq:bdpprp} remain the same, but it is less clear whether the assumption $\mathrm{d}_t \delta \rho = 0$ can be applied to both $\delta \rho^{(1)}$ and $\delta \rho^{(3/2)}$, or if that can only be said of the leading order. This may still be the case as we expect the only other source of density fluctuations to be Alfv\'en wave nonlinearities, which should appear at order $\epsilon^2$, rather than $\epsilon^{3/2}$. However, this is also reliant on discerning to how many orders ion-acoustic wave fluctuations can be ignored, given that some weak mixing may still occur.


\begin{thebibliography}{65}
\expandafter\ifx\csname natexlab\endcsname\relax\def\natexlab#1{#1}\fi

\bibitem[{Arzamasskiy} {\em et~al.\/}(2023){Arzamasskiy}, {Kunz}, {Squire}, {Quataert} \& {Schekochihin}]{arzamasskiy23}
{\sc {Arzamasskiy}, L., {Kunz}, M.W., {Squire}, J., {Quataert}, E. \& {Schekochihin}, A.A.} 2023 {Kinetic turbulence in collisionless high-{\ensuremath{\beta}} plasmas}. {\em \prx\/} {\bf 13}~(2), 021014.

\bibitem[{Bacchini} {\em et~al.\/}(2022){Bacchini}, {Arzamasskiy}, {Zhdankin}, {Werner}, {Begelman} \& {Uzdensky}]{bacchini22}
{\sc {Bacchini}, F., {Arzamasskiy}, L., {Zhdankin}, V., {Werner}, G.R., {Begelman}, M.C. \& {Uzdensky}, D.A.} 2022 {Fully kinetic shearing-box simulations of magnetorotational turbulence in 2D and 3D. I. Pair plasmas}. {\em \apj\/} {\bf 938}~(1), 86.

\bibitem[{Balbus}(2004)]{balbus04}
{\sc {Balbus}, S.A.} 2004 {Viscous shear instability in weakly magnetized, dilute plasmas}. {\em \apj\/} {\bf 616}, 857--864.

\bibitem[{Balbus} \& {Hawley}(1991)]{balbus91}
{\sc {Balbus}, S.A. \& {Hawley}, J.F.} 1991 {A powerful local shear instability in weakly magnetized disks I. Linear analysis}. {\em \apj\/} {\bf 376}, 214.

\bibitem[{Barnes}(1966)]{barnes66}
{\sc {Barnes}, A.} 1966 {Collisionless damping of hydromagnetic waves}. {\em \pof\/} {\bf 9}, 1483.

\bibitem[{Biskamp}(2003)]{biskamp03}
{\sc {Biskamp}, D.} 2003 {\em {Magnetohydrodynamic turbulence}\/}. Cambridge University Press.

\bibitem[{Bott} {\em et~al.\/}(2021){Bott}, {Arzamasskiy}, {Kunz}, {Quataert} \& {Squire}]{bott21}
{\sc {Bott}, A.F.A., {Arzamasskiy}, L., {Kunz}, M.W., {Quataert}, E. \& {Squire}, J.} 2021 {Adaptive critical balance and firehose instability in an expanding, turbulent, collisionless plasma}. {\em \apjl\/} {\bf 922}~(2), L35.

\bibitem[{Bott} {\em et~al.\/}(2024){Bott}, {Cowley} \& {Schekochihin}]{bott24}
{\sc {Bott}, A.F.A., {Cowley}, S.C. \& {Schekochihin}, A.A.} 2024 {Kinetic stability of Chapman-Enskog plasmas}. {\em \jpp\/} {\bf 90}~(2), 975900207.

\bibitem[{Braginskii}(1965)]{braginskii65}
{\sc {Braginskii}, S.I.} 1965 {Transport processes in a plasma}. {\em \ropp\/} {\bf 1}, 205.

\bibitem[{Brandenburg} \& {Subramanian}(2005)]{brandenburg05}
{\sc {Brandenburg}, A. \& {Subramanian}, K.} 2005 {Astrophysical magnetic fields and nonlinear dynamo theory}. {\em \physrep\/} {\bf 417}~(1-4), 1--209.

\bibitem[{Chen} {\em et~al.\/}(2011){Chen}, {Mallet}, {Yousef}, {Schekochihin} \& {Horbury}]{chen2011}
{\sc {Chen}, C.H.K., {Mallet}, A., {Yousef}, T.A., {Schekochihin}, A.A. \& {Horbury}, T.S.} 2011 {Anisotropy of Alfv{\'e}nic turbulence in the solar wind and numerical simulations}. {\em \mnras\/} {\bf 415}~(4), 3219--3226.

\bibitem[{Chew} {\em et~al.\/}(1956){Chew}, {Goldberger} \& {Low}]{cgl56}
{\sc {Chew}, G.F., {Goldberger}, M.L. \& {Low}, F.E.} 1956 {The Boltzmann equation and the one-fluid hydromagnetic equations in the absence of particle collisions}. {\em Proc.~Roy.~Soc.~London Ser.~A\/} {\bf 236}, 112.

\bibitem[{Cho} \& {Lazarian}(2009)]{cho09}
{\sc {Cho}, J. \& {Lazarian}, A.} 2009 {Simulations of electron magnetohydrodynamic turbulence}. {\em \apj\/} {\bf 701}~(1), 236--252.

\bibitem[{Cowie} \& {McKee}(1977)]{cowie77}
{\sc {Cowie}, L.L. \& {McKee}, C.F.} 1977 {The evaporation of spherical clouds in a hot gas. I. Classical and saturated mass loss rates.} {\em \apj\/} {\bf 211}, 135--146.

\bibitem[{Els\"asser}(1950)]{elsasser50}
{\sc {Els\"asser}, W.M.} 1950 {The hydromagnetic equations}. {\em \pr\/} {\bf 79}~(1), 183.

\bibitem[{Goldreich} \& {Sridhar}(1995)]{gs95}
{\sc {Goldreich}, P. \& {Sridhar}, S.} 1995 {Toward a theory of interstellar turbulence. 2: Strong Alfv\'{e}nic turbulence}. {\em \apj\/} {\bf 438}, 763.

\bibitem[{Goldstein} {\em et~al.\/}(1995){Goldstein}, {Roberts} \& {Matthaeus}]{goldstein95}
{\sc {Goldstein}, M.L., {Roberts}, D.A. \& {Matthaeus}, W.H.} 1995 {Magnetohydrodynamic turbulence in the solar wind}. {\em \araa\/} {\bf 33}, 283--326.

\bibitem[{Grete} {\em et~al.\/}(2017){Grete}, {O'Shea}, {Beckwith}, {Schmidt} \& {Christlieb}]{grete17}
{\sc {Grete}, P., {O'Shea}, B.~W., {Beckwith}, K., {Schmidt}, W. \& {Christlieb}, A.} 2017 {Energy transfer in compressible magnetohydrodynamic turbulence}. {\em \pop\/} {\bf 24}~(9), 092311.

\bibitem[{Hasegawa}(1969)]{hasegawa69}
{\sc {Hasegawa}, A.} 1969 {Drift mirror instability of the magnetosphere.} {\em \pof\/} {\bf 12}, 2642.

\bibitem[{Hawley} \& {Balbus}(1991)]{hawley91}
{\sc {Hawley}, J.F. \& {Balbus}, S.A.} 1991 {A powerful local shear instability in weakly magnetized disks II. Nonlinear evolution}. {\em \apj\/} {\bf 376}, 223.

\bibitem[{Heinrich} {\em et~al.\/}(2024){Heinrich}, {Zhuravleva}, {Zhang}, {Churazov}, {Forman} \& {van Weeren}]{heinrich24}
{\sc {Heinrich}, A., {Zhuravleva}, I., {Zhang}, C., {Churazov}, E., {Forman}, W. \& {van Weeren}, R.J.} 2024 {Merger-driven multiscale ICM density perturbations: testing cosmological simulations and constraining plasma physics}. {\em \mnras\/} {\bf 528}~(4), 7274--7299.

\bibitem[{Hellinger} \& {Matsumoto}(2000)]{hm00}
{\sc {Hellinger}, P. \& {Matsumoto}, H.} 2000 {New kinetic instability: oblique Alfv\'{e}n fire hose}. {\em \jgr\/} {\bf 105}, 10519.

\bibitem[{Hollweg}(1974)]{hollweg74}
{\sc {Hollweg}, Joseph~V.} 1974 {On electron heat conduction in the solar wind}. {\em \jgr\/} {\bf 79}~(25), 3845.

\bibitem[{Kawazura} \& {Kimura}(2024)]{kawazura24}
{\sc {Kawazura}, Y. \& {Kimura}, S.S.} 2024 {Inertial range of magnetorotational turbulence}. {\em arXiv e-prints\/} .

\bibitem[{Kawazura} {\em et~al.\/}(2022){Kawazura}, {Schekochihin}, {Barnes}, {Dorland} \& {Balbus}]{kawazura22}
{\sc {Kawazura}, Y., {Schekochihin}, A.A., {Barnes}, M., {Dorland}, W. \& {Balbus}, S.A.} 2022 {Energy partition between Alfv{\'e}nic and compressive fluctuations in magnetorotational turbulence with near-azimuthal mean magnetic field}. {\em \jpp\/} {\bf 88}~(3), 905880311.

\bibitem[{Kazantsev}(1968)]{kazantsev68}
{\sc {Kazantsev}, A.P.} 1968 {Enhancement of a magnetic field by a conducting fluid}. {\em Sov.~J.~Exp.~Theor.~Phys.\/} {\bf 26}, 1031.

\bibitem[{Kempski} {\em et~al.\/}(2019){Kempski}, {Quataert}, {Squire} \& {Kunz}]{kempski19}
{\sc {Kempski}, P., {Quataert}, E., {Squire}, J. \& {Kunz}, M.W.} 2019 {Shearing-box simulations of MRI-driven turbulence in weakly collisional accretion discs}. {\em \mnras\/} {\bf 486}~(3), 4013--4029.

\bibitem[{Kulsrud}(1983)]{kulsrud83}
{\sc {Kulsrud}, R.M.} 1983 {MHD description of plasma}. In {\em Basic Plasma Physics: Selected Chapters, Handbook of Plasma Physics, Volume 1\/} (ed. {A.A.~Galeev \& R.N.~Sudan}), p.~1.

\bibitem[{Kunz} {\em et~al.\/}(2022){Kunz}, {Jones} \& {Zhuravleva}]{kjz22}
{\sc {Kunz}, M.W., {Jones}, T.W. \& {Zhuravleva}, I.} 2022 {Plasma physics of the intracluster medium}. In {\em Handbook of X-ray and Gamma-ray Astrophysics\/}, p.~56. Springer Nature Singapore.

\bibitem[{Kunz} {\em et~al.\/}(2015){Kunz}, {Schekochihin}, {Chen}, {Abel} \& {Cowley}]{kunz15}
{\sc {Kunz}, M.W., {Schekochihin}, A.A., {Chen}, C.H.K., {Abel}, I.G. \& {Cowley}, S.C.} 2015 {Inertial-range kinetic turbulence in pressure-anisotropic astrophysical plasmas}. {\em \jpp\/} {\bf 81}, 325810501.

\bibitem[{Kunz} {\em et~al.\/}(2020){Kunz}, {Squire}, {Schekochihin} \& {Quataert}]{kunz20}
{\sc {Kunz}, M.W., {Squire}, J., {Schekochihin}, A.A. \& {Quataert}, E.} 2020 {Self-sustaining sound in collisionless, high-{\ensuremath{\beta}} plasma}. {\em \jpp\/} {\bf 86}, 905860603.

\bibitem[{Kunz} {\em et~al.\/}(2014){Kunz}, {Stone} \& {Bai}]{kunz14}
{\sc {Kunz}, M.W., {Stone}, J.M. \& {Bai}, X.-N.} 2014 {Pegasus: a new hybrid-kinetic particle-in-cell code for astrophysical plasma dynamics}. {\em \jcp\/} {\bf 259}, 154.

\bibitem[{Kunz} {\em et~al.\/}(2016){Kunz}, {Stone} \& {Quataert}]{kunz16}
{\sc {Kunz}, M.W., {Stone}, J.M. \& {Quataert}, E.} 2016 {Magnetorotational turbulence and dynamo in a collisionless plasma}. {\em \prl\/} {\bf 117}~(23), 235101.

\bibitem[{Ley} {\em et~al.\/}(2024){Ley}, {Zweibel}, {Miller} \& {Riquelme}]{ley24}
{\sc {Ley}, F., {Zweibel}, E.~G., {Miller}, D. \& {Riquelme}, M.} 2024 {Secondary whistler and ion-cyclotron instabilities driven by mirror modes in galaxy clusters}. {\em \apj\/} {\bf 965}~(2), 155.

\bibitem[{Majeski} \& {Kunz}(2024)]{mk23}
{\sc {Majeski}, S. \& {Kunz}, M.W.} 2024 {On hydromagnetic wave interactions in collisionless, high-{\ensuremath{\beta}} plasmas}. {\em \jpp\/} {\bf 90}~(1), 535900101.

\bibitem[{Majeski} {\em et~al.\/}(2023){Majeski}, {Kunz} \& {Squire}]{mks23}
{\sc {Majeski}, S., {Kunz}, M.W. \& {Squire}, J.} 2023 {Microphysically modified magnetosonic modes in collisionless, high-{\ensuremath{\beta}} plasmas}. {\em \jpp\/} {\bf 89}~(3), 905890303.

\bibitem[Marcowith {\em et~al.\/}(2021)Marcowith, van Marle \& Plotnikov]{mmp21}
{\sc Marcowith, A., van Marle, A.J. \& Plotnikov, I.} 2021 {The cosmic ray-driven streaming instability in astrophysical and space plasmas}. {\em \pop\/} {\bf 28}~(8), 080601.

\bibitem[{Marsch}(2006)]{marsch06}
{\sc {Marsch}, E.} 2006 {Kinetic physics of the solar corona and solar wind}. {\em Liv.~Rev.~Sol.~Phys.\/} {\bf 3}~(1), 1.

\bibitem[{Meyrand} {\em et~al.\/}(2019){Meyrand}, {Kanekar}, {Dorland} \& {Schekochihin}]{meyrand19}
{\sc {Meyrand}, R., {Kanekar}, A., {Dorland}, W. \& {Schekochihin}, A.A.} 2019 {Fluidization of collisionless plasma turbulence}. {\em \pnas\/} {\bf 116}~(4), 1185--1194.

\bibitem[{Quataert}(2003)]{quataert03}
{\sc {Quataert}, E.} 2003 {Radiatively inefficient accretion flow models of Sgr~A$^\ast$}. {\em Astron.~Nachrichten Suppl.\/} {\bf 324}, 435--443.

\bibitem[{Quataert} {\em et~al.\/}(2002){Quataert}, {Dorland} \& {Hammett}]{qdh02}
{\sc {Quataert}, E., {Dorland}, W. \& {Hammett}, G.W.} 2002 {The magnetorotational instability in a collisionless plasma}. {\em \apj\/} {\bf 577}, 524--533.

\bibitem[{Reichherzer} {\em et~al.\/}(2023){Reichherzer}, {Bott}, {Ewart}, {Gregori}, {Kempski}, {Kunz} \& {Schekochihin}]{reicherzer23}
{\sc {Reichherzer}, P., {Bott}, A.F.A., {Ewart}, R.J., {Gregori}, G., {Kempski}, P., {Kunz}, M.W. \& {Schekochihin}, A.A.} 2023 {Efficient micromirror confinement of sub-TeV cosmic rays in galaxy clusters}. {\em arXiv e-prints\/} p. arXiv:2311.01497.

\bibitem[{Riquelme} {\em et~al.\/}(2018){Riquelme}, {Quataert} \& {Verscharen}]{riquelme18}
{\sc {Riquelme}, M., {Quataert}, E. \& {Verscharen}, D.} 2018 {PIC simulations of velocity-space instabilities in a decreasing magnetic field: Viscosity and thermal conduction}. {\em \apj\/} {\bf 854}~(2), 132.

\bibitem[{Sandoval} {\em et~al.\/}(2024){Sandoval}, {Riquelme}, {Spitkovsky} \& {Bacchini}]{sandoval24}
{\sc {Sandoval}, A., {Riquelme}, M., {Spitkovsky}, A. \& {Bacchini}, F.} 2024 {Particle-in-cell simulations of the magnetorotational instability in stratified shearing boxes}. {\em \mnras\/} {\bf 530}~(2), 1866--1884.

\bibitem[{Santos-Lima} {\em et~al.\/}(2014){Santos-Lima}, {de Gouveia Dal Pino}, {Kowal}, {Falceta-Gon{\c{c}}alves}, {Lazarian} \& {Nakwacki}]{santoslima14}
{\sc {Santos-Lima}, R., {de Gouveia Dal Pino}, E.M., {Kowal}, G., {Falceta-Gon{\c{c}}alves}, D., {Lazarian}, A. \& {Nakwacki}, M.S.} 2014 {Magnetic field amplification and evolution in turbulent collisionless magnetohydrodynamics: an application to the intracluster medium}. {\em \apj\/} {\bf 781}~(2), 84.

\bibitem[{Schekochihin} \& {Cowley}(2007)]{sc07}
{\sc {Schekochihin}, A.A. \& {Cowley}, S.C.} 2007 {\em Turbulence and magnetic fields in astrophysical plasmas\/}, pp. 85--115. Dordrecht: Springer Netherlands.

\bibitem[{Schekochihin} {\em et~al.\/}(2009){Schekochihin}, {Cowley}, {Dorland}, {Hammett}, {Howes}, {Quataert} \& {Tatsuno}]{schekochihin09}
{\sc {Schekochihin}, A.A., {Cowley}, S.C., {Dorland}, W., {Hammett}, G.W., {Howes}, G.G., {Quataert}, E. \& {Tatsuno}, T.} 2009 {Astrophysical gyrokinetics: kinetic and fluid turbulent cascades in magnetized weakly collisional plasmas}. {\em \apjs\/} {\bf 182}, 310.

\bibitem[{Schekochihin} {\em et~al.\/}(2010){Schekochihin}, {Cowley}, {Rincon} \& {Rosin}]{schekochihin10}
{\sc {Schekochihin}, A.A., {Cowley}, S.C., {Rincon}, F. \& {Rosin}, M.S.} 2010 {Magnetofluid dynamics of magnetized cosmic plasma: firehose and gyrothermal instabilities}. {\em \mnras\/} {\bf 405}~(1), 291--300.

\bibitem[{Schekochihin} {\em et~al.\/}(2004){Schekochihin}, {Cowley}, {Taylor}, {Maron} \& {McWilliams}]{schekochihin04}
{\sc {Schekochihin}, A.A., {Cowley}, S.C., {Taylor}, S.F., {Maron}, J.L. \& {McWilliams}, J.C.} 2004 {Simulations of the small-scale turbulent dynamo}. {\em \apj\/} {\bf 612}~(1), 276--307.

\bibitem[{Sharma} {\em et~al.\/}(2003){Sharma}, {Hammett} \& {Quataert}]{sharma03}
{\sc {Sharma}, P., {Hammett}, G.W. \& {Quataert}, E.} 2003 {Transition from collisionless to collisional magnetorotational instability}. {\em \apj\/} {\bf 596}~(2), 1121--1130.

\bibitem[{Sharma} {\em et~al.\/}(2006){Sharma}, {Hammett}, {Quataert} \& {Stone}]{sharma06}
{\sc {Sharma}, P., {Hammett}, G.W., {Quataert}, E. \& {Stone}, J.M.} 2006 {Shearing box simulations of the MRI in a collisionless plasma}. {\em \apj\/} {\bf 637}, 952.

\bibitem[{Sharma} {\em et~al.\/}(2007){Sharma}, {Quataert}, {Hammett} \& {Stone}]{sharma07}
{\sc {Sharma}, P., {Quataert}, E., {Hammett}, G.W. \& {Stone}, J.M.} 2007 {Electron heating in hot accretion flows}. {\em \apj\/} {\bf 667}, 714.

\bibitem[{Snyder} {\em et~al.\/}(1997){Snyder}, {Hammett} \& {Dorland}]{shd97}
{\sc {Snyder}, P.B., {Hammett}, G.W. \& {Dorland}, W.} 1997 {Landau fluid models of collisionless magnetohydrodynamics}. {\em \pop\/} {\bf 4}, 3974.

\bibitem[{Spitzer}(1962)]{spitzer62}
{\sc {Spitzer}, L.} 1962 {\em {Physics of Fully Ionized Gases}\/}. Dover.

\bibitem[Squire {\em et~al.\/}(2023)Squire, Kunz, Arzamasskiy, Johnston, Quataert \& Schekochihin]{squire23}
{\sc Squire, J., Kunz, M.W., Arzamasskiy, L., Johnston, Z., Quataert, E. \& Schekochihin, A.A.} 2023 Pressure anisotropy and viscous heating in weakly collisional plasma turbulence. {\em \jpp\/} {\bf 89}, 905890417.

\bibitem[{Squire} {\em et~al.\/}(2017{\natexlab{{\em a\/}}}){Squire}, {Kunz}, {Quataert} \& {Schekochihin}]{squire17num}
{\sc {Squire}, J., {Kunz}, M.W., {Quataert}, E. \& {Schekochihin}, A.A.} 2017{\natexlab{{\em a\/}}} {Kinetic simulations of the interruption of large-amplitude shear-Alfv{\'e}n waves in a high-{$\beta$} plasma}. {\em \prl\/} {\bf 119}~(15), 155101.

\bibitem[{Squire} {\em et~al.\/}(2017{\natexlab{{\em b\/}}}){Squire}, {Quataert} \& {Kunz}]{squire17_mri}
{\sc {Squire}, J., {Quataert}, E. \& {Kunz}, M.W.} 2017{\natexlab{{\em b\/}}} {Pressure-anisotropy-induced nonlinearities in the kinetic magnetorotational instability}. {\em \jpp\/} {\bf 83}~(6), 905830613.

\bibitem[{Squire} {\em et~al.\/}(2016){Squire}, {Quataert} \& {Schekochihin}]{squire16}
{\sc {Squire}, J., {Quataert}, E. \& {Schekochihin}, A.A.} 2016 {A stringent limit on the amplitude of Alfv{{\'e}}nic perturbations in high-beta low-collisionality plasmas}. {\em \apj\/} {\bf 830}, L25.

\bibitem[{Squire} {\em et~al.\/}(2017{\natexlab{{\em c\/}}}){Squire}, {Schekochihin} \& {Quataert}]{squire17}
{\sc {Squire}, J., {Schekochihin}, A.A. \& {Quataert}, E.} 2017{\natexlab{{\em c\/}}} {Amplitude limits and nonlinear damping of shear-Alfv{\'e}n waves in high-beta low-collisionality plasmas}. {\em \njp\/} {\bf 19}, 055005.

\bibitem[{Squire} {\em et~al.\/}(2019){Squire}, {Schekochihin}, {Quataert} \& {Kunz}]{squire19}
{\sc {Squire}, J., {Schekochihin}, A.A., {Quataert}, E. \& {Kunz}, M.W.} 2019 {Magneto-immutable turbulence in weakly collisional plasmas}. {\em \jpp\/} {\bf 85}~(1), 905850114.

\bibitem[{St-Onge} {\em et~al.\/}(2020){St-Onge}, {Kunz}, {Squire} \& {Schekochihin}]{St-Onge2020}
{\sc {St-Onge}, D.A., {Kunz}, M.W., {Squire}, J. \& {Schekochihin}, A.A.} 2020 {Fluctuation dynamo in a weakly collisional plasma}. {\em \jpp\/} {\bf 86}~(5), 905860503.

\bibitem[{Stone} {\em et~al.\/}(2020){Stone}, {Tomida}, {White} \& {Felker}]{stone20}
{\sc {Stone}, J.M., {Tomida}, K., {White}, C.J. \& {Felker}, K.G.} 2020 {The Athena++ adaptive mesh refinement framework: design and magnetohydrodynamic solvers}. {\em \apjs\/} {\bf 249}~(1), 4.

\bibitem[{Uhlenbeck} \& {Ornstein}(1930)]{uhlenbeck30}
{\sc {Uhlenbeck}, G.E. \& {Ornstein}, L.S.} 1930 {On the theory of the Brownian motion}. {\em \pr\/} {\bf 36}~(5), 823--841.

\bibitem[{Zank} \& {Matthaeus}(1992)]{zm92}
{\sc {Zank}, G.P. \& {Matthaeus}, W.H.} 1992 {The equations of reduced magnetohydrodynamics}. {\em \jpp\/} {\bf 48}, 85--100.

\bibitem[{Zhuravleva} {\em et~al.\/}(2019){Zhuravleva}, {Churazov}, {Schekochihin}, {Allen}, {Vikhlinin} \& {Werner}]{zhuravleva19}
{\sc {Zhuravleva}, I., {Churazov}, E., {Schekochihin}, A.A., {Allen}, S.W., {Vikhlinin}, A. \& {Werner}, N.} 2019 {Suppressed effective viscosity in the bulk intergalactic plasma}. {\em Nature Astro.\/} {\bf 3}, 832--837.

\end{thebibliography}

\end{document}